\documentclass[fleqn,usenatbib]{mnras}

\usepackage{newtxtext,newtxmath}

\usepackage[T1]{fontenc}

\DeclareRobustCommand{\VAN}[3]{#2}
\let\VANthebibliography\thebibliography
\def\thebibliography{\DeclareRobustCommand{\VAN}[3]{##3}\VANthebibliography}

\usepackage{graphicx}	
\usepackage{amsmath}	

\usepackage{verbatim}
\usepackage{bm}
\usepackage{textcomp}
\usepackage{enumerate}
\usepackage{caption}
\usepackage{subcaption}
\usepackage{comment}
\usepackage{array}
\usepackage{multirow}
\usepackage{CJK}
\usepackage[export]{adjustbox}
\usepackage{soul}

\usepackage[text=black, background=white]{callouts}
\usepackage{tikz}

\newcommand\um{$\mu$m}
\newcommand\msun{M$_\odot$}

\newcommand\meth{CH$_{3}$OH}

\newcommand{\nht}{NH$_3$}

\newcommand{\nhp}{N$_2$H$^+$}
\newcommand{\hdp}{H$_2$D$^+$}
\newcommand{\dcop}{DCO$^+$}

\newcommand{\hden}{$n_{\rm H_2}$}

\newcommand{\lsun}{L$_\odot$}
\newcommand{\HII}{H\,{\sc ii}}
\newcommand{\millimeter}{mm} 
\newcommand{\mjb}{mJy~beam$^{-1}$} 
\newcommand{\kms}{km~s$^{-1}$}

\newcommand{\perccc}{cm$^{-3}$} 
\newcommand{\jykms}{Jy~km~s$^{-1}$}  
 
\newcommand{\mjy}{mJy}
\newcommand{\mjyb}{mJy~beam$^{-1}$}
\newcommand{\mjykms}{mJy~km~s$^{-1}$} 

\newcommand{\mum}{$\mu$m}

\newcommand{\gm}{G11.92-0.61 MM2}
\newcommand{\g}{G11.92-0.61}

\newcommand{\arcdeg}{$^\circ$}

\title[Filamentary mass accretion towards G11.92 MM2]{Filamentary mass accretion towards the high-mass protobinary system G11.92$-$0.61 MM2}

\author[S. Zhang et al.]{
S. Zhang (张遂楠),$^{1,2}$\thanks{Contact e-mail: sz57@st-andrews.ac.uk}
C. J. Cyganowski,$^{1}$
J. D. Henshaw,$^{3,4}$ 
C. L. Brogan,$^{5}$
T. R. Hunter,$^{5}$
R. Friesen,$^{6}$
\newauthor
I. A. Bonnell,$^{1}$
and S. Viti$^{7}$
\\
$^{1}$SUPA, School of Physics and Astronomy, University of St Andrews, St Andrews, Fife, Scotland\\
$^{2}$Shanghai Astronomical Observatory, Chinese Academy of Sciences, 80 Nandan Road, Shanghai 200030, P.\ R.\ China\\
$^{3}$Astrophysics Research Institute, Liverpool John Moores University, 146 Brownlow Hill, Liverpool L3 5RF, UK\\
$^{4}$Max Planck Institute for Astronomy, K̈onigstuhl 17, D-69117 Heidelberg, Germany\\
$^{5}$National Radio Astronomy Observatory, 520 Edgemont Road, Charlottesville, VA 22903, USA\\
$^{6}$University of Toronto, Toronto, ON, Canada\\
$^{7}$Leiden Observatory, Leiden University, P.O. Box 9513, NL-2300 RA Leiden, the Netherlands\\
}
\date{Accepted XXX. Received YYY; in original form ZZZ}

\pubyear{2023}

\begin{document}
\begin{CJK*}{UTF8}{gbsn}
\label{firstpage}
\pagerange{\pageref{firstpage}--\pageref{lastpage}}
\maketitle

\begin{abstract}
We present deep, sub-arcsecond ($\sim$2000 AU) resolution ALMA 0.82\,mm observations of the former high-mass prestellar core candidate \gm, recently shown to be an $\sim$500~AU-separation protobinary.
Our observations show that 
\gm, 
located in the \g\/ protocluster,
lies on a filamentary structure traced by 0.82\,mm continuum and \nhp\/(4-3) emission.
The \nhp\/(4-3) spectra are multi-peaked, indicative of multiple velocity components along the line of sight. 
To analyse the gas kinematics, 
we performed pixel-by-pixel Gaussian decomposition of the \nhp\/ spectra 
using SCOUSEPY and hierarchical clustering of the extracted velocity components using ACORNS.  
Seventy velocity- and position-coherent clusters (called "trees") are identified in the \nhp-emitting gas, with the 8 largest trees accounting for $>$60\% of the fitted velocity components. 
The primary tree, with $\sim$20\% of the fitted velocity components, displays a roughly north-south velocity gradient along the filamentary structure traced by the 0.82~mm continuum. 
Analysing a $\sim$0.17~pc-long substructure, we interpret its velocity gradient of $\sim$10.5~\kms\/ pc$^{-1}$ as tracing filamentary accretion towards MM2 and estimate a mass inflow rate of $\sim$1.8$\times10^{-4}$  to 1.2$\times10^{-3}$ \msun\/ yr$^{-1}$.
Based on the recent detection of a 
bipolar molecular outflow associated with MM2, accretion onto the protobinary 
is ongoing, likely fed by the larger-scale filamentary accretion flows.  
If 50\% of the filamentary inflow reaches the protostars, each member of the protobinary would attain a mass of 8~\msun\/ within $\sim1.6\times$10$^5$ yr, comparable to the combined timescale of the 70~\um\/- and MIR-weak phases derived for ATLASGAL-TOP100 massive clumps using chemical clocks.
\end{abstract}



\section{Introduction}
\label{sec:intro}
How mass aggregation proceeds during the formation of high-mass stars (M$_{\rm ZAMS}$ $\geq$ 8~\msun\/) is a key open question in studies of star formation and the interstellar medium (ISM). In particular, whether and how the large-scale ($\gtrsim$ 1~pc) environments of young high-mass stars affect their mass collection remains unclear: from what spatial scales do high-mass stars assemble their masses? How does the scale of the gas reservoir impact the final stellar mass of a high-mass star?

Core accretion models \citep[e.g.][]{McKee:2002sz,mc2003} propose that the earliest phase of high-mass star formation is the high-mass prestellar core: a massive, gravitationally bound, centrally condensed, starless core with a radius  $\leq$0.1~pc. 
High-mass prestellar cores are predicted to be self-contained gas reservoirs nearly in internal virial equilibrium, and initially in pressure equilibrium with their environment; they are then expected to undergo relatively ordered collapse to form single high-mass stars or small multiple systems (see the review of \citealt{tan2014} and references therein).  
However, cores with masses larger than the critical mass (i.e., the Jeans mass) are expected to fragment (e.g., \citealt{dobbs2005}). It thus remains a challenge for the core accretion theory to explain how a high-mass prestellar core (with $\sim$10$^2$ Jeans masses) is supported against fragmentation (\citealt{tan2014}).

Many numerical simulations have been carried out of the collapse of massive prestellar cores, both with and without magnetic fields. 
Radiation-hydrodynamic simulations have demonstrated that the radiation pressure barrier problem can be overcome by disc accretion and flashlight effects due to outflow cavities (e.g.\ \citealt{krumholz09}, \citealt{cunningham2011}, \citealt{Kuiper2015}, \citealt{rosen2019}), with gas channelled to the star-disc system by gravitational and Rayleigh-Taylor instabilities (e.g.\ \citealt{krumholz09}, \citealt{rosen2016}).
Radiation-magnetohydrodynamic (RMHD) simulations (e.g.\ \citealt{com2011}, \citealt{myers2013}, \citealt{rosen2020}, \citealt{mignon2021}) have shown that the combined effects of magnetic fields and radiation can effectively suppress the initial fragmentation of massive dense cores by increasing the magnetic and thermal Jeans mass of the collapsing gas, contributing to the formation of massive stars. Recent RMHD simulations of the collapse of massive prestellar cores (M = 150~\msun\/, r = 0.1~pc) including isotropic stellar winds show that accretion onto massive protostars with masses $\geq$30~\msun\/ is impeded by their wind feedback, suggesting that larger-scale dynamical effects are required to form >30~\msun\/ stars (\citealt{rosen2022}).

Searching for high-mass prestellar cores in observations has, however, identified very few candidates in the past decades. 
Many initially promising candidates have been ruled out by subsequent observations that revealed signs of active star formation and/or resolved low-mass cores. 
G028C1-S, initially discovered to contain $\sim$60~\msun\/ within $\sim$0.09~pc by \cite{tan2013}, has been resolved into two protostars driving outflows (\citealt{tan2016}). \cite{kong2017} reveal that G028C9A ($\sim$80~\msun\/ within $\sim$0.05~pc) actually consists of two lower-mass cores. 
In addition, though identified as one of the best high-mass prestellar core candidates ($\sim$25~\msun\/ within $\sim$0.025~pc, \citealt{bon2010}), CygXN53-MM2 is located close to CygXN53-MM1, a protostar driving outflows, making it difficult to distinguish if CygXN53-MM2 is not associated with any outflow (\citealt{dc2013}, \citealt{dc2014}).
The most promising high-mass prestellar core candidates identified to date are G11P6-SMA1 ($\sim$30~\msun\/ within $\sim$0.02~pc, \citealt{WANG2014}), W43-MM1 core 6 ($\sim$37~\msun\/, $\sim$1800~AU, \citealt{nony2018}, \citealt{molet2019}), AG354 ($\sim$39~\msun\/, \citealt{redaelli2021}), and G028.37+00.07 C2c1a ($\sim$23-31~\msun\/, \citealt{barnes2023})

In contrast, in competitive accretion models (e.g.\ \citealt{bonnell2001}, \citealt{Bonnell:2003ny}) high-mass star formation is closely connected to cluster formation (e.g., \citealt{Bonnell:2004qk}). The fragmentation of turbulent molecular clouds produce low-mass cores as seeds of protostars (\citealt{Bonnell:2003ny}). Protostars that are located near the centres of protoclusters are more likely to become massive stars, as these protostars have the advantage in being fed by gas infalling into the large-scale gravitational potential well of the cluster (e.g., \citealt{bonnell2006}). 
In this picture, the precursors of massive stars are predicted to accrete gas widely from their large-scale environments, in which case self-contained gas reservoirs (i.e. high-mass prestellar cores) do not necessarily exist.
In numerical simulations spanning a range of initial conditions -- from idealised spheres or cylinders of turbulent molecular clouds (e.g.\ \citealt{Bonnell:2004qk}, \citealt{bonnell2008}, \citealt{smith2009}, \citealt{maschberger2010}) to more realistic spiral arm dynamics (e.g.\ \citealt{bonnell2013}, \citealt{smilgys2016}, \citealt{smilgys2017}) -- and included physics (such as photoionisation, e.g.\ \citealt{dale2011}, \citealt{dale2012}, and photoionisation plus radiation forces, e.g.\ \citealt{kuiper2018}), high-mass stars form via significant accretion from r$\gtrsim$0.2 - 10~pc spatial scales. 
More recently, by considering supernova-driven turbulence in magnetohydrodynamic (MHD) simulations of Giant Molecular Clouds (GMCs), \cite{padoan2020} found that massive stars accumulate mass via inertial gas flows from parsec-scale filaments. 
Additionally, simulations of GMCs formed by supersonic convergent flows of warm diffuse atomic gas have led to the development of the global hierarchical collapse (GHC) scenario, in which multi-scale collapse takes place within GMCs and (sub)structures at all scales (i.e., clouds, clumps, cores) accrete from their larger-scale environments via non-isotropic gas inflows (e.g.\ \citealt{vs2007}, \citealt{gomez2014}, \citealt{vs2017}, \citealt{vs2019}).

Many studies of the accretion reservoirs of forming high-mass stars have focused on Infrared Dark Clouds (IRDCs), 
identified as extinction features against the Galactic mid-infrared (MIR) background (\citealt{bt2009}, \citealt{bt2012}). IRDCs have been found to harbour the early phases of high-mass star and cluster formation (e.g., \citealt{rathborne2006}, \citealt{jackson2010}, \citealt{peretto2010}, \citealt{Kauffmann2010}, \citealt{miettinen2012a,miettinen2012b}, \citealt{ragan2012a}, \citealt{Kainulainen2013}, \citealt{henshaw2013}, \citealt{Busquet2013}, \citealt{peretto2013}). They often display filamentary morphologies (e.g., \citealt{jackson2010}, \citealt{henshaw2013}, \citealt{beuther2015}, \citealt{andre2016}), with kinematic features of dense gas in IRDCs indicating gas flows along filaments (e.g., \citealt{zernickel2013}, \citealt{tackenberg2014}, \citealt{henshaw2014}, \citealt{zhang2015}, \citealt{chen2019}). 
Multiple filaments can merge near the centre of a molecular cloud to form a hub, a phenomenon referred to as a "hub-filament system" (e.g.\ \citealt{myers2009}, \citealt{schneider2012}, \citealt{hennemann2012}). 
Observations over the past decade have pointed to hub-filament systems playing a key role in shaping high-mass star and cluster formation (e.g., \citealt{peretto2013}, \citealt{peretto2014}, \citealt{Henshaw2017}, \citealt{tige2017}, \citealt{dewangan2017}, \citealt{yuan2018}, \citealt{hacar2018}, \citealt{williams2018}, \citealt{keown2019}, \citealt{trevin2019}, \citealt{chen2019}, \citealt{dewangan2020}, \citealt{Kumar2020}, \citealt{wang2020}, \citealt{anderson2021}, \citealt{liu2023}, \citealt{zhou2023}, \citealt{mookerjea2023}), with recent kinematic studies suggesting that hub-filament systems serve as the parsec-scale gas reservoirs feeding  embedded young stars via filamentary gas flows (e.g., \citealt{zhou2022}, \citealt{xu2023}).

\begin{figure}
	\includegraphics[width=\columnwidth]{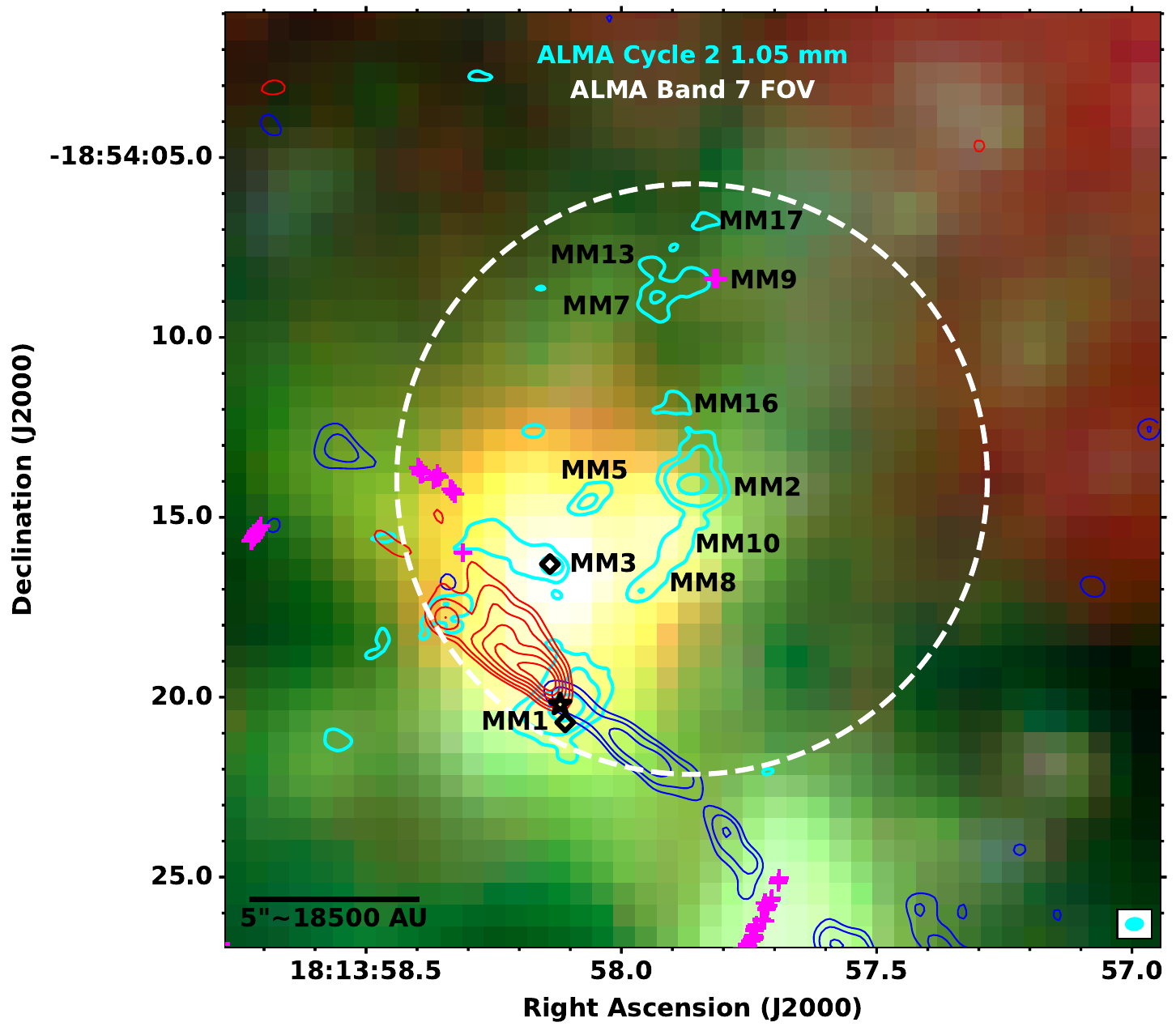}
   \caption{\textit{Spitzer} GLIMPSE three-colour image (red 8.0~\mum\/, green 4.5 ~\mum\/, blue 3.6~\mum\/) 
   overlaid with contours of the ALMA Cycle 2 1.05~mm continuum emission (cyan: [5,15,100] $\times$ 0.35 \mjyb\/, \citealt{cc2017}) 
   and SMA blue- and red-shifted $^{12}$CO (3-2) line emission (blue: [4,6,9] $\times$ 1.0 \jykms\/, red: [4,6,9,12,15,18] $\times$ 1.0 \jykms\/, \citealt{cc2014,cc2017}). 
   The magenta $+$ and black $\Diamond$ mark the locations of Class 
   {\sc i} and Class {\sc ii} \meth\/ masers, respectively, from \citet{cc2009} and 
   the black star in MM1 marks the H$_2$O maser from \citet{breen2011}. 
   The dashed white circle indicates the 50\% response level of the primary beam for our ALMA Band 7 observations. The synthesised beam of the ALMA Cycle 2 1.05~mm continuum image is shown at bottom right.}
    \label{fig:intro}
\end{figure} 

To investigate the properties of a candidate high-mass prestellar core and its relationship to its environment, we proposed ALMA Cycle 3-5 observations of \gm\/ targeting \hdp\/(1$_{1,0}$-1$_{1,1}$) and \nhp\/(4-3), which trace dense and depleted gas (\citealt{cecc2014}) and are commonly observed towards low-mass prestellar cores (\citealt{caselli2003}, \citealt{van2005}, \citealt{vastel2006}, \citealt{caselli2008}, \citealt{friesen2014}, \citealt{miettinen2020}, \citealt{koumpia2020},  \citealt{redaelli2021, redaelli2022}, \citealt{kong2023}). 
\gm\/ (hereafter MM2) is one of three massive millimetre cores associated with the GLIMPSE Extended Green Object \citep[EGO;][]{cc2008} G11.92$-$0.61, which were first detected with the Submillimeter Array (SMA) and the Combined Array for Research in Millimetre-wave Astronomy (CARMA) by \citet{cc2011a}.
Subsequent ALMA observations revealed 16 additional low-mass cores and showed that the  
G11.92$-$0.61 protocluster hosts ongoing, simultaneous high- and low-mass star formation \citep{cc2017}.
MM2 is the second-brightest mm source in the G11.92$-$0.61 protocluster. 
The brightest, MM1, 
has been resolved into 
a Keplerian disc around a proto-O star that is forming a binary system via disc fragmentation \citep{ilee2016,ilee2018}.

Located $\sim$0.12~pc from MM1, MM2 was considered to be a promising high-mass prestellar core candidate because it was a strong and compact dust continuum source that had no molecular line emission or other star formation indicators in SMA and VLA observations \citep[][and references therein]{cc2014}.
In addition to its remarkable lack of (sub)mm line emission \citep[no lines detected across $\sim$24 GHz of bandwidth observed with the SMA at 1.3, 1.1, 0.8~mm;][]{cc2014}, no CH$_3$OH or H$_2$O masers, molecular outflows, or cm continuum emission indicative of ionised gas had been detected in VLA and SMA imaging covering the G11.92$-$0.61 region \citep[][]{hofner1996,cc2009,breen2011,cc2011a,cc2011b,cc2014,hunter2015}.
The physical properties inferred by \citet{cc2014} based on their SMA and VLA data were extreme: M$\gtrsim$30~\msun\/ within R$<$1000~AU and $n_{H_2}> 10^9$~\perccc\/, with $T_{dust}$ $\sim$17-19~K and $L$ $\sim$5-7~\lsun\ from submillimeter SED fitting.
Based on gas-grain astrochemical modeling using MONACO (\citealt{v2009}),  
\citet{cc2014} found that the lack of line emission observed with the SMA could be explained with standard chemistry under cold ($T_{dust}<$20~K) and very dense ($n_{H_2}$ $\gg$ $10^8$~\perccc) conditions.  
Deuterium fractionation is expected to take place at $<$20~K, increasing the abundance of \hdp\/ (e.g.\ \citealt{caselli2012}, \citealt{redaelli2021}). With most C- and O-bearing species frozen out onto dust grain surfaces, \nhp\/ becomes abundant as this molecule forms from the protonation of N$_2$ in cold environments and is destroyed mainly by  reactions with CO (e.g.\ \citealt{oberg2005}, \citealt{vanthoff2017}).
The expectation of extreme depletion motivated our targeting \hdp\/(1$_{1,0}$-1$_{1,1}$) and \nhp\/(4-3) in our ALMA Band 7 observations.

\begin{table*}
	\small
	\centering
	\caption{Summary of ALMA 0.82\,mm observations.}
	\label{tab:obs}
	\begin{tabular}{lcccccccc}
	\hline\hline
    Project code & Observing date & Configuration & Number of antennas & Time on-source & \multicolumn{3}{c}{Calibrators} \\
                 &                &               &                    & & Gain & Bandpass & Flux \\
    \hline
    2015.1.00827.S & 2016 April 9$^{a}$ & C36-2/3 & 43 & 2$\times$ 42.5 min & J1733-1304 &  J1924-2914 & J1924-2914, Titan \\
    2015.1.00827.S & 2017 April 22 & C40-3 & 41 & 42.5 min & J1733-1304 &  J1924-2914 & Titan \\ 
    2015.1.00827.S & 2017 April 26 & C40-3 & 41 & 42.5 min &   J1733-1304 &  J1924-2914 & Titan \\ 
    2017.1.01373.S & 2018 July 10  &  C43-1  &  46  &   46.5 min  &   J1911-2006    & J1924-2914 &   J1924-2914 \\ 
    2017.1.01373.S & 2018 August 16  & C43-2 & 43-44  &  3$\times$ 46.5 min  &  J1911-2006 & J1924-2914 & J1924-2914 \\ 
	
	\hline
	\end{tabular}
	\begin{flushleft}
 		$^a$ Due to a tuning issue, the correlator setup on this date did not cover \nhp\/ (4-3).\\
	\end{flushleft}
\end{table*}

Recent high-resolution ($\lesssim$160 AU) ALMA 1.3~mm observations targeting MM1 have changed the picture of MM2, revealing that MM2 is in fact  
a 505~AU-separation protobinary system \citep{cc22}.  The members of the protobinary  have 1.3~mm brightness temperatures of 68.4~K and 64.6~K, indicative of internal heating; thus, MM2 is not a high-mass prestellar core but rather hosts the very early stages of high-mass binary formation \citep{cc22}. 
Intriguingly, 
the discovery of a low-velocity asymmetric bipolar molecular outflow associated with MM2 in our deep ALMA Band 7 observations 
\citep[traced by
\meth\/ (4$_{-1,3}$-3$_{0,3}$), reported in][]{cc22}
indicates that accretion onto the protobinary system is ongoing.

In this paper, we present our ALMA Band 7 continuum, \hdp\/ and \nhp\/  observations towards MM2, with a focus on the gas kinematics around MM2 traced by the \nhp\/ (4-3) emission. 
In Section~\ref{sec:obs}, we provide details of the observations.  In Section~\ref{sec:results}, we show the results for the dust continuum and molecular line emission. 
In Section~\ref{sec:gaussian_decomposition}, we describe the procedure used to perform Gaussian decomposition on the \nhp\/ data cube and in Section~\ref{sec:hierarchical_clustering}, we describe the hierarchical clustering analysis. Section~\ref{sec:fil_acc_flows} presents our analysis of the physical properties of the filamentary accretion flows, which are further discussed in Section~\ref{sec:discussion}. Section~\ref{sec:conclusions} summarises our main conclusions. Throughout this work, we assume that MM2 is located at the same distance as MM1 and adopt MM1's maser parallax distance of 3.37$^{+0.39}_{-0.32}$ kpc \citep{sato2014}.

\section{ALMA Observations}
\label{sec:obs}

We observed \gm\/ at 0.82\,mm with the ALMA 12-m array in Cycles 3-5 (project codes 2015.1.00827.S and  2017.1.01373.S, PI C.\ Cyganowski).  The phase centre of the single-pointing observations was 18$^{\rm h}$13$^{\rm m}$57$^{\rm s}$.8599 $-$18$^{\circ}$54\arcmin13\farcs958 (ICRS); at 0.82\,mm, the full width half power (FWHP) size of the primary beam is $\sim$17\arcsec\/ (see Figure~\ref{fig:intro}).
Additional observational parameters, including observing dates, configurations, and calibrators, are listed in Table~\ref{tab:obs}.   
The projected baselines of the combined dataset range from $\sim$14--583 k$\lambda$, corresponding to a largest angular scale (LAS)\footnote{Estimated from the fifth percentile shortest baseline using the analysisUtils \citep{au2023} task \texttt{au.estimateMRS}.} of $\sim$4\farcs5 ($\sim$0.07\,pc $\sim$15200 AU at D=3.37\,kpc).

The correlator setup included four spectral windows (spws): three narrow spws tuned to cover \hdp\/(1$_{1,0}$--1$_{1,1}$) ($\nu_{\rm rest}$=372.421\,GHz, $E_{upper}$=104.2\,K), \nhp\/(4-3) ($\nu_{\rm rest}$=372.672\,GHz, $E_{upper}$=44.7\,K), and \dcop\/(5-4) ($\nu_{\rm rest}$=360.170\,GHz, $E_{upper}$=51.9\,K)\footnote{Line rest frequencies are from CDMS \citep{muller2001,muller2005}, accessed via the Splatalogue NRAO spectral line catalogue (\url{https://splatalogue.online/}).}, and a wide spw, centred at $\sim$358.02\,GHz, for continuum sensitivity.  
The observed bandwidths and spectral resolutions (accounting for online Hanning smoothing and channel averaging) were 117.2~MHz ($\sim$94 \kms) and 0.122\,MHz ($\sim$0.098 \kms) for \hdp, 58.6\,MHz ($\sim$47 \kms) and 0.061\,MHz ($\sim$0.049 \kms) for \nhp, 243.4\,MHz ($\sim$195 \kms) and 0.141\,MHz ($\sim$0.117 \kms) for \dcop, and 1875\,MHz ($\sim$1570 \kms) and 1.129\,MHz ($\sim$0.945 \kms) for the wide spw.

The data were calibrated using the CASA 5.4.0 version of the ALMA science pipeline.  Line-free channels were identified using the approach of \citet{brogan2016,cc2017}  and used to construct a pseudo-continuum data set, which has   
an aggregate continuum bandwidth of $\sim$0.44 GHz.  
These continuum data were iteratively self-calibrated, and the solutions were then applied to the line data. 
The final continuum image, made using multi-frequency synthesis, multiscale clean, and Briggs weighting with a robust parameter R = 0.5, has an rms noise level of 1$\sigma$=0.5~\mjyb\/ and a synthesised beam size of 0.575\arcsec\/ $\times$ 0.431\arcsec\/ (P.A.=$-$80$^{\circ}$), equivalent to $\sim$1940~AU $\times$ 1450~AU at D=3.37\,kpc. 
In this paper, we consider only the continuum, \nhp, and \hdp\/data; images of the \meth\/ (4$_{-1,3}$-3$_{0,3}$) line, included in the wide spw, were presented in \citet{cc22} (see also Section~\ref{sec:intro}). 
The \nhp\/ line data were imaged with multiscale clean, Briggs weighting with a robust parameter R=0.5, and a velocity resolution $\Delta$v=0.2 \kms.  The resulting \nhp\/ (4-3) image cube has a synthesised beam size of 0.651\arcsec\/ $\times$ 0.442\arcsec\/ (P.A.=$-$79$^{\circ}$) and an rms noise level of 1$\sigma$=5.9 \mjyb\/.  To maximise sensitivity, the \hdp\/ line data were imaged with multiscale clean, Briggs weighting with a robust parameter R=2, and a velocity resolution $\Delta$v=0.4 \kms.  The resulting \hdp\/(1$_{1,0}$--1$_{1,1}$) image cube has a synthesised beam size of 0\farcs762 $\times$ 0\farcs551 (P.A. $-$75$^{\circ}$) and an rms noise level of 1$\sigma$=2.7 \mjyb\/.    
All measurements were made from images corrected for the primary beam response. 

\begin{figure*}
    \includegraphics[width=1.0\textwidth]{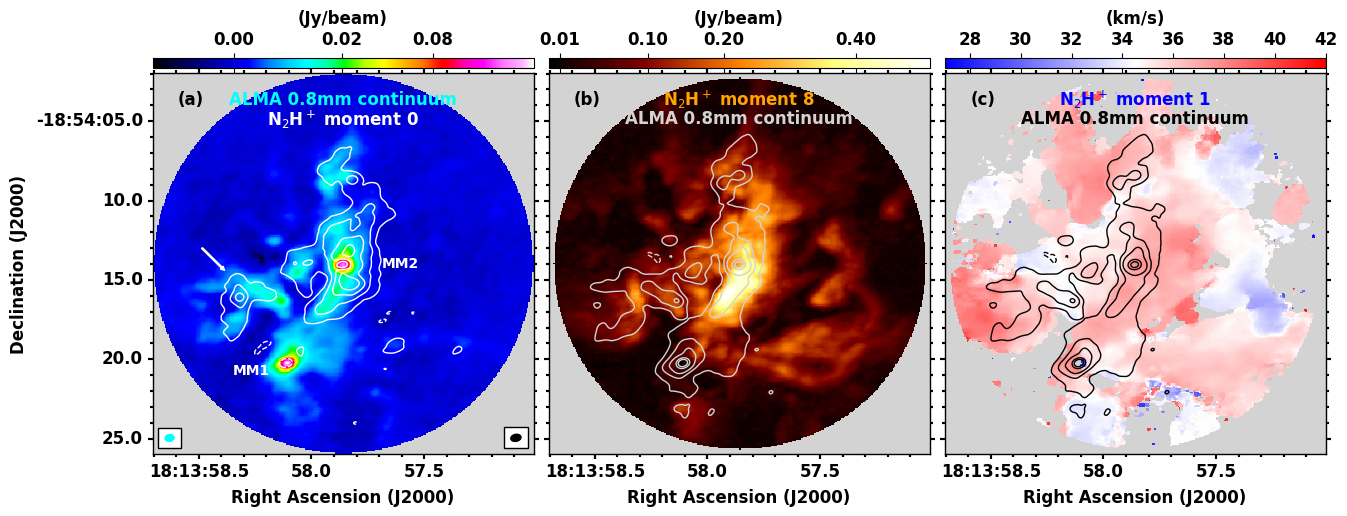}
    \caption{(a): ALMA 0.82~mm continuum (colourscale) overlaid with contours of integrated \nhp\/(4-3) emission (velocity range: 27.0 - 42.0 \kms\/; levels: 0.06 \jykms\/$\times$ [-2,5,10,15,20], negative contours shown as dashed lines). 
    The arrow marks the bow-like feature discussed in Section~\ref{sec:n2hp_results} and the synthesised beams of the 0.82\,mm continuum image and \nhp\/ cube are shown at bottom left (in cyan) and right (in black), respectively. (b): ALMA \nhp\/ peak intensity map (colourscale) overlaid with 0.82~mm continuum contours. (c): \nhp\/(4-3) moment 1 map (colourscale) overlaid with 0.82~mm continuum contours. In the moment 1 map, pixels with line emission $<$4 $\times \sigma_{\rm npb}$, where $\sigma_{\rm npb}$=4.7\mjb\/ is the rms of the \nhp\/ cube before primary beam correction, are masked and shown in grey. 
    $v_{\rm LSR}$(MM1)= 35.2$\pm$0.4 \kms\/ \citep[from compact molecular line emission observed with the SMA;][]{cc2011a}; \citet{cc22} estimate $v_{\rm LSR}$(MM2)$\sim$37 \kms\/ based on the velocity of the peak CH$_3$OH emission from the filament at the protobinary position. The ALMA images shown here have not been corrected for the primary beam response; all measurements were made from corrected images and corrected images are shown in Figure~\ref{fig:nhp_spec}. In (b) and (c), the 0.82~mm continuum contour levels are $\sigma_{\rm npb}$=0.48 \mjyb\/ $\times$ [-2,4,16,40,160,280], where $\sigma_{\rm npb}$=0.48 \mjyb\/ is the rms of the 0.82~mm continuum image before primary beam correction. In all panels, the edge of the colourscale corresponds to the 20$\%$ response level of the ALMA primary beam.}
   \label{fig:obs_res}
\end{figure*}

\section{Results}
\label{sec:results}
\subsection{ALMA 0.82~mm continuum emission}
\label{sec:cont_results}

The ALMA 0.82\,mm continuum image is shown in Figure \ref{fig:obs_res} (in colourscale in (a) and as contours in (b) and (c)).
As expected, the 0.82 \millimeter\/ continuum emission displays a broadly similar morphology to the ALMA 1.05 \millimeter\/ continuum emission observed with similar angular resolution and sensitivity by \citet{cc2017} (1.05\,mm $\theta_{\rm syn}$=0\farcs49 $\times$ 0\farcs34, $\sigma$=0.35 \mjb).
In the 0.82\,mm continuum image, MM2 is located on a filamentary structure that extends to the north and southeast of MM2; this structure encompasses the MM8, MM10, and MM16 cores identified by \citet{cc2017}.
To the southeast, the extended 0.82\,mm continuum emission "bridges" MM2 and the proto-O star MM1 \citep{ilee2016,ilee2018}. 
To the east of MM2, $\sim 5 \sigma$ extended emission also connects the dust continuum sources MM3 and MM5 \citep[][see Figure~\ref{fig:intro}]{cc2017} to the main north-southeast dust filament, with the extended emission joining the main filamentary structure to the northeast of MM2. 
The MM7/MM9/MM13/MM17 group of mm sources \citep[][Figure~\ref{fig:intro}]{cc2017}, located north of MM2, is also detected in 0.82\,mm continuum emission (see Figure~\ref{fig:obs_res}).

At the $\sim$0\farcs5-resolution of our 0.82\,mm image, MM2 
appears as a strong, compact source of submillimetre continuum emission within the larger filamentary structure.
To estimate the 0.82\,mm properties of MM2, we fit a one-component two-dimensional Gaussian model to its 0.82~mm continuum emission using the CASA task \texttt{imfit}. 
From this one-component Gaussian fit, the fitted peak intensity of MM2 is I$_{\rm peak}$=158.3$\pm$0.6 \mjyb\/ and its fitted integrated flux density is S$_{\rm 0.82mm}$=390$\pm$2 \mjy\/. The fitted position of MM2 is 18$^{\rm h}$13$^{\rm m}$57\fs86094$\pm$0\fs00009, $-$18$^{\circ}$54\arcmin14\farcs023$\pm$0\farcs001 (J2000) and its deconvolved size is 614($\pm5$) mas $\times$ 584($\pm$4) mas (P.A.=87$^{\circ}\pm$8$^{\circ}$), equivalent to $\sim$ 2070~AU $\times$ 1970~AU at D=3.37~kpc. 
Notably, the single-component fit underestimates the peak intensity of MM2's 0.82\,mm continuum emission: 
the residual image has a $\sim$27~\mjyb\/ peak, which is $<$0\farcs02 from the fitted centroid position,
surrounded by a negative ring. 
This indicates that a single-component Gaussian does not fully represent the emission, and points to the presence of additional substructure(s).
This is consistent with the results from higher-resolution ALMA observations, where MM2 is resolved to be a protobinary system with a projected separation of $\sim$500~AU \citep[][see also Section~\ref{sec:intro}]{cc22}. 
To check the robustness of the integrated flux density  measurement from the Gaussian fit, we also use the CASA task \texttt{imstat} to measure the integrated flux density within the 30$\sigma$ contour around MM2 and within a 2\farcs0$\times$1\farcs0 ellipse centred on its fitted position. 
These methods yield integrated flux density estimates of 401$\pm$40~\mjy\/ and 327$\pm$33~\mjy, respectively,\footnote{Uncertainties are estimated using Equation (1) of \cite{thwala2019}, assuming 10$\%$ absolute flux calibration uncertainty for ALMA in Band 7 (\citealt{almac10}).} within $<$3\%-16\% of the value from the Gaussian fit. 
This comparison suggests that the single-component Gaussian fit provides a reasonably robust estimate for the 0.82\,mm integrated flux density of MM2, though we note that all estimates potentially include a contribution from the background filament emission (see also Section~\ref{sec:mass}). 

\begin{figure*}
	\includegraphics[width=1.0\textwidth, height=0.6\textheight]{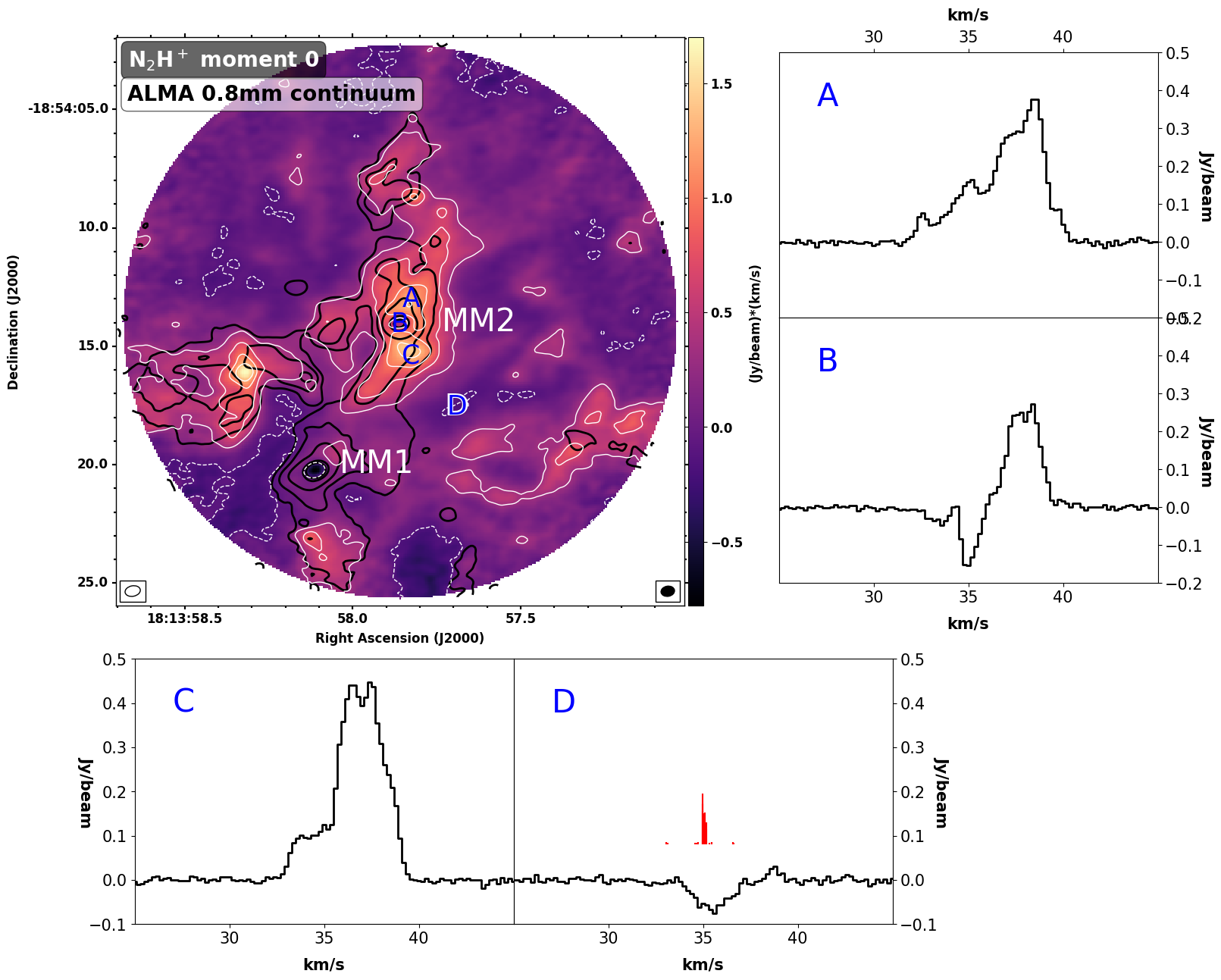} 
    \caption{Top left: ALMA \nhp\/ (4-3) integrated intensity map (integrated over v = 27.0 - 42.0 \kms, colourscale and white contours, levels 0.066 \jykms\/$\times$ [-2,5,10,15,20], negative contours shown as dashed lines) overlaid with contours of the ALMA 0.82~mm continuum emission (black, levels: 0.5 \mjyb $\times$ [5,15,40,160,280]), both corrected for the primary beam response.  The edge of the colourscale corresponds to the 20$\%$ response level of the ALMA primary beam and the synthesised beams of the \nhp\/ cube and the 0.82~mm continuum image are shown at bottom left (unfilled ellipse) and right (filled black ellipse), respectively. Panels labelled with letters show the \nhp\/ spectra extracted from the primary-beam-corrected image cube at the corresponding labelled locations on the integrated intensity map.  Location A is the local peak of the integrated \nhp\/ emission to the north of MM2, B is the 0.82~mm continuum peak of MM2, C is the local peak of the integrated \nhp\/ emission to the south of MM2, and D is a representative negative bowl. The red vertical lines overlaid in Panel D indicate the relative line strengths of the 38 hyperfine components of the \nhp\/ (4-3) transition (see Section~\ref{sec:n2hp_results}) for $v_{\rm lsr}$=35.0~\kms\/. }
   \label{fig:nhp_spec}
\end{figure*}

\subsection{Molecular lines}

\subsubsection{\hdp\/ (1$_{1,0}$-1$_{1,1}$)}
\label{sec:h2dp_results}
\hdp\/ (1$_{1,0}$-1$_{1,1}$) is undetected towards MM2 in our observations, to a 4$\sigma$ limit of 10.8 \mjb\/ (equivalent to a 4$\sigma$ brightness temperature limit of 0.23~K). 
The nondetection of \hdp\/ towards MM2 is unsurprising, since MM2 is now known to host a protobinary system with two internally-heated protostars \citep[][see Section~\ref{sec:intro}]{cc22}. 
\hdp\/ is destroyed by gas-phase reactions with CO, which evaporates from grains at temperatures $>$20~K; thus, \hdp\/(1$_{1,0}$-1${_1,1}$) emission is only expected in the pre-stellar phase \citep[e.g.\ ][]{redaelli2021}. 
Indeed, the few reported interferometric (ALMA) detections of \hdp (1$_{1,0}$-1${_1,1}$) in high-mass star-forming regions in the literature are towards very quiescent, MIR-dark clumps and regions of IRDCs \citep[e.g.][]{redaelli2021,kong2023}.
In the more evolved cluster environment of G11.92$-$0.61, where there are many protostars heating the gas (including the MIR-bright massive protostars MM1 and MM3), \hdp\/ emission would only have been expected towards the cold, dense interior of a 
massive starless core. 

\label{sec:line_results}
\subsubsection{\nhp\/ (4-3)}
\label{sec:n2hp_results}
As shown in Figure \ref{fig:obs_res}, extensive \nhp\/ (4-3) emission is detected around MM2 in our ALMA observations. Notably, the morphology of the integrated \nhp\/ (4-3) emission differs markedly from that of the 0.82\,mm continuum (see Figure~\ref{fig:obs_res}a), with the peaks of the integrated \nhp\/ emission clearly offset from the continuum peak of MM2. 
The integrated \nhp\/(4-3) emission surrounding MM2 peaks to its northwest and southwest, at offsets of $\sim$0\farcs9 ($\sim$3000~AU) and $\sim$1\farcs2 ($\sim$4000~AU) from the continuum peak, respectively.  
Other than the peaks near MM2, the most significant feature in the \nhp\/ (4-3) integrated intensity map is a bow-like structure located towards the red-shifted lobe of the high-velocity bipolar molecular outflow driven by MM1 \citep[Figure~\ref{fig:intro}, see also][]{cc2011a,cc2017}.
This intriguing feature, which may be associated with bow shocks caused by protostellar feedback from the proto-O star MM1, will be discussed in detail in a separate publication. The bow-like feature is also visible in the \nhp\/(4-3) peak intensity map in Figure~\ref{fig:obs_res}b, which shows that the strongest \nhp\/ emission (543~\mjyb\/, SNR$\sim$92) is detected towards the filamentary structure to the southeast of MM2. 

The intensity-weighted velocity (moment 1) map of the \nhp\/ (4-3) emission is shown in Figure~\ref{fig:obs_res}c; however, examination of the data cube indicates that the spectra are in general too complex for the gas kinematics to be captured by a moment analysis. This complexity is illustrated by the sample spectra presented in Figure \ref{fig:nhp_spec}. 
As shown by panels A-C, the line profiles of the \nhp\/ (4-3) emission display significant spatial variation across the mapped region.  
In addition to negative bowls associated with missing short-spacing information (panel D, see also Section~\ref{sec:obs}), \nhp\/ (4-3) absorption is detected against the submillimetre continuum towards MM2 (Figure~\ref{fig:nhp_spec} panel B). As noted above and shown in Figure~\ref{fig:nhp_spec}, the \nhp\/ spectra are in general complex, with multiple emission peaks and/or absorption features. 
\nhp\/ (4-3) has 38 hyperfine structures (hfs) components spaced over $\lesssim 4.3$ \kms\/ (\citealt{pagani2009}). To assess the potential contribution of the hfs to the observed line profiles, we calculate the relative line strengths of the hfs components using the methods described in \cite{daniel2006}, and plot them in Panel D of Figure~\ref{fig:nhp_spec}. 
The strongest hfs components (with relative line strengths $>$0.1) are tightly grouped within $\lesssim$0.2~\kms\/ (i.e.\ within one channel in our image cube, which has $\Delta$v=0.2 \kms; Section~\ref{sec:obs}). 
We therefore argue that while the hfs will have some effect on the line profiles (as discussed further in Section~\ref{sec:future}), the multiple emission peaks observed in the spectra are predominantly caused by multiple velocity components present along the line of sight in the \nhp\/-emitting gas surrounding MM2. 

We note that these observations are, to our knowledge, the first reported interferometric detection of \nhp\/ (4-3) in a high-mass star-forming region.
\nhp\/ (4-3) has been detected with ALMA in a protoplanetary disc \citep[where this transition was used to trace the CO snowline;][]{qioberg2013} and towards low-mass cores in Ophiuchus \citep{friesen2014}. 
Interferometric observations of this transition are, however, relatively uncommon, as it is more challenging to observe than lower-frequency \nhp\/ transitions (see also Section~\ref{sec:future}), which likely accounts for the lack of previous detections in high-mass star-forming regions.
With single dish telescopes, \nhp\/ (4-3) has been detected towards many low- and high-mass star-forming regions, with beams corresponding to physical resolutions of $\sim$2000 - 86000~AU (e.g.\ \citealt{van1992}, \citealt{blake1995}, \citealt{stark1999}, \citealt{friesen2010b}, \citealt{pillai2012}, \citealt{ma2013},  \citealt{kong2016}, \citealt{miettinen2020}, \citealt{koumpia2020}). 
For \nhp\/ (4-3) detections in nearby regions, the physical scale of the JCMT beam is similar to that of the synthesised beam of our ALMA data.
The observations of Ophiuchus B2, L1544 and L1521f by
\citet{friesen2010b} and \citet{koumpia2020}, for example, have a resolution of $\sim$2000~AU and 1$\sigma$ rms noise of $\sim$0.03-0.06~K (compared to $\sigma_{T_b} \sim$ 0.19~K for our data). 
Interestingly, in these observations, while \nhp\/ is generally detected towards regions with submillimetre continuum emission, the strongest \nhp\/ (4-3) emission is found offset from the submillimetre continuum peaks \citep{friesen2010b,koumpia2020}, as we observe in G11.92$-$0.61.  
\cite{friesen2014} also report an offset between the \nhp\/ (4-3) and submillimetre continuum peaks in their ALMA study of low-mass cores in Ophiuchus, though these observations probe much smaller spatial scales ($<$200 AU) and the authors note that the observations are significantly affected by self-absorption and missing short spacing data.

\section{Gaussian Decomposition on the N$_2$H$^+$ (4-3) cube}
\label{sec:gaussian_decomposition}
\subsection{Gaussian Decomposition with SCOUSEPY}
\label{sec:gd_scousepy}
To study the gas kinematics in the vicinity of MM2, we first apply Gaussian Decomposition to the \nhp\/ (4-3) cube using the SCOUSEPY package (\citealt{henshaw2016a}, \citealt{henshaw2019}). SCOUSEPY is a Python implementation of the Semi-Automated multi-COmponent Universal Spectral-line fitting Engine (SCOUSE), designed to apply Gaussian Decomposition to large volumes of complex spectra in a semi-automated way. The SCOUSEPY version used in this work can be downloaded from github\footnote{\url{https://github.com/jdhenshaw/scousepy/tree/delta}}. The workflow consists of four stages, as outlined below.

In stage 1), SCOUSEPY generates Spectral Averaging Areas (SAAs) and the spectra of SAAs 
based on spectral complexity. A 5$\sigma$ mask ($\sim$~54 \mjb\/) was first applied to remove emission-free regions from the analyses that follow. 
As this dataset exhibits strong spatial variation, 
we chose a conservative SAA size of 20 pixels ($\sim$5300~AU) with a filling factor of 0.6\footnote{This filling factor selects SAAs with significant emission in more than 60\% of the enclosed pixels; see SCOUSEPY documentation (linked from github) and \citet{henshaw2016a} for additional details.}.  The velocity range and pixel range are set to include the whole data cube. Other parameters are kept at their default values. 
With these parameters applied to the \nhp\/ data cube, SCOUSEPY generates 452 SAAs along with their spectra; as they are Nyquist sampled, each SAA overlaps with neighbouring SAAs. In the area covered by SAAs, the program identifies a total of 48141 individual spectra (corresponding to individual pixels in the data cube) to be analysed.

In stage 2), SCOUSEPY applies Gaussian Decomposition to the 452 spectra of SAAs generated in stage 1). The SCOUSEPY version used in this work has introduced derivative spectroscopy (\citealt{lindner2015}, \citealt{riener2019}) to assist with multi-component Gaussian fitting. 
The program will first filter spectra with Gaussian kernels and then calculate derivatives of the smoothed spectra up to the fourth order. The derivative spectroscopy technique measures the locations of spectral components by searching for ``bumps'', which mathematically means the local minimum of negative curvature in the filtered spectra (i.e. second derivative $<$ 0, third derivative $=$ 0, and fourth derivative $>$ 0, see details in \citealt{lindner2015}). SCOUSEPY first fits the spectra with initial guesses adopted from the measurements of derivative spectroscopy. 
The fitting is controlled by two input parameters: the Signal-to-Noise Ratio (SNR) and the standard deviation of the Gaussian kernel ($\sigma_{\rm ker}$) used for filtering. We vary the SNR from 3 to 5 and the $\sigma_{\rm ker}$ from 1 to 3 channels to obtain satisfying fits for $\gtrsim 96 \%$ of the SAA spectra; for an additional four SAA spectra, increasing the SNR to 7-9 yields reasonable fits.   
For the remaining SAA spectra, we use the manual fitting option of SCOUSEPY to fit the spectra "by hand" to obtain
physically reasonable fits.  
In some cases, for example, derivative spectroscopy fits artificially broad red-shifted components; we manually fit multiple weak components to the red tails of these spectra to obtain physically reasonable fits. 
There is only one spectrum (out of the 452 SAA spectra) which cannot be fitted well either by the derivative spectroscopy fitting or manual fitting. 
We flag the corresponding SAA at the end of stage (2) so that the problematic fit is not used to provide initial guesses for fits to individual spectra. 
The problematic SAA overlaps with neighbouring SAAs, so SCOUSEPY fits the spectra of the pixels within the flagged SAA based on the fits for the surrounding SAAs. 

\begin{table}
	\centering
	\caption{Summary of tolerance levels used in stage 3 of the Gaussian Decomposition using SCOUSEPY.}
	\label{tab:scouse_input}
	\begin{tabular}{lccr} 
		\hline
		Parameter & version 0 & version 1 & version 2 \\
		\hline
		$T_0$ & 2.0  & 2.0 & 2.0 \\	
		$T_1$ & 3.0  & 3.0 & 3.0 \\
		$T_2$ & 1.0  & 1.0 & 1.0 \\
		$T_3$ & 2.5  & 4.0 & 2.5 \\
		$T_4$ & 2.5  & 2.5 & 1.0 \\
		$T_5$ & 0.5  & 0.5 & 0.5 \\
		\hline
	\end{tabular}
\end{table}

\begin{figure*}
	\includegraphics[width=1.0\textwidth, height=0.5\textheight]{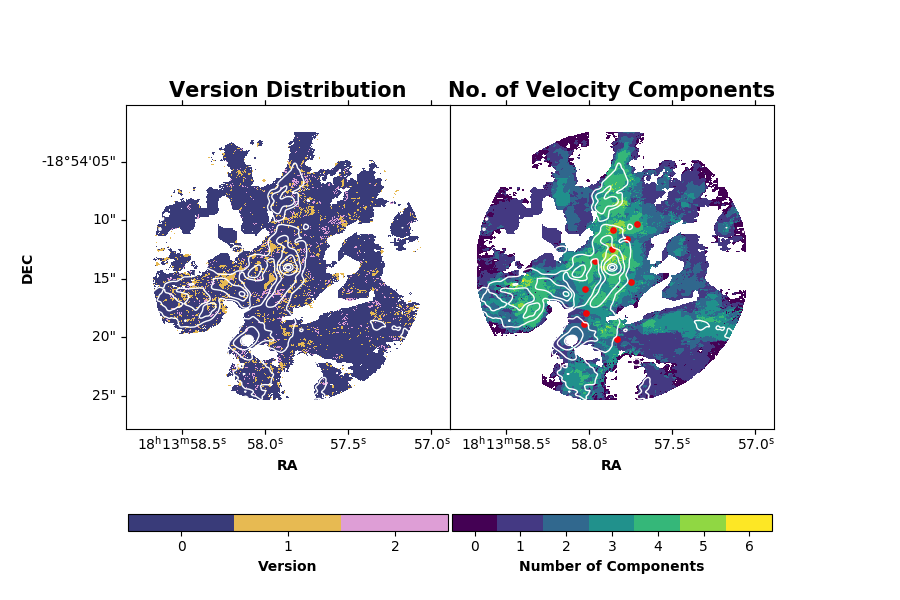}
    \caption{Left: the version of the best-fitting model employed in the final compiled solution for each of the 48141 individual spectra. 
    The numbers on the colourbar correspond to the versions described in Section~\ref{sec:gd_scousepy}. 
    Right: distribution of the number of velocity components in the best-fitting models.  Red circles mark the locations of the 10 spectra selected for hyperfine structure fitting (see Section~\ref{sec:future}).  ALMA 0.82~mm continuum contours  (white, levels: 0.5\mjyb $\times$ [5,15,40,160,280]) are overlaid in both panels. }
   \label{fig:scouse_res}
\end{figure*}

Stage 3) uses the fits for the spectra of SAAs as initial guesses to conduct Gaussian Decomposition for the 48141 individual spectra. SCOUSEPY adopts the fits from stage (2) as initial guesses and applies automated Gaussian Decomposition to the individual spectra within each SAA. To control the process of automated Gaussian Decomposition, users set tolerance levels using six parameters: $T_0$, $T_1$, $T_2$, $T_3$, $T_4$ and $T_5$. 
$T_0$ is newly introduced to the SCOUSEPY version used here. It limits the maximum difference between the number of fitted components for spatially averaged spectra and individual spectra. $T_1$ sets the threshold for amplitude, below which weak components are discarded before subsequent analyses. $T_2$ and $T_3$ control the full width at half-maximum (FWHM) of fitted components; $T_4$ limits the centroid velocity. $T_5$ sets the minimum separation in centroid velocity between two adjacent Gaussian components. Detailed descriptions of these parameters can be found in \citet{henshaw2016a}.

In the final stage, SCOUSEPY selects the best-fitting models with Akaike information criterion (AIC) for the 48141 individual spectra. As SAAs are Nyquist sampled, each of the individual spectra may hold multiple solutions of Gaussian Decomposition. The program identifies locations with multiple models and chooses the models with the lowest AIC values as the best-fitting ones. 

We first run SCOUSEPY with default tolerance levels in stage 3: we refer to these results as "version 0". This version generates reasonable fits for $\sim 80 \%$ of the spectra. However, given the complicated spectral structure and significant spatial variation of this dataset, the automated fitting could miss or overfit components at certain regions. For example, in version 0 SCOUSEPY identifies two adjacent peaks as a single velocity component a few pixels to the north of MM2. Therefore, a single set of tolerance levels may not be ideal for the whole dataset. To test the performance of automated fitting, we run stage 3) with another two sets of tolerance levels labelled as version 1 and version 2. 
In version 1, we relax the $T_3$ constraint on the maximum velocity dispersion of fitted components by 60\% compared with version 0; in version 2, we tighten the $T_4$ constraint on the fitted centroid velocity to 40\% of the value in version 0 (see Table~\ref{tab:scouse_input}). 
We compile all the results from the three versions and select the models with the lowest AIC values as the best-fitting ones. More than $79\%$ of the final best-fitting models are adopted from version 0; while $8\%$ are taken from version 1 and $3\%$ from version 2. 
For $\sim 9.7\%$ of the pixels within the mask applied in stage 1, SCOUSEPY could not find best-fitting models which satisfy all the conditions mentioned above. The tolerance levels applied in each version are summarized in Table \ref{tab:scouse_input}. The contribution of each version to the final results is shown in the left panel of Fig \ref{fig:scouse_res}. Examples of the SCOUSEPY fits to the \nhp\/ spectra are shown in Fig.~\ref{fig:fitted_spectra_sample}.

We note that SCOUSEPY is not designed to reproduce absorption and that Gaussian Decomposition is applied only to the \nhp\/ emission. The analyses described below also focus only on the \nhp\/ emission features.

\subsection{Results and basic statistics}
\label{sec:gd_results}
We obtain best-fitting models for 43079 spectra, and a total of 96157 Gaussian velocity components. 
This corresponds to an average of $\sim 2.2$ components per pixel. 
Indeed, more than $66\%$ of the spectra are fitted with models containing multiple velocity components. The right panel of Figure~\ref{fig:scouse_res} shows the distribution of the number of velocity components at each pixel. Pixels with zero components do not have accepted models (see Section~\ref{sec:gd_scousepy}). 
SCOUSEPY fits from 1 to 6 components to individual spectra (with no maximum number of components imposed), with a median number of components of 2. 
As illustrated by the right-hand panel of Figure~\ref{fig:scouse_res},
the complexity of the velocity structure varies spatially.
The vicinity of MM2 is among the regions with the most complex spectra: to the north of MM2, the median number of components/pixel reaches 4, and to the south of MM2 the median number of components/pixel is 3.

\begin{figure}
	\includegraphics[width=\columnwidth]{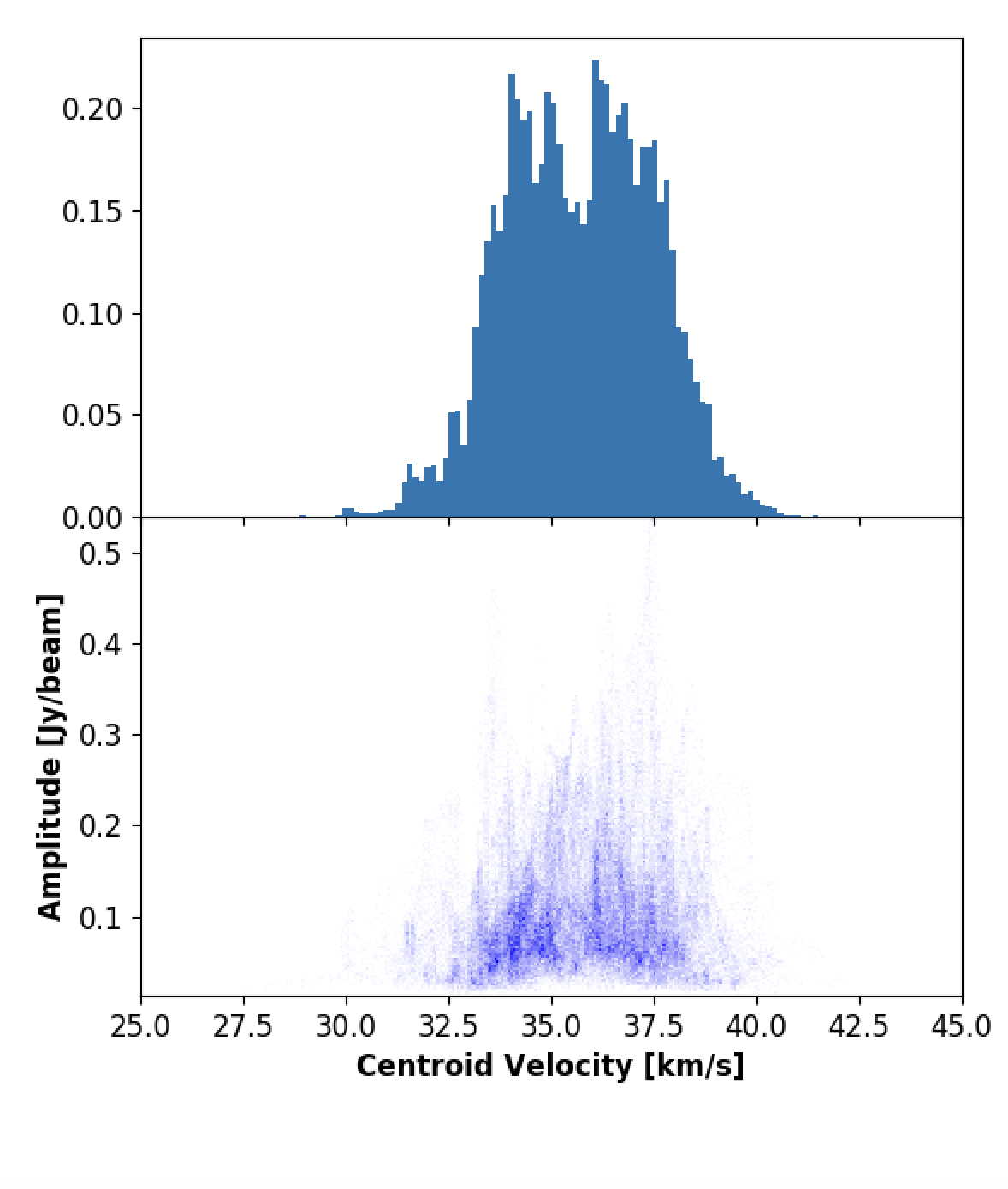}
    \caption{Top: histogram of centroid velocities for the best-fitting components (bin size: 0.25 \kms\/). Bottom: fitted amplitudes of velocity components versus centroid velocity.}
   \label{fig:densityv}
\end{figure}

\begin{figure}
	\includegraphics[width=\columnwidth]{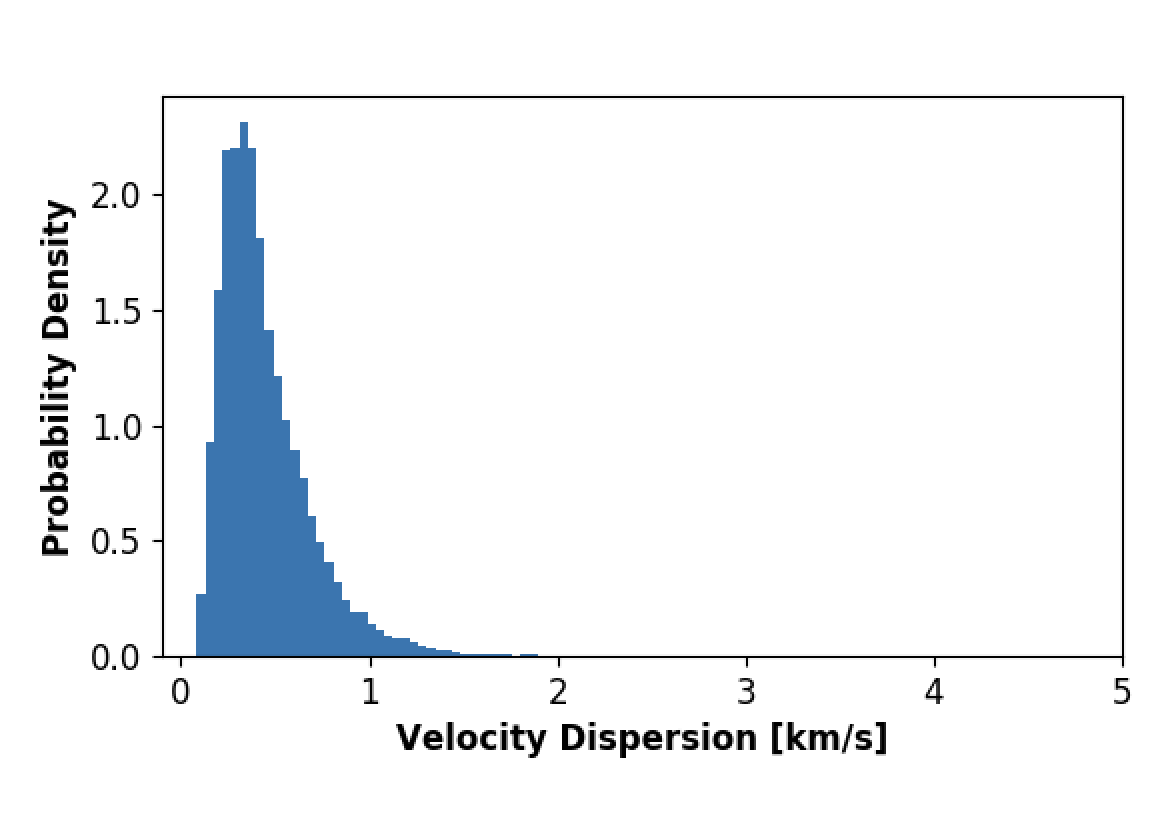}
    \caption{Histogram of velocity dispersion for the best-fitting components (bin size: 0.06 \kms\/). }
   \label{fig:lwv}
\end{figure}

The top panel of Figure \ref{fig:densityv} presents a histogram of the centroid velocities of the best-fit components. 
The centroid velocities range from 27.6 to 42.1 \kms\/, with a mean of 35.7 \kms\/, a standard deviation of 1.8 \kms\/, and an interquartile range of 2.7 \kms\/. 
The velocity histogram deviates from a Gaussian-like distribution in having 
two main
groups of 
peaks, at $\sim 34.5$ and $\sim 37$ \kms\/,  
possibly indicative of 
large-scale velocity gradients across the region.  
Sharp spikes and shoulders are also  
visible,  
reflecting complex velocity substructures on small spatial scales. 
The bottom panel of Figure \ref{fig:densityv} plots amplitude against centroid velocity for all fitted components. The fitted amplitudes range from $\sim$12.5 \mjyb\/ to $\sim$539.4 \mjyb\/, with a median of $\sim$ 93.3 \mjyb\/. 
The strongest components generally fall within the interquartile range in velocity, while the components on the tails of velocity distribution are generally weaker.

\begin{figure}
	\includegraphics[width=\columnwidth]{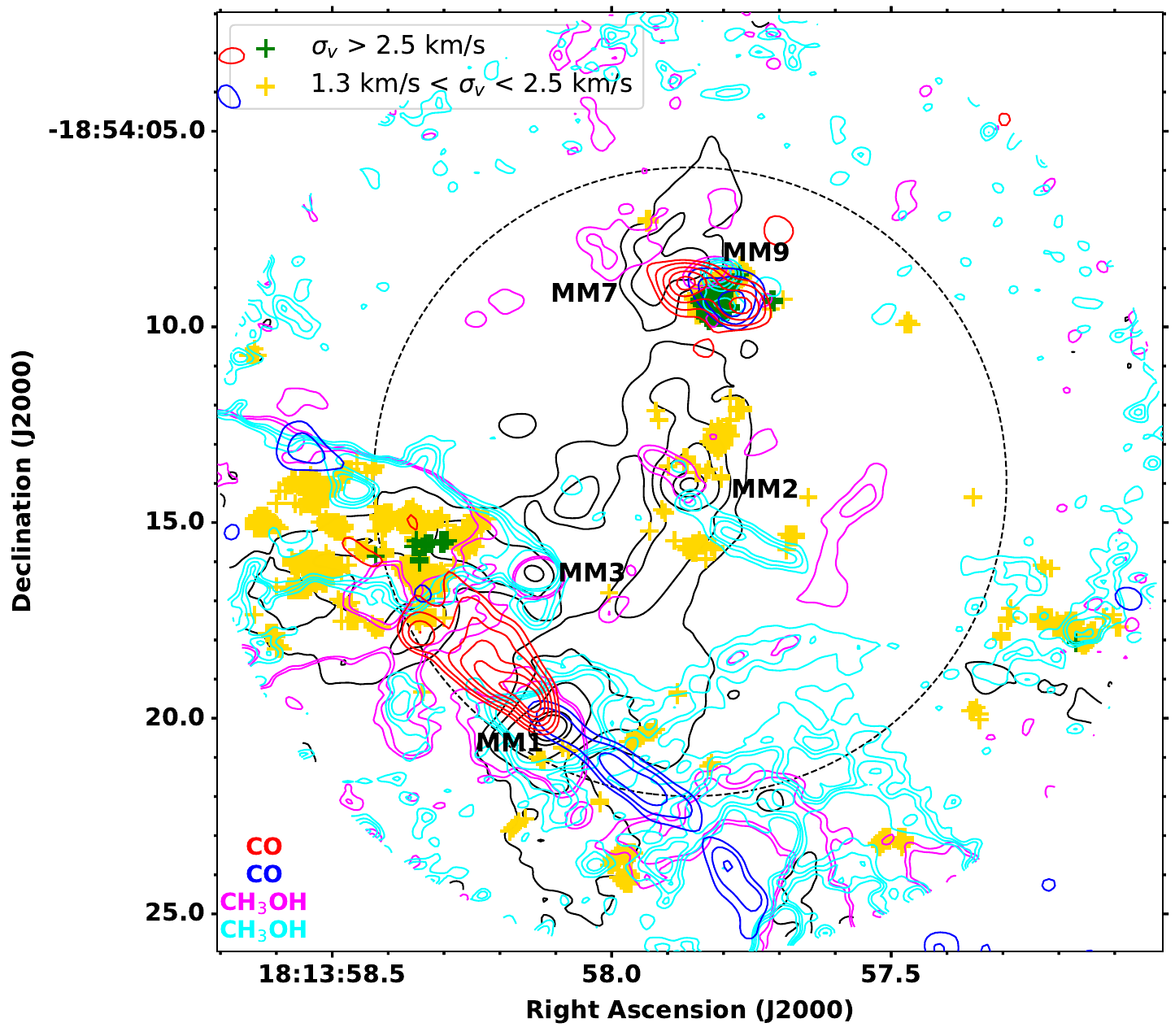}
    \caption{Spatial distribution of fitted velocity components with high velocity dispersions: yellow crosses mark 1.3~\kms\/ < $\sigma_v$ < 2.5~\kms\/ and green crosses mark $\sigma_v$ > 2.5~\kms\/, where 1.3~\kms\/ and 2.5~\kms\/ are the 99$^{th}$ and 99.9$^{th}$ percentile of the velocity dispersion distribution. Contours show the ALMA 0.82~mm continuum (black, levels: 0.5\mjyb $\times$ [5,15,40,160,280]), SMA blue- and red-shifted $^{12}$CO (3-2) line emission from \citet{cc2014,cc2017} (blue: [4,6,9] $\times$ 1.0 \jykms\/, red: [4,6,9,12,15,18] $\times$ 1.0 \jykms), and ALMA blue- and red-shifted \meth\/ (4$_{-1,3}$-3$_{0,3}$) line emission from \citet{cc22} (cyan: [4,7,10,15] $\times$ $\sigma$ = 5.5~\mjykms\/, magenta: [4,7] $\times$ $\sigma$ = 3.3~\mjykms). Contours are from images corrected for the primary beam response.  The dashed black circle shows the FWHP (50\% response) level of the ALMA primary beam, and the ALMA images are masked at its 20\% response level.}
   \label{fig:dv_tail}
\end{figure}

The fitted velocity dispersion ranges from 0.1 to 4.6 \kms\/, with a median of 0.4 \kms\/ and an interquartile range of 0.3 \kms\/. As shown in Figure \ref{fig:lwv}, the histogram of $\sigma_v$ is highly asymmetric, with a tail extending to 12$\times$ the median value. This asymmetry is also reflected in a positive skewness of $\sim 2.6$. 
We therefore present median and interquartile values instead of mean and standard deviation values to better describe the concentration and dispersion of $\sigma_v$. Investigating the spatial distribution of the high-velocity-dispersion tail, we find that the highest-dispersion components (99.9th percentile, $\sigma_v$ > 2.5~\kms\/), shown as green crosses in Figure~\ref{fig:dv_tail}, are 
associated with the protostellar outflows from MM7/9 and from MM1 and/or MM3 \citep{cc2011a,cc2014,cc2017}.
As shown in Figure \ref{fig:dv_tail}, many components with 1.3~\kms\/ < $\sigma_v$ < 2.5~\kms\/ (99th to 99.9th percentile, shown as yellow crosses) are also spatially associated with the outflows from MM1/3/7/9 or with MM2's outflow, though some broad components are also distributed along the north-southeast filamentary structure near MM2.    
(We focus here on high-dispersion points within the FWHP extent of the ALMA primary beam, where the data are most sensitive, see Fig.~\ref{fig:dv_tail}.) 
The spatial correlation between many high-velocity-dispersion components and known protostellar outflows indicates that we are observing the imprint of protostellar feedback 
on the dense molecular gas, which will be discussed in detail, in conjunction with the bow-like feature mentioned in Section~\ref{sec:n2hp_results}, in a separate publication.  

\begin{figure*}
	\includegraphics[width=1.0\textwidth, height=0.7\textheight]{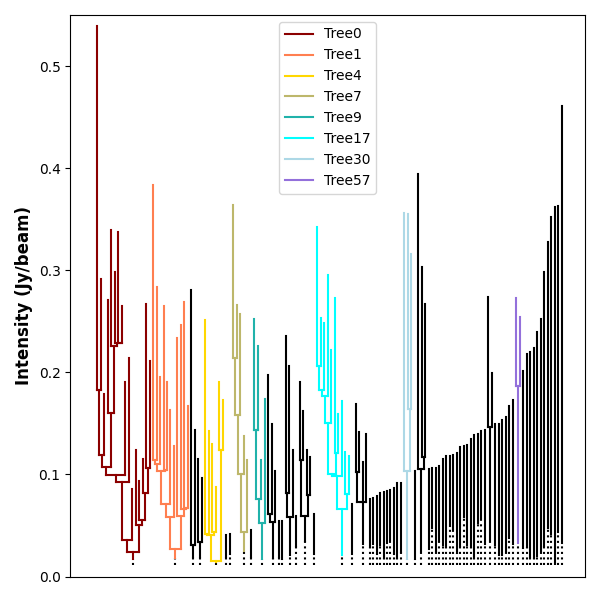}
    \caption{Dendrogram of the hierarchical system in the \nhp\/ (4-3)-emitting gas identified by ACORNS, referred to a ``forest''. The forest consists of 70 trees, which are position- and velocity-coherent clusters. $\sim$23$\%$ (16/70) of the trees have branches or leaves (i.e. substructure). The 8 most dominant trees (Section~\ref{sec:hc_results}) are plotted in colour. Substructure is indicated by short horizontal lines and the highest points of the leaves represent their peak intensities.}
   \label{fig:dendro}
\end{figure*}

\section{Hierarchical Clustering of velocity components}
\label{sec:hierarchical_clustering}
\subsection{Hierarchical clustering with ACORNS}
\label{sec:hc_acorns}
To further analyse the velocity structure within the \nhp\/-emitting gas, we conduct hierarchical clustering of the velocity components extracted in Section~\ref{sec:gaussian_decomposition} using  ACORNS\footnote{ACORNS is a publicly available Python package, and can be downloaded from https://github.com/jdhenshaw/acorns.} (Agglomerative  Clustering  for  Organising  Nested  Structures, \citealt{henshaw2019}). ACORNS is designed to characterise the hierarchical structure within discrete spectroscopic data (e.g. data points in position-position-velocity space) based on the hierarchical agglomerative clustering technique. A comprehensive description of the philosophy and parameters of this package can be found in \cite{henshaw2019} Appendix B.

Before running the ACORNS clustering, we perform a cleaning of the velocity components extracted by SCOUSEPY. We discard all pixels without fits ($\sim5 \%$ of the total SCOUSEPY data), remove components with peak intensities $<$3$\sigma$ (where $\sigma$ is the rms, measured individually for each spectrum), and remove components where either the peak intensity or the velocity dispersion is smaller than its associated uncertainty. We then conduct the ACORNS clustering only on the remaining velocity components, which represent $\sim$94\% of the total number of extracted components.

We first perform ACORNS clustering of the SCOUSEPY fits in position-position-velocity space, with parameters based on the spatial and spectral resolution of our \nhp\/ data. 
The minimum radius of a cluster is set to 4$\times$ the pixel size to ensure that the minimum number of data points in a cluster is comparable to the number of pixels in a synthesised beam ($\sim$52).  
The minimum height above the merge level is set to be 3$\times$ the mean rms and
the maximum absolute velocity difference between two components identified as linked is set to 0.2 \kms, the channel width of the \nhp\/ image cube (Section~\ref{sec:obs}). 
We add an additional clustering criterion based on 
the velocity dispersion, requiring the absolute difference in velocity dispersion between two linked components to be smaller than 0.2 \kms.  
With these parameters, ACORNS identifies a total of 198 clusters, which contain $\sim95 \%$ of the input velocity components. 

To investigate the sensitivity of the ACORNS results to the choice of input parameters, we performed additional clustering runs varying the peak intensity cutoff and the relaxation of linking lengths.  In these tests, we found that applying a 5$\sigma$ (rather than 3$\sigma$) threshold to the velocity components excluded a significant amount of data ($\sim$23\%) and so did not reflect the extended structure of clusters.  We then tested the clustering criteria for position, velocity, and velocity dispersion with different relaxation levels of linking lengths, relaxing the linking criteria by 20\% and 50\% to enable clusters to grow. 
These tests showed that the clustering results are relatively robust against relaxation: the percentage of data points assigned to clusters grew by only $\sim$0.3\% and $\sim$0.8\% in the 20\% and 50\% relaxed versions respectively, indicating that the vast majority of pixels are clustered without the need for relaxing the criteria. 
Based on these results, we adopt the original clustering run described above, with a 3$\sigma$ cutoff and without relaxing the linking criteria, for the remainder of our analysis. 

\subsection{Results and statistics}
\label{sec:hc_results}
Figure~\ref{fig:dendro} shows the hierarchical system (referred to as a ``forest'') identified by the ACORNS PPV clustering. 
In general, a dendrogram "forest" consists of a number of "trees"; each tree may have substructure(s), which are referred to as ``branches'' if they have further substructure(s), or ``leaves'' if they do not \citep[see e.g][]{rosolowsky2008,goodman2009}. 
Trees with no substructure are also categorised as leaves. 
Our ACORNS PPV clustering identifies a total of 70 trees, 
of which 16 ($\sim$23\%)  have substructure, i.e.\ branches or leaves.  
Eight of the identified trees together comprise $\gtrsim$60\% of the data points used for the clustering analysis. 
These trees, shown in colour in Figure~\ref{fig:dendro}, are designated Tree 0, Tree 17, Tree 1, Tree 4, Tree 9, Tree 7, Tree 30, Tree 57 (listed in order of decreasing number of assigned data points). 
In this paper, we focus our analysis on Tree 0, the dominant feature of the hierarchical system. 
Tree 9 will also be discussed in Section~\ref{sec:multi_fil_acc_flow} as supporting evidence for the physical scenario proposed in this work. Analysis of other trees and their substructures will be presented in separate publications focusing on different scientific topics.

\begin{figure*}
	\includegraphics[width=1.0\textwidth, height=0.3\textheight]{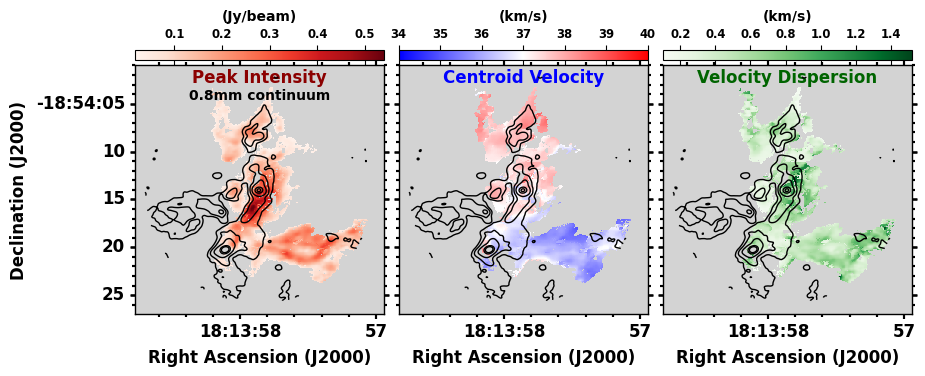}
    \caption{Left: distribution of peak intensities of the velocity components of Tree 0.  Middle: centroid velocity map of Tree 0. Right: velocity dispersion map of Tree 0. ALMA 0.82~mm continuum contours (black, levels: 0.5\mjyb $\times$ [5,15,40,160,280]) are overlaid in all panels.}
   \label{fig:tree0_sum}
\end{figure*}

Tree 0 contains $\approx20\%$ of the total velocity components, the most of any tree, and the components with the highest peak intensity. 
Figure~\ref{fig:tree0_sum} shows maps of the spatial distribution of the peak intensities, centroid velocities, and velocity dispersions of the velocity components in Tree 0. 
The maximum peak intensity in Tree 0 is  539.4~\mjyb\/ and the median peak intensity is 122.7~\mjyb\/. 
The centroid velocities of the components in Tree 0 span a relatively narrow range of 35.0 \kms\/ to 38.8 \kms\/, compared to the range of 27.6 \kms\/ to 42.1 \kms\/ for all fitted components (see Section~\ref{sec:gd_results}).
The median velocity of Tree 0, which is 37.0 \kms\/, falls within the higher-velocity of the two main peaks seen in Figure~\ref{fig:densityv}. 
The median velocity dispersion of Tree 0 agrees with that of the full dataset ($\sim$0.4~\kms), but Tree 0 lacks the high-dispersion tail seen in Figure~\ref{fig:lwv}, with a maximum velocity dispersion of 1.5~\kms.  
Spatially, Tree 0 appears to trace the major features in the moment 8 map (Fig.~\ref{fig:obs_res}b), including the main north-southeast filamentary structure and the extended emission that lies to the west of the southern end of the main filament. 
The strongest emission in Tree 0 is located in the filamentary structure to the southeast of MM2 (see Fig.~\ref{fig:tree0_sum}). 
Notably, the centroid velocity map of Tree 0 (shown in the middle panel of Fig.~\ref{fig:tree0_sum}) reveals a roughly north-south velocity gradient along the main filamentary structure, which we consider further in Section~\ref{sec:velocity_gradient}.  
The most notable feature of the velocity dispersion map 
(right panel of Fig.~\ref{fig:tree0_sum}) is an area of enhanced velocity dispersion ($\sim$ 50-70\%) around MM2, within a radius of $\sim$3\arcsec. 

\begin{figure}
	\includegraphics[width=1.0\columnwidth, height=0.35\textheight]{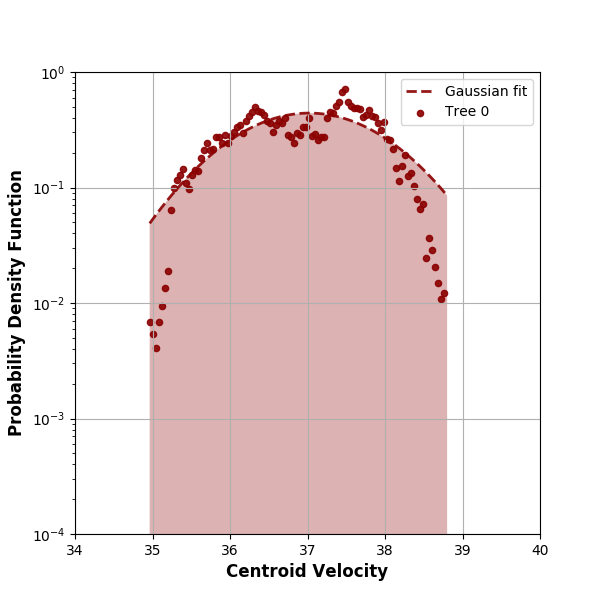}
    \caption{Probability Distribution Function (PDF) of the centroid velocities of the components in Tree 0 (filled circles), overlaid with the best-fit Gaussian model to the velocity PDF (dashed line).}
   \label{fig:tree0_vpdf}
\end{figure}

Figure \ref{fig:tree0_vpdf} shows the Probability Distribution Function (PDF) of the centroid velocities of the components in Tree 0.  
As illustrated by Figure~\ref{fig:tree0_vpdf}, the velocity PDF is not well-represented by a Gaussian model, with excesses at $\sim$36.3 \kms\/ and $\sim$37.8 \kms\/, and moderation in the tails. 
The kurtosis of the velocity PDF of Tree 0 is estimated to be 2.3; a value $<$3  (the kurtosis of the Gaussian distribution) implies lighter tails, consistent with the features of the PDF. While simulations of turbulence suggest Gaussian-like velocity PDFs (e.g. \citealt{fed2013}), observations do not always agree (e.g. \citealt{fed2016,henshaw2019}). In their study of the Galactic Centre molecular cloud G0.253+0.016, \cite{fed2016} interpret double-peaked deviations from a Gaussian distribution as the result of a large-scale velocity gradient, and note that observational noise, the excitation of the molecular tracer, and smaller-scale systematic motions may also contribute to deviations from a Gaussian model \citep[see also][]{henshaw2019}. 
We suggest that the double-peak feature in the Tree 0 velocity PDF may be caused by the large-scale velocity gradient seen in Figure \ref{fig:tree0_sum}, similar to the case in \citet{fed2016}. 

\begin{figure*}
	\includegraphics[width=0.8\textwidth, height=0.4\textheight]{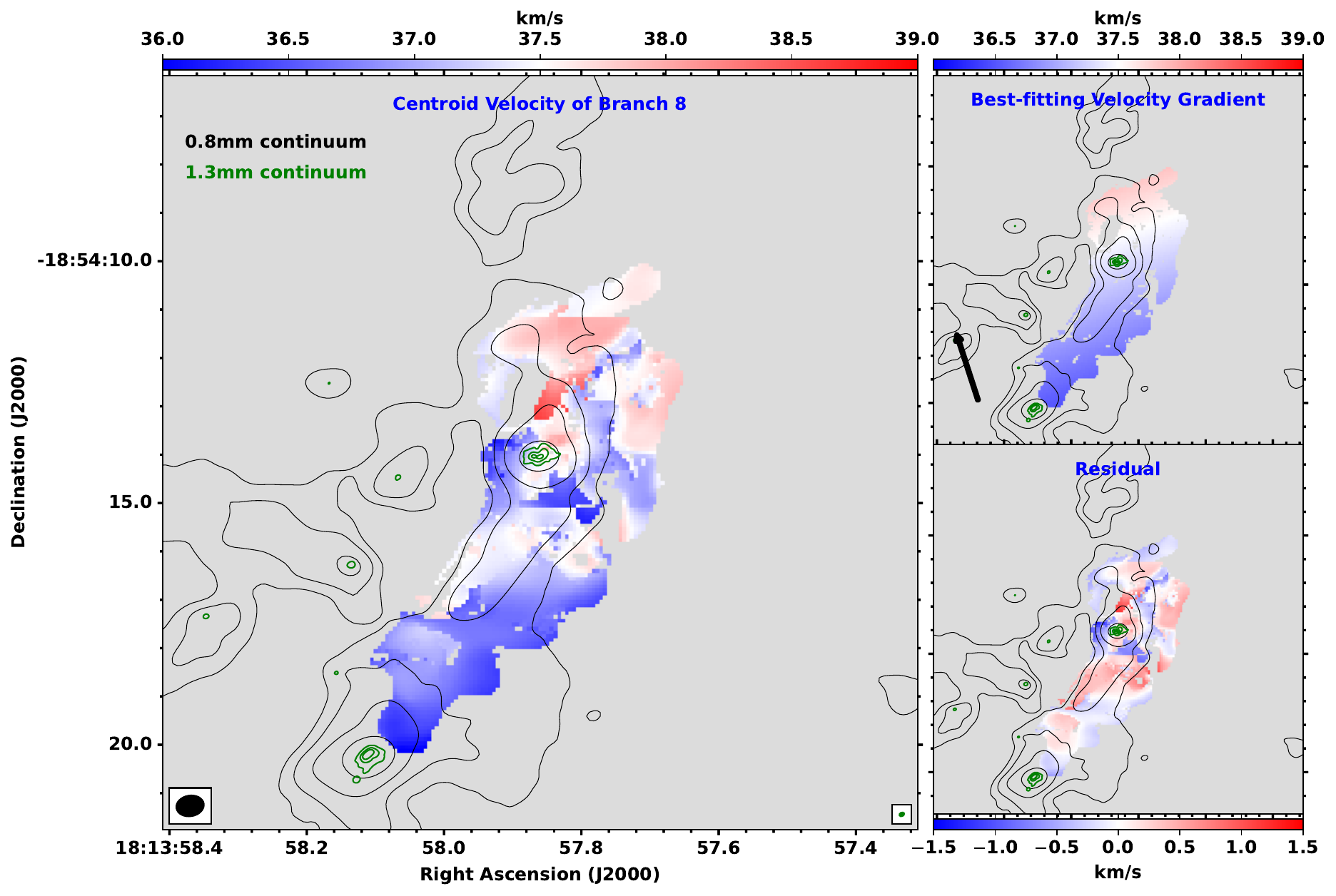}
    \caption{Left: centroid velocity map of Branch 8. Top right: map of best-fit velocity gradient; its direction is indicated by the black arrow. Bottom right: residual map, generated by subtracting the best-fitting velocity gradient from the observations.  In all panels, ALMA 0.8~mm continuum contours (black; contour levels: 0.5 $\times$ [5,15,40,160,280] \mjyb) and ALMA 1.3~mm continuum contours from \citet{ilee2018} (green; contour levels: 0.06 $\times$ [10,50,150] \mjyb) are overlaid.  The synthesised beams of the 0.8~mm and 1.3~mm data are shown in the left-hand panel, at bottom left and right, respectively.}
   \label{fig:branch8_v}
\end{figure*}

\section{Filamentary accretion flows}
\label{sec:fil_acc_flows}
\subsection{Velocity gradient}
\label{sec:velocity_gradient}
As discussed in Section~\ref{sec:hc_results}, the centroid velocity map of Tree 0 shows a $\sim$north-south velocity gradient, consistent with 
this tree's double-peaked velocity PDF. 
As shown in Figure \ref{fig:tree0_sum} (middle panel), this velocity gradient goes through MM2 and extends to MM1 to the south  
and to the MM7/MM9/MM13/MM17 group of sources to the north of MM2. 
To analyse the gas kinematics in the surroundings of MM2, we select a position- and velocity-coherent substructure (a ``branch'') of Tree 0 that 
encompasses the immediate environment of MM2, designated Branch 8. 
The centroid velocity map of this branch is shown in  Figure~\ref{fig:branch8_v}. 
On the plane of the sky, Branch 8 is $\sim$0.17~pc long and $\sim$0.05~pc wide. 
It hosts the brightest velocity components of Tree 0 (with a peak intensity of 539.4~\mjyb\/), and has a median intensity of 209.1~\mjyb\/, which is higher than the median intensity of Tree 0 by a factor of $\sim$1.7. 
The centroid velocities in Branch 8 range from 35.9 \kms\/ to 38.6 \kms\/ with a median value of 37.3 \kms\/, covering $\sim$71$\%$ of the velocity range of Tree 0. Branch 8 also contains the broadest velocity components of Tree 0, with a maximum velocity dispersion of 1.5~\kms\/ and a median of 0.5~\kms\/. 

We measure the magnitude and direction of the velocity gradient of Branch 8 using the methods proposed by \cite{goodman1993}. A linear gradient of centroid velocities is described by the equation
\begin{equation}
    v = v_0 + a\Delta\alpha + b\Delta\delta ,
\end{equation}
where $\Delta\alpha$ and $\Delta\delta$ are the variations in right ascension and declination in units of radians; $a$ and $b$ are the projections of the velocity gradient per radian on the axes of right ascension and declination, respectively; and $v_0$ is the systemic velocity. The magnitude of the velocity gradient $|\nabla v|$ is given by
\begin{equation}
    |\nabla v| = \frac{\sqrt{a^2 + b^2}}{D} , 
\end{equation}
where $D$ is the distance to the source (3.37~kpc, Section~\ref{sec:intro}). The direction of increasing velocity (east of north) is
\begin{equation}
    \Theta = {\rm arctan}(\frac{a}{b}).
\end{equation}
We estimate the values of $a$ and $b$ by fitting a first-degree bivariate polynomical to the centroid velocity map of Branch 8 using LMFIT\footnote{LMFIT is a Python package designed for non-linear least-squares minimisation and curve-fitting based on scipy.optimize. The package is publicly available via \url{https://github.com/lmfit/lmfit-py}.}. The uncertainties associated with the centroid velocities from SCOUSEPY are included to constrain the fitting. 
We estimate the velocity gradient exhibited by Branch 8 to be 10.5$\pm$0.2 km s$^{-1}$ pc$^{-1}$ in a direction of $\sim$18.9\arcdeg\/ (east of north) over a spatial scale of $\sim$0.17~pc. The best-fitting model is shown in the top right panel of Figure~\ref{fig:branch8_v}, with the black arrow indicating the direction of the fitted velocity gradient. 
The bottom right panel of Figure~\ref{fig:branch8_v} shows the residual map, calculated by subtracting the best-fitting model from the centroid velocity map. The residuals range from $-$1.3~\kms\/ to 1.2~\kms\/.

In other filamentary high-mass star-forming regions, 
the magnitudes of velocity gradients measured along (sub)filaments 
vary by an order of magnitude, with a trend of larger velocity gradients at smaller spatial scales: $\sim$0.2-0.9 km s$^{-1}$ pc$^{-1}$ at $>$1-8~pc scales (e.g., \citealt{bally1987}, \citealt{peretto2014}, \citealt{tackenberg2014}, \citealt{zhang2015}, \citealt{hacar2017}, \citealt{yuan2018}, \citealt{chen2019}, \citealt{trevin2019}, \citealt{hu2021}, \citealt{beltran2022}), $\sim$0.7-0.8 km s$^{-1}$ pc$^{-1}$ at $\sim$1~pc scales (e.g., \citealt{henshaw2014}, \citealt{fl2014}), and $\gtrsim$1-20 km s$^{-1}$ pc$^{-1}$ at $\sim$0.1-0.5~pc scales (\citealt{henshaw2013}, \citealt{liu2016}, \citealt{williams2018}, \citealt{hacar2018}, \citealt{Li2022}). 
This finding is reinforced by a recent systematic study of the gas kinematics of a large sample of filaments using ALMA H$^{13}$CO$^+$ data, where the magnitude of the velocity gradients increases from a few $\times$ $\sim$ 1 km s$^{-1}$ pc$^{-1}$ to a few $\times$ $\sim$ 10 km s$^{-1}$ pc$^{-1}$ as the extent over which the velocity gradients are measured decreases from $\sim$1~pc to $\sim$0.1~pc (\citealt{zhou2022}). 

\citet{zhou2022} argue that the smaller velocity gradients at larger scales are likely caused by pressure-driven inertial inflows shaped by either turbulence or gravity at large spatial scales, while the larger velocity gradients at smaller scales are likely attributable to  the gravity of dense structures (e.g., cores and/or the hubs of hub-filament systems). The latter scenario agrees well with our measurements for Branch 8 ($\sim$10.5 km s$^{-1}$ pc$^{-1}$ over $\sim$0.17~pc). For comparison, we also estimate the velocity gradient of Tree 0 using the same technique, which gives 9.3$\pm$0.1 km s$^{-1}$ pc$^{-1}$ in a direction of $\sim$34.7\arcdeg\/ (east of north) over $\sim$0.47~pc. The magnitude of the velocity gradient increases moderately (1.2 km s$^{-1}$ pc$^{-1}$, $\sim$13$\%$) when the spatial scale decreases by 0.3~pc ($\sim$60$\%$). 
We note that the variation in observed velocity gradients along filaments may also be influenced by projection effects: 
the observed magnitude of the velocity gradient along a filament will be smaller if the orientation of the filament is closer to the plane of sky. 

Simulations and observations suggest that gravity is dynamically more important at higher densities and smaller spatial scales \citep[e.g.][]{smilgys2016,chen2019}. \cite{williams2018} find that $>$60\% of the dense cores in the IRDC SDC13 are associated with the peaks of velocity gradients traced by \nht\/, which is interpreted as a signature of cores accreting material from their parental filaments. We note that the southeast end of Branch 8 reaches the 160$\sigma$ 0.8~mm continuum contour around the proto-O star MM1, roughly $\sim$1700~AU from the centre of MM1. 
The gravity of MM1 (with an enclosed mass of $\sim$40~\msun\/, \citealt{ilee2018}) is expected to dominate over pressure support in its local ($\lesssim$0.1-0.6~pc) environment (\citealt{williams2018}, \citealt{chen2019}). 
As a test of minimising the effects of MM1, we measured the velocity gradient adopting the southern end of the 15$\sigma$ contour extending from MM2 to MM1 as a boundary, and obtain a 
value of 8.4$\pm$0.2 km s$^{-1}$ pc$^{-1}$ in the direction of $\sim$14\arcdeg\/ (east of north). This suggests that introducing an artificial boundary has a limited ($\sim$20\%) effect on the velocity gradient measurement. The Global Hierarchical Collapse model (\citealt{vs2017, vs2019}) also suggests that global collapse driven by the large-scale gravitational potential well and local collapse driven by small-scale overdensities take place concurrently. 
In this scenario, the observed velocity gradient of Branch 8 would most likely be caused by a combination of large-scale gas flows towards the centre of the massive cluster and small-scale gas flows towards both MM1 and MM2 (as seen in \citealt{peretto2014}). 
In this context, it would be difficult to define a physically meaningful boundary within a continuity of gas flows. We thus adopt the measurement using the full extent of Branch 8 in our analysis.

\subsection{Mass inflow rate}
\label{sec:mass}
Velocity gradients along filaments are commonly interpreted as signatures of longitudinal gas flows (e.g., \citealt{kirk2013}, \citealt{peretto2014}, \citealt{liu2016}, \citealt{lu2018}, \citealt{yuan2018}, \citealt{chen2019}, \citealt{trevin2019}, \citealt{chenm2020}, \citealt{hu2021}). 
If the filamentary structure observed in G11.92$-$0.61 is oriented such that its southeast end is further from the observer than its northwest end, the observed velocity gradient in Branch 8 is consistent with  accelerating accretion flows towards MM2\footnote{\cite{henshaw2014} interpret the symmetrical velocity gradients (with a mean magnitude of $\sim$2.5 km s$^{-1}$ pc$^{-1}$) towards the SW core in IRDC G035.39-00.33 
as either accretion flows along filaments towards the SW core, or an expanding shell of dense gas driven by 
protostellar outflows and/or stellar winds. The expanding shell scenario is supported by the detection of 8 and 24~\um\/ emission (indicative of the presence of young stars) and a high-velocity redshifted wing traced by CO (1-0) (indicative of protostellar outflows) towards the SW core. 
As introduced in Section~\ref{sec:intro}, \gm\/ hosts an early-stage high-mass protobinary system that drives a low-velocity bipolar outflow. 
Compared with the SW core, MM2 is on an earlier evolutionary stage; its protostellar feedback is thus less likely to drive an expanding shell of dense gas with a velocity gradient of $\sim$10.5 km s$^{-1}$ pc$^{-1}$, which is higher than that reported by \cite{henshaw2014} by a factor of $\sim$4. In addition, the observed $\sim$south-north velocity gradient in Branch 8 is unlikely to be caused by the $\sim$west-east bipolar outflow associated with MM2 (see Figure \ref{fig:dv_tail}). 
}. 

Based on the hypothesis of filamentary accretion flows, we can derive the mass inflow rate using the velocity gradient and mass of Branch 8 following the procedure in \cite{kirk2013}.
We estimate the gas mass of Branch 8 from its 0.82~mm  
integrated flux density, which is measured by integrating the intensity within the boundary of Branch 8 using the CASA task \texttt{imstat}, under the simple assumptions of optically thin, isothermal dust emission as \citep[see e.g][]{cc2011a,cc2017}:
\begin{equation}
    M_{\rm gas}({\rm M_{\odot}})=\frac{4.79\times10^{-14}RS_{\nu}({\rm Jy})D^2({\rm kpc})}{B(\nu, T_{\rm dust})\kappa_{\nu}},
	\label{eq:mass}
\end{equation}
where $R$ is the gas-to-dust mass ratio (assumed to be 100), $S_{\nu}$ is the integrated flux density, and $D$ is the distance to G11.92$-$0.61.  We adopt $\kappa_{\rm 0.82~mm}=2.3~{\rm cm^2~g^{-1}}$, based on linear interpolation of the \citet{ossenkopf1994} values for MRN grains with thick ice mantles and \hden\/ = $10^6$ \perccc\/ (column 7 of their Table 1), and a dust temperature range $T_{\rm dust}=$15--25~K. 
Our adopted dust temperature range is informed by clump-scale NH$_3$ temperature measurements for G11.92$-$0.61 (T$_{\rm kin}\sim$26~K from single-component fitting and T$_{\rm kin}\sim$14~K and 54~K from two-component fitting, where the 54~K-component is attributable to the immediate environment of MM1; \citealt{cyganowski13}) and by temperature maps derived from VLA NH$_3$ observations of other EGOs \citep[e.g.][their Figure 3]{brogan11}.
With our adopted parameters, we estimate the mass of Branch 8 as $\sim$30-65~\msun. 
For a simple cylindrical model with a length of $L$, a mass of $M$, a velocity gradient observed along the main axis of the filament of $\nabla v$, and an inclination angle to the plane of sky of $\alpha$, the rate of mass flow $\dot{ M}$ is given by \citep{kirk2013}: 
\begin{equation}
    \dot{M} = \frac{|\nabla v| M}{\rm tan (\alpha)},
    	\label{eq:mdot}
\end{equation}
where $|\nabla v|$ is the magnitude of the velocity gradient. Based on our estimates of the branch mass and velocity gradient (above and Section~\ref{sec:velocity_gradient}) and considering a range of plausible inclination angles,
we derive mass accretion rates of $1.8 \times 10^{-4}$ \msun\/ yr$^{-1}$ (for $T_{\rm dust}$=25~K and an inclination angle $\alpha$=60$^{\circ}$) to $1.2\times 10^{-3}$ \msun\/ yr$^{-1}$ ($T_{\rm dust}$=15~K and $\alpha$=30$^{\circ}$). 

We note that in addition to the uncertainties in inclination angle and dust temperature reflected in the range quoted above, our estimates of $\dot{ M}$ are also affected by other sources of uncertainty in our mass estimates.  We estimate a factor of $\sim$2 systematic uncertainty, encompassing the uncertainties in the adopted $\kappa$ and gas-to-dust mass ratio and the uncertainty associated with the isothermal assumption \citep[see also e.g.][]{brogan2009,beuther2021}. 
While assuming optically thin dust emission can lead to underestimating masses on core scales \citep[e.g.][]{cc2014,brogan2016}, the optically thin assumption is likely to be valid over the vast majority of the area of Branch 8, which has a mean brightness temperature\footnote{Measured from the brightness temperature map computed using the beamsize and the \texttt{tt.brightnessImage} function of \texttt{toddTools}.} of only 0.35~K \citep[see also e.g.][]{beuther2018}. 
We also note that we have not attempted to impose a boundary between 0.82~mm continuum emission associated with the filament and with the MM2 core.  The integrated flux density reported for MM2 in Section~\ref{sec:cont_results} is a fit to the total emission at the core position, so includes a contribution from the background filament, and our measurement of the integrated flux density of Branch 8 includes the area of the compact core. 

\begin{figure*}
	\includegraphics[width=1.0\textwidth, height=0.3\textheight]{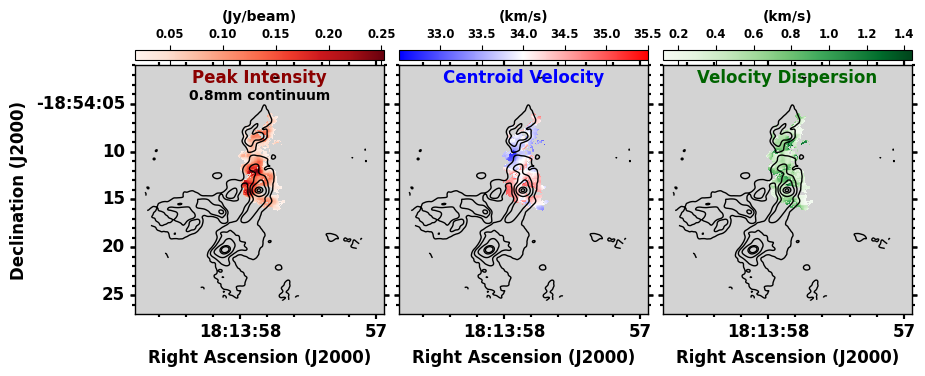}
    \caption{Left: distribution of peak intensities of the velocity components of Tree 9.  Middle: centroid velocity map of Tree 9. Right: velocity dispersion map of Tree 9. ALMA 0.82~mm continuum contours (black, levels: 0.5\mjyb $\times$ [5,15,40,160,280]) are overlaid in all panels.}
   \label{fig:tree9_sum}
\end{figure*}

Reported mass accretion rates along filaments vary by two orders of magnitude, from $\sim10^{-5}$ \msun\/ yr$^{-1}$ (e.g., \citealt{kirk2013}, \citealt{peretto2014}, \citealt{henshaw2014}, \citealt{trevin2019}, \citealt{Li2022}) to $\sim 10^{-4}$ \msun\/ yr$^{-1}$ (e.g., \citealt{chen2019}, \citealt{lu2018}, \citealt{yuan2018}) to $\sim 10^{-3}$ \msun\/ yr$^{-1}$ (e.g., \citealt{liu2016}, \citealt{hu2021}). 
This range is likely attributable to 
the different total masses of these star-forming regions (e.g., low-mass or high-mass) combined with  
variations in velocity gradient measurements (e.g., tracer, spatial scale), mass calculations (e.g., based on molecular line or dust emission, clump mass or filament mass), and inclination angle assumptions. 
Our estimated $\dot{ M}$ of $1.8 \times 10^{-4}$ \msun\/ yr$^{-1}$ to $1.2\times 10^{-3}$ \msun\/ yr$^{-1}$ is higher than the values reported in early studies of filamentary accretion flows 
in high-mass IRDCs \citep[$\sim$(2.5-7)$\times 10^{-5}$ \msun\/ yr$^{-1}$, e.g.][]{henshaw2014,peretto2014}.  The typical accretion rates of (0.5-3.5) $\times 10^{-4}$ \msun\/ yr$^{-1}$ found by \citet{lu2018} in their study of NH$_3$ filaments in high-mass star-forming clouds also fall at the low end of our estimated range.
We note that our estimated velocity gradient is also higher ($\sim$ 10.5 km s$^{-1}$ pc$^{-1}$ compared to $\sim$ 1 - 2.5 km s$^{-1}$ pc$^{-1}$ for the clouds studied by \citealt{henshaw2014,peretto2014,lu2018}), which would naturally lead to higher estimated accretion rates for filaments of roughly comparable mass (equation~\ref{eq:mdot}).  \citet{liu2016} report a similarly large velocity gradient ($\sim$10 km s$^{-1}$ pc$^{-1}$) in filaments identified in NH$_3$(1,1) in the high-mass cluster-forming region AFGL 5142.  Adopting the clump mass as a lower limit to the total mass of filaments, \citet{liu2016} derive a filamentary mass accretion rate of $\sim 2.1\times 10^{-3}$ \msun\/ yr$^{-1}$, within a factor of 2 of the high end of our estimates for G11.92$-$0.61 MM2. 

\subsection{Multiple filamentary accretion flows}
\label{sec:multi_fil_acc_flow}
In addition to Tree 0, Tree 9 also displays a velocity gradient along the main filamentary structure in a $\sim$north-south direction. 
Figure \ref{fig:tree9_sum} shows maps of the spatial distribution of the peak intensities, centroid velocities, and velocity dispersions of the velocity components in Tree 9 (as shown for Tree 0 in Figure~\ref{fig:tree0_sum}). 
The maximum peak intensity in Tree 9 is 252.1~\mjyb\/ and the median is 75.0~\mjyb\/. 
The centroid velocities of the components in Tree 9 fall within the lower-velocity of the two main peaks in Figure~\ref{fig:densityv}, ranging from 32.7 \kms\/ to 35.2 \kms\/ with a median of 34.0 \kms (in contrast to Tree 0, see Section~\ref{sec:hc_results}). 
The maximum velocity dispersion in Tree 9 is 1.4~\kms\/ and the median is 0.5~\kms\/. 
Though the brightest components of Tree 0 are about a factor of two stronger than those of Tree 9, the maximum and median velocity dispersions of the two trees are similar. 
Notably, the distinct centroid velocity ranges of the two trees (32.7 \kms\/ to 35.2 \kms\/ for Tree 9 and 35.0 \kms\/ to 38.8 \kms\/ for Tree 0) suggest that they are kinematically separate structures. 
We measure the velocity gradient of Tree 9 following the procedure described in Section~\ref{sec:velocity_gradient}, and obtain a velocity gradient with a magnitude of $\sim$9.0~\kms\/pc$^{-1}$ and a direction of $\sim$149\arcdeg\/ (east of north). 
If the structure traced by Tree 9 is oriented such that its northern end is further from the observer than its southern end, the observed velocity gradient is consistent with a
stream of gas flowing towards MM2. 
The distinct ranges of centroid velocities and the comparable velocity dispersions of Tree 0 and Tree 9 suggest a scenario in which MM2 is fed by multiple filamentary gas flows that partially overlap along the line of sight. This implies that the filamentary structure seen in the 0.82~mm dust continuum contains multiple kinematic substructures. 

Velocity-coherent substructures within filaments (often known as "fibres") have been identified in other high-mass star-forming regions, including Orion (where \citealt{hacar2018} identified 55 position- and velocity-coherent fibres within the integral-shaped filament (ISF) using \nhp\/ (1-0)), NGC 6334 (\citealt{shimajiri2019b}, \citealt{Li2022}) and the DR21 ridge in Cygnus X (\citealt{cao2022}). 
Hydrodynamic simulations with turbulence can reproduce filament substructures similar to those seen in observations, and demonstrate that they play a crucial role in channelling mass onto star-forming cores embedded within filaments (e.g., \citealt{smith2014}, \citealt{smith2016}, \citealt{clarke2020}).
However, simulation-based studies of the correspondence between fibres identified in Position-Position-Velocity space and coherent substructures in Position-Position-Position space have also emphasised the need for caution in interpreting observed filament substructures due to line-of-sight confusion and projection effects (e.g., \citealt{za2017} and \citealt{clarke2018}, with the latter  analysing synthetic C$^{18}$O observations). 

\section{Discussion}
\label{sec:discussion}
\subsection{Mass inflow feeding the protobinary}
\label{sec:binary_growth}
Our analysis of the gas kinematics in the surroundings of MM2 reveals position- and velocity-coherent substructures in the \nhp-emitting gas that we interpret as tracing filamentary accretion flows onto MM2 (Section~\ref{sec:fil_acc_flows}). As MM2 is now known to host a young protobinary system undergoing ongoing accretion \citep[Section~\ref{sec:intro}, see also][]{cc22}, we can use our estimates of the filamentary mass inflow rate to consider the timescales for filament-fed growth of the embedded protostars. 
\citet{cc22} estimate the current protostellar masses of the binary members to be $\sim$1\msun\/ each. 
If the overall efficiency factor for transporting the gas carried by the filamentary inflows onto the protostars is 50\%, the members of the protobinary will double their masses in $\sim$3.3$\times$10$^{3}$ to $\sim$2.2$\times$10$^{4}$ yr (where the range corresponds to the range in our  estimated filamentary accretion rate: $1.8 \times 10^{-4}$ \msun\/ yr$^{-1}$ to $1.2\times 10^{-3}$ \msun\/ yr$^{-1}$, Section~\ref{sec:mass}), comparable to the dynamical timescale of the CH$_3$OH outflow \citep[$t_{\rm dyn}\sim$4600 yr and 12,100 yr for the blue and red lobes, respectively;][]{cc22}.
Similarly, the time for each member of the protobinary to reach a mass of 8 \msun\/ would be $\sim$2.3$\times$10$^{4}$ to $\sim$1.6$\times$10$^{5}$ yrs, and the time for each member of the protobinary to reach 10 \msun\/  would be $\sim$3.0$\times$10$^{4}$ to $\sim$2.0$\times$10$^{5}$ yrs. Interestingly, using chemical clocks, \cite{sabatini21} derive timescales of $\sim5\times 10^{4}$ yr for the 70$\mu$m-weak phase and of $\sim1.2\times 10^{5}$ yr for the subsequent MIR-weak phase in the evolutionary stages of the ATLASGAL-TOP100 sample of massive star-forming clumps (\citealt{konig2017}).
While the uncertainties are considerable, we note that the combined timescale of the 70$\mu$m-weak and MIR-weak phases ($\sim$ 1.7$\times 10^{5}$ yr) is comparable to our upper estimates of the timescale for the binary members to become massive protostars (M$\gtrsim$8\msun). 

We note that the timescale estimates above consider mass aggregation processes involving multiple spatial scales: a) from filaments to cores and b) from cores to stars. 
The efficiencies of both processes are poorly constrained. 
Analysis of the relationship between the prestellar core population and the column density shows that only a small fraction of dense gas is contained in prestellar cores, for example, $\sim$15$\%$ in the Aquila Cloud Complex (\citealt{andre2014}) and $\sim$22$\%$ in Orion B (\citealt{konyves2020}). 
In the Global Hierarchical Collapse (GHC) scenario, filaments themselves are assembling mass from their surroundings while they transport material onto embedded star-forming cores (\citealt{vs2017, vs2019}). The efficiency of the mass transport from filaments to cores will depend on the dynamical timescales of the two structures. 

\subsection{Velocity gradient across filament}
\label{sec:perp_velocity_gradient}
The direction of the best-fitting velocity gradient for the centroid velocity map of Branch 8 is $\sim$19\arcdeg\/ east of north, offset by $\sim$40\arcdeg\/ from the main axis of Branch 8 (see top right panel of Fig. \ref{fig:branch8_v}). 
This offset suggests that the observed velocity gradient consists of a component that is aligned parallel to the main axis of Branch 8 ($\sim$southeast-northwest) and a perpendicular component ($\sim$southwest-northeast). 
As discussed above, the parallel component is most likely associated with longitudinal filamentary accretion flows towards MM2. 
Velocity gradients that are perpendicular to the main axes of filamentary structures have been observed in (sub)filaments in nearby (\citealt{palmeirim2013}, \citealt{kirk2013}, \citealt{fl2014}, \citealt{dhabal2018}, \citealt{dhabal2019}, \citealt{chen2020}, \citealt{chenm2020}) and high-mass (\citealt{schneider2010}, \citealt{beuther2015}, \citealt{williams2018}, \citealt{zhou2021}) star-forming regions. These perpendicular velocity gradients have been interpreted as signatures of asymmetric mass accretion flows onto (sub)filaments (\citealt{palmeirim2013}, \citealt{chenm2020}), which can be induced by the compression of shocked gas (\citealt{shimajiri2019a}, \citealt{dhabal2018}, \citealt{chen2020}), as direct results of compression flows (\citealt{schneider2010}, \citealt{williams2018}, \citealt{dhabal2019}, \citealt{zhou2021}), and as the combined result of mergers/collisions of velocity-coherent subfilaments and the accretion and rotation of filaments (\citealt{beuther2015}). 

As \g\/ is a young embedded massive protocluster
where UCHII and \HII\/ regions have not yet developed \citep{ilee2016} and is located $\sim$1~pc away from the ultra-compact HII region G11.94-00.62 (\citealt{churchwell1992}), 
the kinematics of dense filamentary structures in \g\/ will not yet be shaped by the feedback from 
either internal or external 
expanding \HII\/ regions. 
Branch 8 has a mass of $\sim$30-65~\msun\/ and a length of $\sim$0.17pc, yielding a mass per unit length of $M_{\rm line}\approx$174-380~\msun\/ pc$^{-1}$. These values are higher than the thermal critical values $M_{\rm line,crit}$ calculated with Equation (11) of \cite{kirk2013} for T$\sim$15-25~K by a factor of $\sim$4-15, suggesting that thermal pressure is insufficient to support Branch 8 against gravitational collapse. Branch 8 thus is likely to undergo radial collapse, which, if asymmetric, can also contribute to the observed perpendicular velocity gradient. 
We speculate that the perpendicular velocity gradient could be the combined result of the radial collapse, mass accretion and rotation of the filamentary structure, while accretion may dominate over rotation as it evolves (e.g., \citealt{kirk2013}). 
This is a picture consistent with the GHC scenario (\citealt{vs2017, vs2019}), where the mass aggregation processes of filaments and of the cores within them take place concurrently.  
In this scenario, gas is accreted onto filaments and then continuously flows along them, as posited by \cite{chenm2020} 
based on their NH$_3$ observations of NGC~1333. 

\begin{figure*}
     \centering
     \begin{subfigure}[b]{0.49\textwidth}
         \centering
         \includegraphics[width=\textwidth]{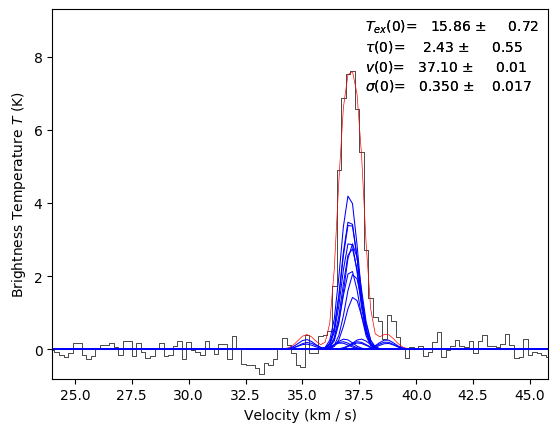}
         \label{fig:hfs_one}
     \end{subfigure}
     \hfill
     \begin{subfigure}[b]{0.49\textwidth}
         \centering
         \includegraphics[width=\textwidth]{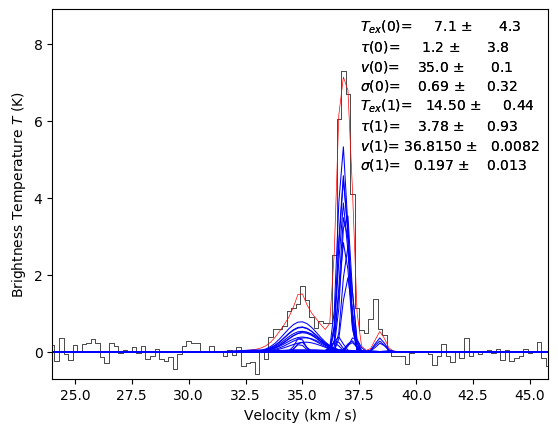}
         \label{fig:hfs_two}
     \end{subfigure}
     \hfill
     \caption{Examples of one-component (left) and two-component (right) \nhp (4-3) hyperfine structure fits (red) overlaid on the observed spectra (black). The individual hyperfine components are shown in blue. The best-fit parameters are listed at top right in each panel.}
     \label{fig:hfs_example}
\end{figure*}

\subsection{Limitations and Future Work}
\label{sec:future}
We note that the line profiles in the \nhp\/(4-3) data cube are shaped by three main factors: multiple velocity components, hyperfine structure (hfs), and missing short spacing information. 
In addition, self-absorption might also affect the line profiles in the densest region. 
In this work, we have used Gaussian Decomposition to disentangle the multiple velocity components and study the gas kinematics. 
Of the remaining factors, we can quantitatively investigate the effects of hfs 
and self-absorption 
using our existing data, 
while missing short spacings are an area for future work, as discussed below.

\begin{figure}
	\includegraphics[width=0.9\columnwidth, height=0.3\textheight]{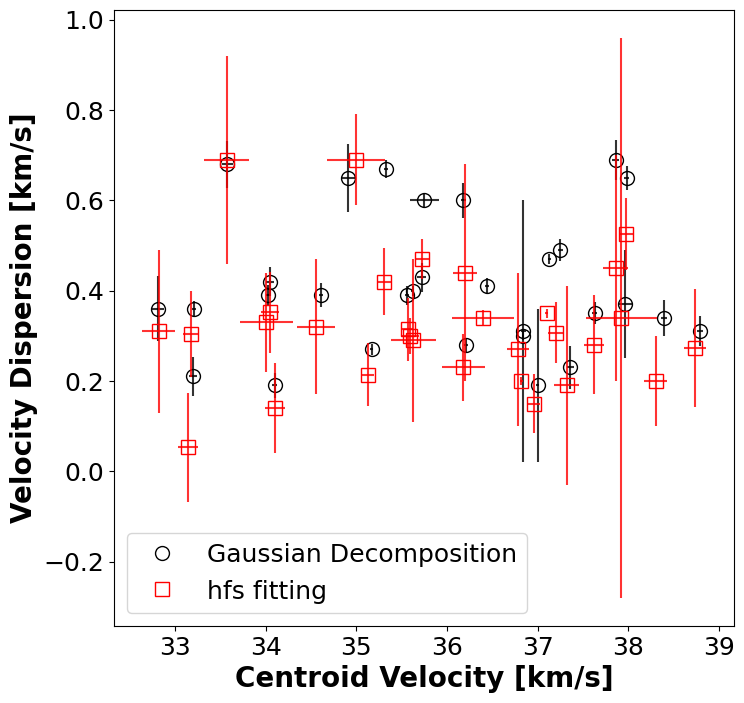}
    \caption{Comparison of fitted parameters from Gaussian Decomposition (black circles) and hyperfine structure fitting (red squares) for selected spectra (see Section~\ref{sec:future}).
    }
   \label{fig:hfs_gd}
\end{figure}

As noted in Section~\ref{sec:n2hp_results}, the hfs of the \nhp\/ (4-3) transition can contribute to shaping the line profile \citep[see also][]{pagani2009} and \citet{friesen2010b} found that Gaussian fitting overestimated the intrinsic \nhp\/ line width by $\sim$10$\%$-25$\%$.  To quantify this effect in our data, we carry out multi-component hfs fitting for a selection of 10 spectra using pyspeckit (\citealt{pyspeckit}, \citealt{pyspeckit2022}). 
The spectra were selected to span a range in the number of fitted Gaussian components (from 1 to 5); their locations are marked in the right-hand panel of Figure \ref{fig:scouse_res}. 
The pyspeckit package applies a uniform excitation temperature (T$_{ex}$) to all fitted hyperfine lines. A one-component hfs fit thus has four free parameters: the centroid velocity, the velocity dispersion, T$_{ex}$, and the optical depth $\tau$. 
An example of a one-component hfs fit is shown in Figure \ref{fig:hfs_example}.  Since the number of fitted parameters in a multi-component hfs fit is 4$\times$ the number of velocity components, we use the best-fitting Gaussian models as initial guesses for the centroid velocity and the velocity dispersion to better constrain the hfs fitting. 
This approach achieves good fits for spectra with one or two velocity components identified by Gaussian decomposition, but T$_{ex}$ and $\tau$ are poorly constrained for many components in more complex spectra. 
To obtain reasonable fits for these more complex cases, we fix T$_{ex}$ for poorly constrained components to the mean value from the initial good fits (14~K). 
Figure \ref{fig:hfs_gd} presents a comparison of the results of our Gaussian Decomposition and hyperfine structure fitting for the selected spectra. 
In this limited sample, we find that in our data Gaussian Decomposition can overestimate the line width by $37\%$ on average, while the fitted centroid velocities are consistent to within 0.1\%.

The hfs fits can also be used to investigate the optical depth of the \nhp\/ emission, and so the potential effects of self-absorption. Inspecting the aforementioned initial good hfs fits, we see no indications of self-absorption in the line profiles for components with moderate optical depths $>$1 (e.g., the components with $\tau\sim$2-4 in Figure \ref{fig:hfs_example}).  As the multi-component hfs fits are in general poorly constrained, we use the radiative transfer code RADEX (\citealt{vandertak2007}) to estimate the optical depth for all velocity components fit in our SCOUSEPY analysis. We use the molecular data and collisional rate coefficients for \nhp\/ from the Leiden Atomic and Molecular Database (LAMDA, \citealt{schoier2005}); the collisional rate coefficients are for \nhp\/ (including hfs) in collisions with H$_2$ for a temperature range of 5 - 2000~K, and are extrapolated and scaled from those for \nhp\/ in collisions with He \citep{daniel2005}. 
The other required inputs to RADEX are the H$_2$ density \hden, the gas kinetic temperature T$_{kin}$, the line width (FWHM), the molecular column density $N(\rm N_2H^+)$, and the background temperature T$_{bg}$. 
The mean gas density of Branch 8 is estimated to be \hden\/ $\sim (1-3)\times10^6$~\perccc\/ assuming a cylinder with a height of $\sim$0.17~pc, a diameter of $\sim$0.05~pc, and a mass of $\sim$30-65~\msun\/ (see Section~\ref{sec:fil_acc_flows}). 
The \nhp\/ column density is estimated following equation A4 in \cite{caselli2002}, using the integrated intensity of each fitted velocity component and T$_{ex} = 14$~K (see the previous paragraph). The partition function for T$_{ex} = 14$~K is interpolated using values from the Cologne Database for Molecular Spectroscopy (CDMS, \citealt{muller2001, muller2005}). 
We estimate the optical depth for each fitted velocity component using the \nhp\/ column density estimated as described above, the fitted velocity dispersion converted to FWHM line width, \hden\/=$10^6$ \perccc, T$_{bg}$=2.73~K, and T$_{kin}$=$15-25$~K (see Section~\ref{sec:mass}) using the Python package SpectralRadex\footnote{\url{ https://spectralradex.readthedocs.io}} \citep{holdship2021}. 
From this analysis, we find that the optical depth of the hyperfine line (\textit{JF$_1$F}) = (456 $-$ 345), the strongest hyperfine line with the largest optical depth, ranges from 0.02 to 1.5 for T$_{kin}$=15~K and from 0.03 to 2.2 for T$_{kin}$=25~K. 
The total optical depth of the \nhp\/ (4-3) transition ranges from 0.1 to 7.2 for T$_{kin}$=15~K, with $\sim$99\% of fitted components having $\tau<$4, and from 0.2 to 11 for for T$_{kin}$=25~K, with $\sim$93\% of fitted components having $\tau<$4. We emphasise that the optical depths estimated from our RADEX modelling are sensitive to the assumed T$_{kin}$ and \hden\/ and to the column density, which depends on the assumed T$_{ex}$ (for example, 
for the single-component spectrum shown in the left-hand panel of Fig.~\ref{fig:hfs_example}, our RADEX modelling yields a total optical depth of 3.0 for T$_{kin}$=15~K and 4.8 for T$_{kin}$=25~K, compared to 2.4 from the hfs fit) and that T$_{ex}$, T$_{kin}$, and \hden\/ are physically expected to vary spatially, which we cannot constrain with our ALMA data. 
Notably, the number of fitted velocity components does not track the strength of the \nhp\/ emission (compare Fig.~\ref{fig:obs_res}b and \ref{fig:scouse_res}), e.g. the number of fitted components is higher to the north of MM2, while the \nhp(4-3) emission is strongest to the south. 
Combined with the generally moderate optical depths, this suggests that self-absorption is not a major contributor to the observed line profiles.

The effects of missing short-spacing information cannot be quantified with existing data (since
no short-spacing \nhp\/ (4-3) observations are available) and this is a limitation of the present study. 
Hacar et al.\footnote{Hacar, A., The need for data combination in the ALMA era, EAS2020 - SS13a: Eight years of ALMA ground-breaking results: A joint venture between the ALMA user community and the ALMA Regional Centres, June 29 - July 3, 2020.} have shown that missing short- and zero-spacings can affect the line profiles and relative intensities of the hyperfine components for \nhp\/ (1-0), and that the effects are highly non-linear. However, these effects have not been investigated for \nhp\/ (4-3), and the impacts may be lessened by the much higher E$_{\rm upper}$ and critical density of the (4-3) transition: E$_{\rm upper}$=44.7~K and n$_{\rm crit}$=2.8$\times$10$^6$~\perccc\/ at 20~K for \nhp (4-3) compared to E$_{\rm upper}$=4.5~K and n$_{\rm crit}$=4.1$\times$10$^4$~\perccc\/ at 20~K for \nhp (1-0).

The most straightforward way to address the lack of short spacing data in this study, and a possible area for future work, would be additional observations of \nhp (4-3) towards G11.92$-$0.61, e.g.\ Atacama Compact Array (ACA) and Total Power (TP) observations.  \nhp (4-3) observations are, however, challenging and observationally expensive, as the poor atmospheric transmission at the line frequency means that observations require excellent weather and/or very large amounts of observing time. The \nhp (4-3) observations presented here total 4.52 hrs of on-source ALMA 12-m time, for a single pointing (Table~\ref{tab:obs}), and ACA 7-m observations suitable for combining with the existing data would require $>$20 hours on-source\footnote{Based on the time multipliers in the ALMA Cycle 10 Proposer's Guide, \url{https://almascience.eso.org/proposing/proposers-guide}.}. Alternately, high-spatial-dynamic-range observations of the G11.92$-$0.61 region in other lines known to trace filamentary accretion flows \citep[e.g.\ lower-J transitions of \nhp or lines of NH$_3$ or H$^{13}$CO;][]{hacar2018,chenm2020,cao2022} may offer a more practical path to confirming our results with additional observations and further investigating kinematics and multi-scale mass accretion in the G11.92$-$0.61 region. 

\section{Conclusions}
\label{sec:conclusions}
We have observed the former high-mass prestellar core candidate \gm\/, which has recently been revealed to host a protobinary system driving a low-velocity \meth\/ outflow. 
Our deep, sub-arcsecond ($\sim$2000 AU) resolution ALMA observations targeted the dense and depleted gas tracers \hdp\/(1$_1,0$-1$_1,1$) and \nhp\/(4-3).  Our main findings are summarised below. 

\smallskip

(i) MM2, which appears as a strong, compact source of 0.82~mm continuum emission, lies on a  $\sim$north-south filamentary structure within the G11.92$-$0.61 massive protocluster.  The filament is traced by both 0.82~mm continuum and \nhp\/ (4-3) emission, but there are differences in the morphologies of the two tracers, with the peaks of the integrated \nhp\/ emission offset by 0\farcs9 and 1\farcs2 ($\sim$3000 and 4000 AU) from the 0.82~mm continuum peak. 

\smallskip

(ii) \hdp\/ is undetected towards MM2, to 4$\sigma$ limits of 10.8 \mjb / 0.23~K.  The non-detection of \hdp\/ is likely due to internal heating from the recently-discovered protobinary system.

\smallskip

(iii) The \nhp\/ spectra are complex, with multiple emission peaks and, towards the MM2 core, absorption features.  The complexity and spatial variation of the \nhp\/ spectra indicate that multiple velocity components are present along the line of sight. 

\smallskip

(iv) Our analysis of the \nhp\/ gas kinematics, using pixel-by-pixel Gaussian decomposition with SCOUSEPY and hierarchical clustering of the extracted velocity components with ACORNS, reveals a hierarchical system in the \nhp-emitting gas comprised of 70 velocity- and position-coherent clusters, known as "trees".  The eight largest trees 
describe $>$60\% of the fitted velocity components, with $\sim$20\% of fitted velocity components in the largest (primary) tree.

\smallskip

(v) The primary tree exhibits a $\sim$north-south velocity gradient along the filamentary structure traced by the 0.82~mm continuum emission.  Analysing a $\sim$0.17~pc-long substructure within this tree to focus on gas kinematics around MM2, we find a best-fit velocity gradient of $\sim$10.5~\kms\/ pc$^{-1}$.  Interpreting this velocity gradient as tracing filamentary accretion flows towards MM2, we estimate a mass inflow rate of $\sim$1.8$\times10^{-4}$ to $\sim$ 1.2 $\times10^{-3}$ \msun\/ yr$^{-1}$. 

\smallskip

Our analysis of the dense gas kinematics in the surroundings of MM2 indicates that rather than being an isolated core, MM2 is connected to its larger-scale environment. 
In particular, the ongoing accretion onto the protobinary system -- indicated by its outflow activity -- is likely fed by the larger-scale filamentary accretion flows.  If 50\% of the mass inflow estimated above reaches the protostars, each will reach a mass of 8 \msun\/ within $\sim$1.6$\times$10$^5$ yrs, which is intriguingly comparable to the combined timescale of the 70$\mu$m-weak+MIR-weak phases 
of high-mass star formation 
estimated from chemical clocks \citep[e.g.][]{sabatini21}.  In addition to the primary tree that is the focus of our analysis, we also identify a velocity gradient consistent with a filamentary accretion flow onto MM2 in a second tree, which is kinematically distinct but partially overlaps the primary tree on the plane of the sky.  This finding suggests that the filamentary dust continuum structure contains multiple kinematic substructures, and that MM2 may be fed by multiple filamentary accretion flows.  Additional observations of the G11.92$-$0.61 region, including in lower-J and more easily observed \nhp\/ transitions, are needed to explore this possibility and expand the investigation of mass accretion in G11.92$-$0.61 to larger spatial scales.

\section*{Acknowledgements}
We thank J.~D.\ Ilee for helpful discussions and for providing the ALMA 1.3~mm continuum image from \citet{ilee2018}.  We also thank K. {\"O}berg, T.~P.\  Robitaille, and C.~M. Koepferl for their input at an early stage of this project.  The National Radio Astronomy Observatory is a facility of the National Science Foundation operated under cooperative agreement by Associated Universities, Inc. 
This work made use of the following ALMA data: ADS/JAO.ALMA$\#$2015.1.00827.S, and ADS/JAO.ALMA$\#$2017.1.01373.S. ALMA is a partnership of ESO (representing its member states), NSF (USA) and NINS (Japan), together with NRC (Canada), MOST and ASIAA (Taiwan), and KASI (Republic of Korea), in cooperation with the Republic of Chile. The Joint ALMA Observatory is operated by ESO, AUI/ NRAO and NAOJ. S.Z. is funded by the China Scholarship Council-University of St Andrews Scholarship (PhD programmes, No. 201806190010). C.J.C. acknowledges support from the University of St Andrews Restarting Research Funding Scheme (SARRF), which is funded through the SFC grant reference SFC/AN/08/020. 
J.D.H gratefully acknowledges financial support from the Royal Society (University Research Fellowship; URF\textbackslash R1\textbackslash 221620).
This work made use of NASA Astrophysics Data System Bibliographic Services and Python packages 
NumPy (an array programming library, \cite{numpy2020}), SciPy (an open-source scientific computing library, \cite{scipy2020}), Astropy (an open-source package with commonly-needed functionality to the astronomical community, \cite{astropy2022}),
 Matplotlib (a 2D graphics package, \cite{Hunter2007}), APLpy (an open-source plotting package for Python, \cite{robitaille2012}) and {\sc analysisUtils} \citep{au2023}.

\section*{Data Availability}
The ALMA 0.82\,mm continuum image and \nhp\/ (4-3) and \hdp\/(1$_{1,0}$-1$_{1,1}$) line cubes are available at \url{https://doi.org/10.5281/zenodo.12640479}.



\bibliographystyle{mnras}
\bibliography{main} 

\begin{thebibliography}{}
\makeatletter
\relax
\def\mn@urlcharsother{\let\do\@makeother \do\$\do\&\do\#\do\^\do\_\do\%\do\~}
\def\mn@doi{\begingroup\mn@urlcharsother \@ifnextchar [ {\mn@doi@} {\mn@doi@[]}}
\def\mn@doi@[#1]#2{\def\@tempa{#1}\ifx\@tempa\@empty \href {http://dx.doi.org/#2} {doi:#2}\else \href {http://dx.doi.org/#2} {#1}\fi \endgroup}
\def\mn@eprint#1#2{\mn@eprint@#1:#2::\@nil}
\def\mn@eprint@arXiv#1{\href {http://arxiv.org/abs/#1} {{\tt arXiv:#1}}}
\def\mn@eprint@dblp#1{\href {http://dblp.uni-trier.de/rec/bibtex/#1.xml} {dblp:#1}}
\def\mn@eprint@#1:#2:#3:#4\@nil{\def\@tempa {#1}\def\@tempb {#2}\def\@tempc {#3}\ifx \@tempc \@empty \let \@tempc \@tempb \let \@tempb \@tempa \fi \ifx \@tempb \@empty \def\@tempb {arXiv}\fi \@ifundefined {mn@eprint@\@tempb}{\@tempb:\@tempc}{\expandafter \expandafter \csname mn@eprint@\@tempb\endcsname \expandafter{\@tempc}}}

\bibitem[\protect\citeauthoryear{{Anderson} et~al.,}{{Anderson} et~al.}{2021}]{anderson2021}
{Anderson} M.,  et~al., 2021, \mn@doi [\mnras] {10.1093/mnras/stab2674}, \href {https://ui.adsabs.harvard.edu/abs/2021MNRAS.508.2964A} {508, 2964}

\bibitem[\protect\citeauthoryear{{Andr{\'e}}, {Di Francesco}, {Ward-Thompson}, {Inutsuka}, {Pudritz}  \& {Pineda}}{{Andr{\'e}} et~al.}{2014}]{andre2014}
{Andr{\'e}} P.,  {Di Francesco} J.,  {Ward-Thompson} D.,  {Inutsuka} S.~I.,  {Pudritz} R.~E.,   {Pineda} J.~E.,  2014, in {Beuther} H.,  {Klessen} R.~S.,  {Dullemond} C.~P.,   {Henning} T.,  eds, Protostars and Planets VI. pp 27--51 (\mn@eprint {arXiv} {1312.6232}), \mn@doi{10.2458/azu_uapress_9780816531240-ch002}

\bibitem[\protect\citeauthoryear{{Andr{\'e}} et~al.,}{{Andr{\'e}} et~al.}{2016}]{andre2016}
{Andr{\'e}} P.,  et~al., 2016, \mn@doi [\aap] {10.1051/0004-6361/201628378}, \href {https://ui.adsabs.harvard.edu/abs/2016A&A...592A..54A} {592, A54}

\bibitem[\protect\citeauthoryear{{Astropy Collaboration} et~al.,}{{Astropy Collaboration} et~al.}{2022}]{astropy2022}
{Astropy Collaboration} et~al., 2022, \mn@doi [\apj] {10.3847/1538-4357/ac7c74}, \href {https://ui.adsabs.harvard.edu/abs/2022ApJ...935..167A} {935, 167}

\bibitem[\protect\citeauthoryear{{Bally}, {Langer}, {Stark}  \& {Wilson}}{{Bally} et~al.}{1987}]{bally1987}
{Bally} J.,  {Langer} W.~D.,  {Stark} A.~A.,   {Wilson} R.~W.,  1987, \mn@doi [\apjl] {10.1086/184817}, \href {https://ui.adsabs.harvard.edu/abs/1987ApJ...312L..45B} {312, L45}

\bibitem[\protect\citeauthoryear{{Barnes} et~al.,}{{Barnes} et~al.}{2023}]{barnes2023}
{Barnes} A.~T.,  et~al., 2023, \mn@doi [\aap] {10.1051/0004-6361/202245668}, \href {https://ui.adsabs.harvard.edu/abs/2023A&A...675A..53B} {675, A53}

\bibitem[\protect\citeauthoryear{{Beltr{\'a}n}, {Rivilla}, {Kumar}, {Cesaroni}  \& {Galli}}{{Beltr{\'a}n} et~al.}{2022}]{beltran2022}
{Beltr{\'a}n} M.~T.,  {Rivilla} V.~M.,  {Kumar} M.~S.~N.,  {Cesaroni} R.,   {Galli} D.,  2022, \mn@doi [\aap] {10.1051/0004-6361/202243361}, \href {https://ui.adsabs.harvard.edu/abs/2022A&A...660L...4B} {660, L4}

\bibitem[\protect\citeauthoryear{{Beuther}, {Ragan}, {Johnston}, {Henning}, {Hacar}  \& {Kainulainen}}{{Beuther} et~al.}{2015}]{beuther2015}
{Beuther} H.,  {Ragan} S.~E.,  {Johnston} K.,  {Henning} T.,  {Hacar} A.,   {Kainulainen} J.~T.,  2015, \mn@doi [\aap] {10.1051/0004-6361/201527108}, \href {https://ui.adsabs.harvard.edu/abs/2015A&A...584A..67B} {584, A67}

\bibitem[\protect\citeauthoryear{{Beuther} et~al.,}{{Beuther} et~al.}{2018}]{beuther2018}
{Beuther} H.,  et~al., 2018, \mn@doi [\aap] {10.1051/0004-6361/201833021}, \href {https://ui.adsabs.harvard.edu/abs/2018A&A...617A.100B} {617, A100}

\bibitem[\protect\citeauthoryear{{Beuther} et~al.,}{{Beuther} et~al.}{2021}]{beuther2021}
{Beuther} H.,  et~al., 2021, \mn@doi [\aap] {10.1051/0004-6361/202040106}, \href {https://ui.adsabs.harvard.edu/abs/2021A&A...649A.113B} {649, A113}

\bibitem[\protect\citeauthoryear{{Blake}, {Sandell}, {van Dishoeck}, {Groesbeck}, {Mundy}  \& {Aspin}}{{Blake} et~al.}{1995}]{blake1995}
{Blake} G.~A.,  {Sandell} G.,  {van Dishoeck} E.~F.,  {Groesbeck} T.~D.,  {Mundy} L.~G.,   {Aspin} C.,  1995, \mn@doi [\apj] {10.1086/175392}, \href {https://ui.adsabs.harvard.edu/abs/1995ApJ...441..689B} {441, 689}

\bibitem[\protect\citeauthoryear{{Bonnell} \& {Bate}}{{Bonnell} \& {Bate}}{2006}]{bonnell2006}
{Bonnell} I.~A.,  {Bate} M.~R.,  2006, \mn@doi [\mnras] {10.1111/j.1365-2966.2006.10495.x}, \href {https://ui.adsabs.harvard.edu/abs/2006MNRAS.370..488B} {370, 488}

\bibitem[\protect\citeauthoryear{{Bonnell}, {Bate}, {Clarke}  \& {Pringle}}{{Bonnell} et~al.}{2001}]{bonnell2001}
{Bonnell} I.~A.,  {Bate} M.~R.,  {Clarke} C.~J.,   {Pringle} J.~E.,  2001, \mn@doi [\mnras] {10.1046/j.1365-8711.2001.04270.x}, \href {https://ui.adsabs.harvard.edu/abs/2001MNRAS.323..785B} {323, 785}

\bibitem[\protect\citeauthoryear{Bonnell, Bate  \& Vine}{Bonnell et~al.}{2003}]{Bonnell:2003ny}
Bonnell I.~A.,  Bate M.~R.,   Vine S.~G.,  2003, \mn@doi [Mon. Not. Roy. Astron. Soc.] {10.1046/j.1365-8711.2003.06687.x}, 343, 413

\bibitem[\protect\citeauthoryear{Bonnell, Vine  \& Bate}{Bonnell et~al.}{2004}]{Bonnell:2004qk}
Bonnell I.~A.,  Vine S.~G.,   Bate M.~R.,  2004, \mn@doi [Mon. Not. Roy. Astron. Soc.] {10.1111/j.1365-2966.2004.07543.x}, 349, 735

\bibitem[\protect\citeauthoryear{{Bonnell}, {Clark}  \& {Bate}}{{Bonnell} et~al.}{2008}]{bonnell2008}
{Bonnell} I.~A.,  {Clark} P.,   {Bate} M.~R.,  2008, \mn@doi [\mnras] {10.1111/j.1365-2966.2008.13679.x}, \href {https://ui.adsabs.harvard.edu/abs/2008MNRAS.389.1556B} {389, 1556}

\bibitem[\protect\citeauthoryear{{Bonnell}, {Dobbs}  \& {Smith}}{{Bonnell} et~al.}{2013}]{bonnell2013}
{Bonnell} I.~A.,  {Dobbs} C.~L.,   {Smith} R.~J.,  2013, \mn@doi [\mnras] {10.1093/mnras/stt004}, \href {https://ui.adsabs.harvard.edu/abs/2013MNRAS.430.1790B} {430, 1790}

\bibitem[\protect\citeauthoryear{{Bontemps}, {Motte}, {Csengeri}  \& {Schneider}}{{Bontemps} et~al.}{2010}]{bon2010}
{Bontemps} S.,  {Motte} F.,  {Csengeri} T.,   {Schneider} N.,  2010, \mn@doi [\aap] {10.1051/0004-6361/200913286}, \href {https://ui.adsabs.harvard.edu/abs/2010A&A...524A..18B} {524, A18}

\bibitem[\protect\citeauthoryear{{Breen} \& {Ellingsen}}{{Breen} \& {Ellingsen}}{2011}]{breen2011}
{Breen} S.~L.,  {Ellingsen} S.~P.,  2011, \mn@doi [\mnras] {10.1111/j.1365-2966.2011.19020.x}, \href {https://ui.adsabs.harvard.edu/abs/2011MNRAS.416..178B} {416, 178}

\bibitem[\protect\citeauthoryear{{Brogan}, {Hunter}, {Cyganowski}, {Indebetouw}, {Beuther}, {Menten}  \& {Thorwirth}}{{Brogan} et~al.}{2009}]{brogan2009}
{Brogan} C.~L.,  {Hunter} T.~R.,  {Cyganowski} C.~J.,  {Indebetouw} R.,  {Beuther} H.,  {Menten} K.~M.,   {Thorwirth} S.,  2009, \mn@doi [\apj] {10.1088/0004-637X/707/1/1}, \href {https://ui.adsabs.harvard.edu/abs/2009ApJ...707....1B} {707, 1}

\bibitem[\protect\citeauthoryear{{Brogan}, {Hunter}, {Cyganowski}, {Friesen}, {Chandler}  \& {Indebetouw}}{{Brogan} et~al.}{2011}]{brogan11}
{Brogan} C.~L.,  {Hunter} T.~R.,  {Cyganowski} C.~J.,  {Friesen} R.~K.,  {Chandler} C.~J.,   {Indebetouw} R.,  2011, \mn@doi [\apjl] {10.1088/2041-8205/739/1/L16}, \href {https://ui.adsabs.harvard.edu/abs/2011ApJ...739L..16B} {739, L16}

\bibitem[\protect\citeauthoryear{{Brogan}, {Hunter}, {Cyganowski}, {Chandler}, {Friesen}  \& {Indebetouw}}{{Brogan} et~al.}{2016}]{brogan2016}
{Brogan} C.~L.,  {Hunter} T.~R.,  {Cyganowski} C.~J.,  {Chandler} C.~J.,  {Friesen} R.,   {Indebetouw} R.,  2016, \mn@doi [\apj] {10.3847/0004-637X/832/2/187}, \href {https://ui.adsabs.harvard.edu/abs/2016ApJ...832..187B} {832, 187}

\bibitem[\protect\citeauthoryear{{Busquet} et~al.,}{{Busquet} et~al.}{2013}]{Busquet2013}
{Busquet} G.,  et~al., 2013, \mn@doi [\apjl] {10.1088/2041-8205/764/2/L26}, \href {https://ui.adsabs.harvard.edu/abs/2013ApJ...764L..26B} {764, L26}

\bibitem[\protect\citeauthoryear{{Butler} \& {Tan}}{{Butler} \& {Tan}}{2009}]{bt2009}
{Butler} M.~J.,  {Tan} J.~C.,  2009, \mn@doi [\apj] {10.1088/0004-637X/696/1/484}, \href {https://ui.adsabs.harvard.edu/abs/2009ApJ...696..484B} {696, 484}

\bibitem[\protect\citeauthoryear{{Butler} \& {Tan}}{{Butler} \& {Tan}}{2012}]{bt2012}
{Butler} M.~J.,  {Tan} J.~C.,  2012, \mn@doi [\apj] {10.1088/0004-637X/754/1/5}, \href {https://ui.adsabs.harvard.edu/abs/2012ApJ...754....5B} {754, 5}

\bibitem[\protect\citeauthoryear{{Cao}, {Qiu}, {Zhang}  \& {Li}}{{Cao} et~al.}{2022}]{cao2022}
{Cao} Y.,  {Qiu} K.,  {Zhang} Q.,   {Li} G.-X.,  2022, \mn@doi [\apj] {10.3847/1538-4357/ac4696}, \href {https://ui.adsabs.harvard.edu/abs/2022ApJ...927..106C} {927, 106}

\bibitem[\protect\citeauthoryear{{Caselli} \& {Ceccarelli}}{{Caselli} \& {Ceccarelli}}{2012}]{caselli2012}
{Caselli} P.,  {Ceccarelli} C.,  2012, \mn@doi [\aapr] {10.1007/s00159-012-0056-x}, \href {https://ui.adsabs.harvard.edu/abs/2012A&ARv..20...56C} {20, 56}

\bibitem[\protect\citeauthoryear{{Caselli}, {Walmsley}, {Zucconi}, {Tafalla}, {Dore}  \& {Myers}}{{Caselli} et~al.}{2002}]{caselli2002}
{Caselli} P.,  {Walmsley} C.~M.,  {Zucconi} A.,  {Tafalla} M.,  {Dore} L.,   {Myers} P.~C.,  2002, \mn@doi [\apj] {10.1086/324302}, \href {https://ui.adsabs.harvard.edu/abs/2002ApJ...565..344C} {565, 344}

\bibitem[\protect\citeauthoryear{{Caselli}, {van der Tak}, {Ceccarelli}  \& {Bacmann}}{{Caselli} et~al.}{2003}]{caselli2003}
{Caselli} P.,  {van der Tak} F.~F.~S.,  {Ceccarelli} C.,   {Bacmann} A.,  2003, \mn@doi [\aap] {10.1051/0004-6361:20030526}, \href {https://ui.adsabs.harvard.edu/abs/2003A&A...403L..37C} {403, L37}

\bibitem[\protect\citeauthoryear{{Caselli}, {Vastel}, {Ceccarelli}, {van der Tak}, {Crapsi}  \& {Bacmann}}{{Caselli} et~al.}{2008}]{caselli2008}
{Caselli} P.,  {Vastel} C.,  {Ceccarelli} C.,  {van der Tak} F.~F.~S.,  {Crapsi} A.,   {Bacmann} A.,  2008, \mn@doi [\aap] {10.1051/0004-6361:20079009}, \href {https://ui.adsabs.harvard.edu/abs/2008A&A...492..703C} {492, 703}

\bibitem[\protect\citeauthoryear{{Ceccarelli}, {Caselli}, {Bockel{\'e}e-Morvan}, {Mousis}, {Pizzarello}, {Robert}  \& {Semenov}}{{Ceccarelli} et~al.}{2014}]{cecc2014}
{Ceccarelli} C.,  {Caselli} P.,  {Bockel{\'e}e-Morvan} D.,  {Mousis} O.,  {Pizzarello} S.,  {Robert} F.,   {Semenov} D.,  2014, in {Beuther} H.,  {Klessen} R.~S.,  {Dullemond} C.~P.,   {Henning} T.,  eds, Protostars and Planets VI. p.~859 (\mn@eprint {arXiv} {1403.7143}), \mn@doi{10.2458/azu_uapress_9780816531240-ch037}

\bibitem[\protect\citeauthoryear{{Chen} et~al.,}{{Chen} et~al.}{2019}]{chen2019}
{Chen} H.-R.~V.,  et~al., 2019, \mn@doi [\apj] {10.3847/1538-4357/ab0f3e}, \href {https://ui.adsabs.harvard.edu/abs/2019ApJ...875...24C} {875, 24}

\bibitem[\protect\citeauthoryear{{Chen}, {Mundy}, {Ostriker}, {Storm}  \& {Dhabal}}{{Chen} et~al.}{2020a}]{chen2020}
{Chen} C.-Y.,  {Mundy} L.~G.,  {Ostriker} E.~C.,  {Storm} S.,   {Dhabal} A.,  2020a, \mn@doi [\mnras] {10.1093/mnras/staa960}, \href {https://ui.adsabs.harvard.edu/abs/2020MNRAS.494.3675C} {494, 3675}

\bibitem[\protect\citeauthoryear{{Chen} et~al.,}{{Chen} et~al.}{2020b}]{chenm2020}
{Chen} M. C.-Y.,  et~al., 2020b, \mn@doi [\apj] {10.3847/1538-4357/ab7378}, \href {https://ui.adsabs.harvard.edu/abs/2020ApJ...891...84C} {891, 84}

\bibitem[\protect\citeauthoryear{{Churchwell}, {Walmsley}  \& {Wood}}{{Churchwell} et~al.}{1992}]{churchwell1992}
{Churchwell} E.,  {Walmsley} C.~M.,   {Wood} D.~O.~S.,  1992, \aap, \href {https://ui.adsabs.harvard.edu/abs/1992A&A...253..541C} {253, 541}

\bibitem[\protect\citeauthoryear{{Clarke}, {Whitworth}, {Spowage}, {Duarte-Cabral}, {Suri}, {Jaffa}, {Walch}  \& {Clark}}{{Clarke} et~al.}{2018}]{clarke2018}
{Clarke} S.~D.,  {Whitworth} A.~P.,  {Spowage} R.~L.,  {Duarte-Cabral} A.,  {Suri} S.~T.,  {Jaffa} S.~E.,  {Walch} S.,   {Clark} P.~C.,  2018, \mn@doi [\mnras] {10.1093/mnras/sty1675}, \href {https://ui.adsabs.harvard.edu/abs/2018MNRAS.479.1722C} {479, 1722}

\bibitem[\protect\citeauthoryear{{Clarke}, {Williams}  \& {Walch}}{{Clarke} et~al.}{2020}]{clarke2020}
{Clarke} S.~D.,  {Williams} G.~M.,   {Walch} S.,  2020, \mn@doi [\mnras] {10.1093/mnras/staa2298}, \href {https://ui.adsabs.harvard.edu/abs/2020MNRAS.497.4390C} {497, 4390}

\bibitem[\protect\citeauthoryear{{Commer{\c{c}}on}, {Hennebelle}  \& {Henning}}{{Commer{\c{c}}on} et~al.}{2011}]{com2011}
{Commer{\c{c}}on} B.,  {Hennebelle} P.,   {Henning} T.,  2011, \mn@doi [\apjl] {10.1088/2041-8205/742/1/L9}, \href {https://ui.adsabs.harvard.edu/abs/2011ApJ...742L...9C} {742, L9}

\bibitem[\protect\citeauthoryear{Cortes et~al.,}{Cortes et~al.}{2023}]{almac10}
Cortes P.,  et~al., 2023, ALMA Cycle 10 Technical Handbook, \mn@doi{10.5281/zenodo.7822943.
}, \url {https://doi.org/10.5281/zenodo.7822943}

\bibitem[\protect\citeauthoryear{{Cunningham}, {Klein}, {Krumholz}  \& {McKee}}{{Cunningham} et~al.}{2011}]{cunningham2011}
{Cunningham} A.~J.,  {Klein} R.~I.,  {Krumholz} M.~R.,   {McKee} C.~F.,  2011, \mn@doi [\apj] {10.1088/0004-637X/740/2/107}, \href {https://ui.adsabs.harvard.edu/abs/2011ApJ...740..107C} {740, 107}

\bibitem[\protect\citeauthoryear{{Cyganowski} et~al.,}{{Cyganowski} et~al.}{2008}]{cc2008}
{Cyganowski} C.~J.,  et~al., 2008, \mn@doi [\aj] {10.1088/0004-6256/136/6/2391}, \href {https://ui.adsabs.harvard.edu/abs/2008AJ....136.2391C} {136, 2391}

\bibitem[\protect\citeauthoryear{{Cyganowski}, {Brogan}, {Hunter}  \& {Churchwell}}{{Cyganowski} et~al.}{2009}]{cc2009}
{Cyganowski} C.~J.,  {Brogan} C.~L.,  {Hunter} T.~R.,   {Churchwell} E.,  2009, \mn@doi [\apj] {10.1088/0004-637X/702/2/1615}, \href {https://ui.adsabs.harvard.edu/abs/2009ApJ...702.1615C} {702, 1615}

\bibitem[\protect\citeauthoryear{{Cyganowski}, {Brogan}, {Hunter}, {Churchwell}  \& {Zhang}}{{Cyganowski} et~al.}{2011a}]{cc2011a}
{Cyganowski} C.~J.,  {Brogan} C.~L.,  {Hunter} T.~R.,  {Churchwell} E.,   {Zhang} Q.,  2011a, \mn@doi [\apj] {10.1088/0004-637X/729/2/124}, \href {https://ui.adsabs.harvard.edu/abs/2011ApJ...729..124C} {729, 124}

\bibitem[\protect\citeauthoryear{{Cyganowski}, {Brogan}, {Hunter}  \& {Churchwell}}{{Cyganowski} et~al.}{2011b}]{cc2011b}
{Cyganowski} C.~J.,  {Brogan} C.~L.,  {Hunter} T.~R.,   {Churchwell} E.,  2011b, \mn@doi [\apj] {10.1088/0004-637X/743/1/56}, \href {https://ui.adsabs.harvard.edu/abs/2011ApJ...743...56C} {743, 56}

\bibitem[\protect\citeauthoryear{{Cyganowski}, {Koda}, {Rosolowsky}, {Towers}, {Donovan Meyer}, {Egusa}, {Momose}  \& {Robitaille}}{{Cyganowski} et~al.}{2013}]{cyganowski13}
{Cyganowski} C.~J.,  {Koda} J.,  {Rosolowsky} E.,  {Towers} S.,  {Donovan Meyer} J.,  {Egusa} F.,  {Momose} R.,   {Robitaille} T.~P.,  2013, \mn@doi [\apj] {10.1088/0004-637X/764/1/61}, \href {https://ui.adsabs.harvard.edu/abs/2013ApJ...764...61C} {764, 61}

\bibitem[\protect\citeauthoryear{{Cyganowski} et~al.,}{{Cyganowski} et~al.}{2014}]{cc2014}
{Cyganowski} C.~J.,  et~al., 2014, \mn@doi [\apj] {10.1088/2041-8205/796/1/L2}, \href {https://ui.adsabs.harvard.edu/abs/2014ApJ...796L...2C} {796, L2}

\bibitem[\protect\citeauthoryear{{Cyganowski}, {Brogan}, {Hunter}, {Smith}, {Kruijssen}, {Bonnell}  \& {Zhang}}{{Cyganowski} et~al.}{2017}]{cc2017}
{Cyganowski} C.~J.,  {Brogan} C.~L.,  {Hunter} T.~R.,  {Smith} R.,  {Kruijssen} J.~M.~D.,  {Bonnell} I.~A.,   {Zhang} Q.,  2017, \mn@doi [\mnras] {10.1093/mnras/stx043}, \href {https://ui.adsabs.harvard.edu/abs/2017MNRAS.468.3694C} {468, 3694}

\bibitem[\protect\citeauthoryear{{Cyganowski}, {Ilee}, {Brogan}, {Hunter}, {Zhang}, {Harries}  \& {Haworth}}{{Cyganowski} et~al.}{2022}]{cc22}
{Cyganowski} C.~J.,  {Ilee} J.~D.,  {Brogan} C.~L.,  {Hunter} T.~R.,  {Zhang} S.,  {Harries} T.~J.,   {Haworth} T.~J.,  2022, \mn@doi [\apjl] {10.3847/2041-8213/ac69ca}, \href {https://ui.adsabs.harvard.edu/abs/2022ApJ...931L..31C} {931, L31}

\bibitem[\protect\citeauthoryear{{Dale} \& {Bonnell}}{{Dale} \& {Bonnell}}{2011}]{dale2011}
{Dale} J.~E.,  {Bonnell} I.,  2011, \mn@doi [\mnras] {10.1111/j.1365-2966.2011.18392.x}, \href {https://ui.adsabs.harvard.edu/abs/2011MNRAS.414..321D} {414, 321}

\bibitem[\protect\citeauthoryear{{Dale}, {Ercolano}  \& {Bonnell}}{{Dale} et~al.}{2012}]{dale2012}
{Dale} J.~E.,  {Ercolano} B.,   {Bonnell} I.~A.,  2012, \mn@doi [\mnras] {10.1111/j.1365-2966.2012.21205.x}, \href {https://ui.adsabs.harvard.edu/abs/2012MNRAS.424..377D} {424, 377}

\bibitem[\protect\citeauthoryear{{Daniel}, {Dubernet}, {Meuwly}, {Cernicharo}  \& {Pagani}}{{Daniel} et~al.}{2005}]{daniel2005}
{Daniel} F.,  {Dubernet} M.~L.,  {Meuwly} M.,  {Cernicharo} J.,   {Pagani} L.,  2005, \mn@doi [\mnras] {10.1111/j.1365-2966.2005.09542.x}, \href {https://ui.adsabs.harvard.edu/abs/2005MNRAS.363.1083D} {363, 1083}

\bibitem[\protect\citeauthoryear{{Daniel}, {Cernicharo}  \& {Dubernet}}{{Daniel} et~al.}{2006}]{daniel2006}
{Daniel} F.,  {Cernicharo} J.,   {Dubernet} M.~L.,  2006, \mn@doi [\apj] {10.1086/505738}, \href {https://ui.adsabs.harvard.edu/abs/2006ApJ...648..461D} {648, 461}

\bibitem[\protect\citeauthoryear{{Dewangan}, {Ojha}  \& {Baug}}{{Dewangan} et~al.}{2017}]{dewangan2017}
{Dewangan} L.~K.,  {Ojha} D.~K.,   {Baug} T.,  2017, \mn@doi [\apj] {10.3847/1538-4357/aa79a5}, \href {https://ui.adsabs.harvard.edu/abs/2017ApJ...844...15D} {844, 15}

\bibitem[\protect\citeauthoryear{{Dewangan}, {Ojha}, {Sharma}, {Palacio}, {Bhadari}  \& {Das}}{{Dewangan} et~al.}{2020}]{dewangan2020}
{Dewangan} L.~K.,  {Ojha} D.~K.,  {Sharma} S.,  {Palacio} S.~d.,  {Bhadari} N.~K.,   {Das} A.,  2020, \mn@doi [\apj] {10.3847/1538-4357/abb827}, \href {https://ui.adsabs.harvard.edu/abs/2020ApJ...903...13D} {903, 13}

\bibitem[\protect\citeauthoryear{{Dhabal}, {Mundy}, {Rizzo}, {Storm}  \& {Teuben}}{{Dhabal} et~al.}{2018}]{dhabal2018}
{Dhabal} A.,  {Mundy} L.~G.,  {Rizzo} M.~J.,  {Storm} S.,   {Teuben} P.,  2018, \mn@doi [\apj] {10.3847/1538-4357/aaa76b}, \href {https://ui.adsabs.harvard.edu/abs/2018ApJ...853..169D} {853, 169}

\bibitem[\protect\citeauthoryear{{Dhabal}, {Mundy}, {Chen}, {Teuben}  \& {Storm}}{{Dhabal} et~al.}{2019}]{dhabal2019}
{Dhabal} A.,  {Mundy} L.~G.,  {Chen} C.-y.,  {Teuben} P.,   {Storm} S.,  2019, \mn@doi [\apj] {10.3847/1538-4357/ab15d3}, \href {https://ui.adsabs.harvard.edu/abs/2019ApJ...876..108D} {876, 108}

\bibitem[\protect\citeauthoryear{{Dobbs}, {Bonnell}  \& {Clark}}{{Dobbs} et~al.}{2005}]{dobbs2005}
{Dobbs} C.~L.,  {Bonnell} I.~A.,   {Clark} P.~C.,  2005, \mn@doi [\mnras] {10.1111/j.1365-2966.2005.08941.x}, \href {https://ui.adsabs.harvard.edu/abs/2005MNRAS.360....2D} {360, 2}

\bibitem[\protect\citeauthoryear{{Duarte-Cabral}, {Bontemps}, {Motte}, {Hennemann}, {Schneider}  \& {Andr{\'e}}}{{Duarte-Cabral} et~al.}{2013}]{dc2013}
{Duarte-Cabral} A.,  {Bontemps} S.,  {Motte} F.,  {Hennemann} M.,  {Schneider} N.,   {Andr{\'e}} P.,  2013, \mn@doi [\aap] {10.1051/0004-6361/201321393}, \href {https://ui.adsabs.harvard.edu/abs/2013A&A...558A.125D} {558, A125}

\bibitem[\protect\citeauthoryear{{Duarte-Cabral}, {Bontemps}, {Motte}, {Gusdorf}, {Csengeri}, {Schneider}  \& {Louvet}}{{Duarte-Cabral} et~al.}{2014}]{dc2014}
{Duarte-Cabral} A.,  {Bontemps} S.,  {Motte} F.,  {Gusdorf} A.,  {Csengeri} T.,  {Schneider} N.,   {Louvet} F.,  2014, \mn@doi [\aap] {10.1051/0004-6361/201423677}, \href {https://ui.adsabs.harvard.edu/abs/2014A&A...570A...1D} {570, A1}

\bibitem[\protect\citeauthoryear{{Federrath}}{{Federrath}}{2013}]{fed2013}
{Federrath} C.,  2013, \mn@doi [\mnras] {10.1093/mnras/stt1644}, \href {https://ui.adsabs.harvard.edu/abs/2013MNRAS.436.1245F} {436, 1245}

\bibitem[\protect\citeauthoryear{{Federrath} et~al.,}{{Federrath} et~al.}{2016}]{fed2016}
{Federrath} C.,  et~al., 2016, \mn@doi [\apj] {10.3847/0004-637X/832/2/143}, \href {https://ui.adsabs.harvard.edu/abs/2016ApJ...832..143F} {832, 143}

\bibitem[\protect\citeauthoryear{{Fern{\'a}ndez-L{\'o}pez} et~al.,}{{Fern{\'a}ndez-L{\'o}pez} et~al.}{2014}]{fl2014}
{Fern{\'a}ndez-L{\'o}pez} M.,  et~al., 2014, \mn@doi [\apjl] {10.1088/2041-8205/790/2/L19}, \href {https://ui.adsabs.harvard.edu/abs/2014ApJ...790L..19F} {790, L19}

\bibitem[\protect\citeauthoryear{{Friesen}, {Di Francesco}, {Myers}, {Belloche}, {Shirley}, {Bourke}  \& {Andr{\'e}}}{{Friesen} et~al.}{2010}]{friesen2010b}
{Friesen} R.~K.,  {Di Francesco} J.,  {Myers} P.~C.,  {Belloche} A.,  {Shirley} Y.~L.,  {Bourke} T.~L.,   {Andr{\'e}} P.,  2010, \mn@doi [\apj] {10.1088/0004-637X/718/2/666}, \href {https://ui.adsabs.harvard.edu/abs/2010ApJ...718..666F} {718, 666}

\bibitem[\protect\citeauthoryear{{Friesen}, {Di Francesco}, {Bourke}, {Caselli}, {J{\o}rgensen}, {Pineda}  \& {Wong}}{{Friesen} et~al.}{2014}]{friesen2014}
{Friesen} R.~K.,  {Di Francesco} J.,  {Bourke} T.~L.,  {Caselli} P.,  {J{\o}rgensen} J.~K.,  {Pineda} J.~E.,   {Wong} M.,  2014, \mn@doi [\apj] {10.1088/0004-637X/797/1/27}, \href {https://ui.adsabs.harvard.edu/abs/2014ApJ...797...27F} {797, 27}

\bibitem[\protect\citeauthoryear{{Ginsburg} \& {Mirocha}}{{Ginsburg} \& {Mirocha}}{2011}]{pyspeckit}
{Ginsburg} A.,  {Mirocha} J.,  2011, {PySpecKit: Python Spectroscopic Toolkit} (\mn@eprint {ascl} {1109.001})

\bibitem[\protect\citeauthoryear{{Ginsburg}, {Sokolov}, {de Val-Borro}, {Rosolowsky}, {Pineda}, {Sip{\H{o}}cz}  \& {Henshaw}}{{Ginsburg} et~al.}{2022}]{pyspeckit2022}
{Ginsburg} A.,  {Sokolov} V.,  {de Val-Borro} M.,  {Rosolowsky} E.,  {Pineda} J.~E.,  {Sip{\H{o}}cz} B.~M.,   {Henshaw} J.~D.,  2022, \mn@doi [\aj] {10.3847/1538-3881/ac695a}, \href {https://ui.adsabs.harvard.edu/abs/2022AJ....163..291G} {163, 291}

\bibitem[\protect\citeauthoryear{{G{\'o}mez} \& {V{\'a}zquez-Semadeni}}{{G{\'o}mez} \& {V{\'a}zquez-Semadeni}}{2014}]{gomez2014}
{G{\'o}mez} G.~C.,  {V{\'a}zquez-Semadeni} E.,  2014, \mn@doi [\apj] {10.1088/0004-637X/791/2/124}, \href {https://ui.adsabs.harvard.edu/abs/2014ApJ...791..124G} {791, 124}

\bibitem[\protect\citeauthoryear{{Goodman}, {Benson}, {Fuller}  \& {Myers}}{{Goodman} et~al.}{1993}]{goodman1993}
{Goodman} A.~A.,  {Benson} P.~J.,  {Fuller} G.~A.,   {Myers} P.~C.,  1993, \mn@doi [\apj] {10.1086/172465}, \href {https://ui.adsabs.harvard.edu/abs/1993ApJ...406..528G} {406, 528}

\bibitem[\protect\citeauthoryear{{Goodman}, {Rosolowsky}, {Borkin}, {Foster}, {Halle}, {Kauffmann}  \& {Pineda}}{{Goodman} et~al.}{2009}]{goodman2009}
{Goodman} A.~A.,  {Rosolowsky} E.~W.,  {Borkin} M.~A.,  {Foster} J.~B.,  {Halle} M.,  {Kauffmann} J.,   {Pineda} J.~E.,  2009, \mn@doi [\nat] {10.1038/nature07609}, \href {https://ui.adsabs.harvard.edu/abs/2009Natur.457...63G} {457, 63}

\bibitem[\protect\citeauthoryear{{Hacar}, {Alves}, {Tafalla}  \& {Goicoechea}}{{Hacar} et~al.}{2017}]{hacar2017}
{Hacar} A.,  {Alves} J.,  {Tafalla} M.,   {Goicoechea} J.~R.,  2017, \mn@doi [\aap] {10.1051/0004-6361/201730732}, \href {https://ui.adsabs.harvard.edu/abs/2017A&A...602L...2H} {602, L2}

\bibitem[\protect\citeauthoryear{{Hacar}, {Tafalla}, {Forbrich}, {Alves}, {Meingast}, {Grossschedl}  \& {Teixeira}}{{Hacar} et~al.}{2018}]{hacar2018}
{Hacar} A.,  {Tafalla} M.,  {Forbrich} J.,  {Alves} J.,  {Meingast} S.,  {Grossschedl} J.,   {Teixeira} P.~S.,  2018, \mn@doi [\aap] {10.1051/0004-6361/201731894}, \href {https://ui.adsabs.harvard.edu/abs/2018A&A...610A..77H} {610, A77}

\bibitem[\protect\citeauthoryear{{Harris} et~al.,}{{Harris} et~al.}{2020}]{numpy2020}
{Harris} C.~R.,  et~al., 2020, \mn@doi [\nat] {10.1038/s41586-020-2649-2}, \href {https://ui.adsabs.harvard.edu/abs/2020Natur.585..357H} {585, 357}

\bibitem[\protect\citeauthoryear{{Hennemann} et~al.,}{{Hennemann} et~al.}{2012}]{hennemann2012}
{Hennemann} M.,  et~al., 2012, \mn@doi [\aap] {10.1051/0004-6361/201219429}, \href {https://ui.adsabs.harvard.edu/abs/2012A&A...543L...3H} {543, L3}

\bibitem[\protect\citeauthoryear{{Henshaw}, {Caselli}, {Fontani}, {Jim{\'e}nez-Serra}, {Tan}  \& {Hernandez}}{{Henshaw} et~al.}{2013}]{henshaw2013}
{Henshaw} J.~D.,  {Caselli} P.,  {Fontani} F.,  {Jim{\'e}nez-Serra} I.,  {Tan} J.~C.,   {Hernandez} A.~K.,  2013, \mn@doi [\mnras] {10.1093/mnras/sts282}, \href {https://ui.adsabs.harvard.edu/abs/2013MNRAS.428.3425H} {428, 3425}

\bibitem[\protect\citeauthoryear{{Henshaw}, {Caselli}, {Fontani}, {Jim{\'e}nez-Serra}  \& {Tan}}{{Henshaw} et~al.}{2014}]{henshaw2014}
{Henshaw} J.~D.,  {Caselli} P.,  {Fontani} F.,  {Jim{\'e}nez-Serra} I.,   {Tan} J.~C.,  2014, \mn@doi [\mnras] {10.1093/mnras/stu446}, \href {https://ui.adsabs.harvard.edu/abs/2014MNRAS.440.2860H} {440, 2860}

\bibitem[\protect\citeauthoryear{{Henshaw} et~al.,}{{Henshaw} et~al.}{2016}]{henshaw2016a}
{Henshaw} J.~D.,  et~al., 2016, \mn@doi [\mnras] {10.1093/mnras/stw121}, \href {https://ui.adsabs.harvard.edu/abs/2016MNRAS.457.2675H} {457, 2675}

\bibitem[\protect\citeauthoryear{{Henshaw} et~al.,}{{Henshaw} et~al.}{2017}]{Henshaw2017}
{Henshaw} J.~D.,  et~al., 2017, \mn@doi [\mnras] {10.1093/mnrasl/slw154}, \href {https://ui.adsabs.harvard.edu/abs/2017MNRAS.464L..31H} {464, L31}

\bibitem[\protect\citeauthoryear{{Henshaw} et~al.,}{{Henshaw} et~al.}{2019}]{henshaw2019}
{Henshaw} J.~D.,  et~al., 2019, \mn@doi [\mnras] {10.1093/mnras/stz471}, \href {https://ui.adsabs.harvard.edu/abs/2019MNRAS.485.2457H} {485, 2457}

\bibitem[\protect\citeauthoryear{{Hofner} \& {Churchwell}}{{Hofner} \& {Churchwell}}{1996}]{hofner1996}
{Hofner} P.,  {Churchwell} E.,  1996, \aaps, \href {https://ui.adsabs.harvard.edu/abs/1996A&AS..120..283H} {120, 283}

\bibitem[\protect\citeauthoryear{{Holdship} et~al.,}{{Holdship} et~al.}{2021}]{holdship2021}
{Holdship} J.,  et~al., 2021, \mn@doi [\aap] {10.1051/0004-6361/202141233}, \href {https://ui.adsabs.harvard.edu/abs/2021A&A...654A..55H} {654, A55}

\bibitem[\protect\citeauthoryear{{Hu} et~al.,}{{Hu} et~al.}{2021}]{hu2021}
{Hu} B.,  et~al., 2021, \mn@doi [\apj] {10.3847/1538-4357/abd03a}, \href {https://ui.adsabs.harvard.edu/abs/2021ApJ...908...70H} {908, 70}

\bibitem[\protect\citeauthoryear{Hunter}{Hunter}{2007}]{Hunter2007}
Hunter J.~D.,  2007, \mn@doi [Computing in Science \& Engineering] {10.1109/MCSE.2007.55}, 9, 90

\bibitem[\protect\citeauthoryear{{Hunter}, {Brogan}, {Cyganowski}  \& {Schnee}}{{Hunter} et~al.}{2015}]{hunter2015}
{Hunter} T.~R.,  {Brogan} C.~L.,  {Cyganowski} C.~J.,   {Schnee} S.,  2015, in EAS Publications Series. pp 285--286, \mn@doi{10.1051/eas/1575057}

\bibitem[\protect\citeauthoryear{{Hunter}, {Petry}, {Barkats}, {Corder}  \& {Indebetouw}}{{Hunter} et~al.}{2023}]{au2023}
{Hunter} T.~R.,  {Petry} D.,  {Barkats} D.,  {Corder} S.,   {Indebetouw} R.,  2023, {analysisUtils}, Zenodo, \mn@doi{10.5281/zenodo.7502160}

\bibitem[\protect\citeauthoryear{{Ilee}, {Cyganowski}, {Nazari}, {Hunter}, {Brogan}, {Forgan}  \& {Zhang}}{{Ilee} et~al.}{2016}]{ilee2016}
{Ilee} J.~D.,  {Cyganowski} C.~J.,  {Nazari} P.,  {Hunter} T.~R.,  {Brogan} C.~L.,  {Forgan} D.~H.,   {Zhang} Q.,  2016, \mn@doi [\mnras] {10.1093/mnras/stw1912}, \href {https://ui.adsabs.harvard.edu/abs/2016MNRAS.462.4386I} {462, 4386}

\bibitem[\protect\citeauthoryear{{Ilee}, {Cyganowski}, {Brogan}, {Hunter}, {Forgan}, {Haworth}, {Clarke}  \& {Harries}}{{Ilee} et~al.}{2018}]{ilee2018}
{Ilee} J.~D.,  {Cyganowski} C.~J.,  {Brogan} C.~L.,  {Hunter} T.~R.,  {Forgan} D.~H.,  {Haworth} T.~J.,  {Clarke} C.~J.,   {Harries} T.~J.,  2018, \mn@doi [\apj] {10.3847/2041-8213/aaeffc}, \href {https://ui.adsabs.harvard.edu/abs/2018ApJ...869L..24I} {869, L24}

\bibitem[\protect\citeauthoryear{{Jackson}, {Finn}, {Chambers}, {Rathborne}  \& {Simon}}{{Jackson} et~al.}{2010}]{jackson2010}
{Jackson} J.~M.,  {Finn} S.~C.,  {Chambers} E.~T.,  {Rathborne} J.~M.,   {Simon} R.,  2010, \mn@doi [\apjl] {10.1088/2041-8205/719/2/L185}, \href {https://ui.adsabs.harvard.edu/abs/2010ApJ...719L.185J} {719, L185}

\bibitem[\protect\citeauthoryear{{Kainulainen}, {Ragan}, {Henning}  \& {Stutz}}{{Kainulainen} et~al.}{2013}]{Kainulainen2013}
{Kainulainen} J.,  {Ragan} S.~E.,  {Henning} T.,   {Stutz} A.,  2013, \mn@doi [\aap] {10.1051/0004-6361/201321760}, \href {https://ui.adsabs.harvard.edu/abs/2013A&A...557A.120K} {557, A120}

\bibitem[\protect\citeauthoryear{{Kauffmann} \& {Pillai}}{{Kauffmann} \& {Pillai}}{2010}]{Kauffmann2010}
{Kauffmann} J.,  {Pillai} T.,  2010, \mn@doi [\apjl] {10.1088/2041-8205/723/1/L7}, \href {https://ui.adsabs.harvard.edu/abs/2010ApJ...723L...7K} {723, L7}

\bibitem[\protect\citeauthoryear{{Keown} et~al.,}{{Keown} et~al.}{2019}]{keown2019}
{Keown} J.,  et~al., 2019, \mn@doi [\apj] {10.3847/1538-4357/ab3e76}, \href {https://ui.adsabs.harvard.edu/abs/2019ApJ...884....4K} {884, 4}

\bibitem[\protect\citeauthoryear{{Kirk}, {Myers}, {Bourke}, {Gutermuth}, {Hedden}  \& {Wilson}}{{Kirk} et~al.}{2013}]{kirk2013}
{Kirk} H.,  {Myers} P.~C.,  {Bourke} T.~L.,  {Gutermuth} R.~A.,  {Hedden} A.,   {Wilson} G.~W.,  2013, \mn@doi [\apj] {10.1088/0004-637X/766/2/115}, \href {https://ui.adsabs.harvard.edu/abs/2013ApJ...766..115K} {766, 115}

\bibitem[\protect\citeauthoryear{{Kong} et~al.,}{{Kong} et~al.}{2016}]{kong2016}
{Kong} S.,  et~al., 2016, \mn@doi [\apj] {10.3847/0004-637X/821/2/94}, \href {https://ui.adsabs.harvard.edu/abs/2016ApJ...821...94K} {821, 94}

\bibitem[\protect\citeauthoryear{{Kong}, {Tan}, {Caselli}, {Fontani}, {Liu}  \& {Butler}}{{Kong} et~al.}{2017}]{kong2017}
{Kong} S.,  {Tan} J.~C.,  {Caselli} P.,  {Fontani} F.,  {Liu} M.,   {Butler} M.~J.,  2017, \mn@doi [\apj] {10.3847/1538-4357/834/2/193}, \href {https://ui.adsabs.harvard.edu/abs/2017ApJ...834..193K} {834, 193}

\bibitem[\protect\citeauthoryear{{Kong}, {Arce}, {Tobin}, {Zhang}, {Maureira}, {Kratter}  \& {Pillai}}{{Kong} et~al.}{2023}]{kong2023}
{Kong} S.,  {Arce} H.~G.,  {Tobin} J.~J.,  {Zhang} Y.,  {Maureira} M.~J.,  {Kratter} K.~M.,   {Pillai} T. G.~S.,  2023, \mn@doi [\apj] {10.3847/1538-4357/acd252}, \href {https://ui.adsabs.harvard.edu/abs/2023ApJ...950..187K} {950, 187}

\bibitem[\protect\citeauthoryear{{K{\"o}nig} et~al.,}{{K{\"o}nig} et~al.}{2017}]{konig2017}
{K{\"o}nig} C.,  et~al., 2017, \mn@doi [\aap] {10.1051/0004-6361/201526841}, \href {https://ui.adsabs.harvard.edu/abs/2017A&A...599A.139K} {599, A139}

\bibitem[\protect\citeauthoryear{{K{\"o}nyves} et~al.,}{{K{\"o}nyves} et~al.}{2020}]{konyves2020}
{K{\"o}nyves} V.,  et~al., 2020, \mn@doi [\aap] {10.1051/0004-6361/201834753}, \href {https://ui.adsabs.harvard.edu/abs/2020A&A...635A..34K} {635, A34}

\bibitem[\protect\citeauthoryear{{Koumpia}, {Evans}, {Di Francesco}, {van der Tak}  \& {Oudmaijer}}{{Koumpia} et~al.}{2020}]{koumpia2020}
{Koumpia} E.,  {Evans} L.,  {Di Francesco} J.,  {van der Tak} F.~F.~S.,   {Oudmaijer} R.~D.,  2020, \mn@doi [\aap] {10.1051/0004-6361/202038457}, \href {https://ui.adsabs.harvard.edu/abs/2020A&A...643A..61K} {643, A61}

\bibitem[\protect\citeauthoryear{{Krumholz}, {Klein}, {McKee}, {Offner}  \& {Cunningham}}{{Krumholz} et~al.}{2009}]{krumholz09}
{Krumholz} M.~R.,  {Klein} R.~I.,  {McKee} C.~F.,  {Offner} S. S.~R.,   {Cunningham} A.~J.,  2009, \mn@doi [Science] {10.1126/science.1165857}, \href {https://ui.adsabs.harvard.edu/abs/2009Sci...323..754K} {323, 754}

\bibitem[\protect\citeauthoryear{{Kuiper} \& {Hosokawa}}{{Kuiper} \& {Hosokawa}}{2018}]{kuiper2018}
{Kuiper} R.,  {Hosokawa} T.,  2018, \mn@doi [\aap] {10.1051/0004-6361/201832638}, \href {https://ui.adsabs.harvard.edu/abs/2018A&A...616A.101K} {616, A101}

\bibitem[\protect\citeauthoryear{{Kuiper}, {Yorke}  \& {Turner}}{{Kuiper} et~al.}{2015}]{Kuiper2015}
{Kuiper} R.,  {Yorke} H.~W.,   {Turner} N.~J.,  2015, \mn@doi [\apj] {10.1088/0004-637X/800/2/86}, \href {https://ui.adsabs.harvard.edu/abs/2015ApJ...800...86K} {800, 86}

\bibitem[\protect\citeauthoryear{{Kumar}, {Palmeirim}, {Arzoumanian}  \& {Inutsuka}}{{Kumar} et~al.}{2020}]{Kumar2020}
{Kumar} M.~S.~N.,  {Palmeirim} P.,  {Arzoumanian} D.,   {Inutsuka} S.~I.,  2020, \mn@doi [\aap] {10.1051/0004-6361/202038232}, \href {https://ui.adsabs.harvard.edu/abs/2020A&A...642A..87K} {642, A87}

\bibitem[\protect\citeauthoryear{{Li} et~al.,}{{Li} et~al.}{2022}]{Li2022}
{Li} S.,  et~al., 2022, \mn@doi [\apj] {10.3847/1538-4357/ac3df8}, \href {https://ui.adsabs.harvard.edu/abs/2022ApJ...926..165L} {926, 165}

\bibitem[\protect\citeauthoryear{{Lindner} et~al.,}{{Lindner} et~al.}{2015}]{lindner2015}
{Lindner} R.~R.,  et~al., 2015, \mn@doi [\aj] {10.1088/0004-6256/149/4/138}, \href {https://ui.adsabs.harvard.edu/abs/2015AJ....149..138L} {149, 138}

\bibitem[\protect\citeauthoryear{{Liu} et~al.,}{{Liu} et~al.}{2016}]{liu2016}
{Liu} T.,  et~al., 2016, \mn@doi [\apj] {10.3847/0004-637X/824/1/31}, \href {https://ui.adsabs.harvard.edu/abs/2016ApJ...824...31L} {824, 31}

\bibitem[\protect\citeauthoryear{{Liu} et~al.,}{{Liu} et~al.}{2023}]{liu2023}
{Liu} H.-L.,  et~al., 2023, \mn@doi [\mnras] {10.1093/mnras/stad047}, \href {https://ui.adsabs.harvard.edu/abs/2023MNRAS.522.3719L} {522, 3719}

\bibitem[\protect\citeauthoryear{{Lu} et~al.,}{{Lu} et~al.}{2018}]{lu2018}
{Lu} X.,  et~al., 2018, \mn@doi [\apj] {10.3847/1538-4357/aaad11}, \href {https://ui.adsabs.harvard.edu/abs/2018ApJ...855....9L} {855, 9}

\bibitem[\protect\citeauthoryear{{Ma}, {Zhou}, {Esimbek}, {Ji}, {Wu}  \& {Yuan}}{{Ma} et~al.}{2013}]{ma2013}
{Ma} Y.,  {Zhou} J.,  {Esimbek} J.,  {Ji} W.,  {Wu} G.,   {Yuan} Y.,  2013, \mn@doi [\apss] {10.1007/s10509-013-1405-6}, \href {https://ui.adsabs.harvard.edu/abs/2013Ap&SS.345..297M} {345, 297}

\bibitem[\protect\citeauthoryear{{Maschberger}, {Clarke}, {Bonnell}  \& {Kroupa}}{{Maschberger} et~al.}{2010}]{maschberger2010}
{Maschberger} T.,  {Clarke} C.~J.,  {Bonnell} I.~A.,   {Kroupa} P.,  2010, \mn@doi [\mnras] {10.1111/j.1365-2966.2010.16346.x}, \href {https://ui.adsabs.harvard.edu/abs/2010MNRAS.404.1061M} {404, 1061}

\bibitem[\protect\citeauthoryear{McKee \& Tan}{McKee \& Tan}{2002}]{McKee:2002sz}
McKee C.~F.,  Tan J.~C.,  2002, \mn@doi [Nature] {10.1038/416059a}, 416, 59

\bibitem[\protect\citeauthoryear{{McKee} \& {Tan}}{{McKee} \& {Tan}}{2003}]{mc2003}
{McKee} C.~F.,  {Tan} J.~C.,  2003, \mn@doi [\apj] {10.1086/346149}, \href {https://ui.adsabs.harvard.edu/abs/2003ApJ...585..850M} {585, 850}

\bibitem[\protect\citeauthoryear{{Miettinen}}{{Miettinen}}{2012a}]{miettinen2012a}
{Miettinen} O.,  2012a, \mn@doi [\aap] {10.1051/0004-6361/201118552}, \href {https://ui.adsabs.harvard.edu/abs/2012A&A...540A.104M} {540, A104}

\bibitem[\protect\citeauthoryear{{Miettinen}}{{Miettinen}}{2012b}]{miettinen2012b}
{Miettinen} O.,  2012b, \mn@doi [\aap] {10.1051/0004-6361/201219144}, \href {https://ui.adsabs.harvard.edu/abs/2012A&A...542A.101M} {542, A101}

\bibitem[\protect\citeauthoryear{{Miettinen}}{{Miettinen}}{2020}]{miettinen2020}
{Miettinen} O.,  2020, \mn@doi [\aap] {10.1051/0004-6361/201936730}, \href {https://ui.adsabs.harvard.edu/abs/2020A&A...634A.115M} {634, A115}

\bibitem[\protect\citeauthoryear{{Mignon-Risse}, {Gonz{\'a}lez}  \& {Commer{\c{c}}on}}{{Mignon-Risse} et~al.}{2021}]{mignon2021}
{Mignon-Risse} R.,  {Gonz{\'a}lez} M.,   {Commer{\c{c}}on} B.,  2021, \mn@doi [\aap] {10.1051/0004-6361/202141648}, \href {https://ui.adsabs.harvard.edu/abs/2021A&A...656A..85M} {656, A85}

\bibitem[\protect\citeauthoryear{{Molet} et~al.,}{{Molet} et~al.}{2019}]{molet2019}
{Molet} J.,  et~al., 2019, \mn@doi [\aap] {10.1051/0004-6361/201935497}, \href {https://ui.adsabs.harvard.edu/abs/2019A&A...626A.132M} {626, A132}

\bibitem[\protect\citeauthoryear{{Mookerjea}, {Veena}, {G{\"u}sten}, {Wyrowski}  \& {Lasrado}}{{Mookerjea} et~al.}{2023}]{mookerjea2023}
{Mookerjea} B.,  {Veena} V.~S.,  {G{\"u}sten} R.,  {Wyrowski} F.,   {Lasrado} A.,  2023, \mn@doi [\mnras] {10.1093/mnras/stad215}, \href {https://ui.adsabs.harvard.edu/abs/2023MNRAS.520.2517M} {520, 2517}

\bibitem[\protect\citeauthoryear{{M{\"u}ller}, {Thorwirth}, {Roth}  \& {Winnewisser}}{{M{\"u}ller} et~al.}{2001}]{muller2001}
{M{\"u}ller} H.~S.~P.,  {Thorwirth} S.,  {Roth} D.~A.,   {Winnewisser} G.,  2001, \mn@doi [\aap] {10.1051/0004-6361:20010367}, \href {https://ui.adsabs.harvard.edu/abs/2001A&A...370L..49M} {370, L49}

\bibitem[\protect\citeauthoryear{{M{\"u}ller}, {Schl{\"o}der}, {Stutzki}  \& {Winnewisser}}{{M{\"u}ller} et~al.}{2005}]{muller2005}
{M{\"u}ller} H. S.~P.,  {Schl{\"o}der} F.,  {Stutzki} J.,   {Winnewisser} G.,  2005, \mn@doi [Journal of Molecular Structure] {10.1016/j.molstruc.2005.01.027}, \href {https://ui.adsabs.harvard.edu/abs/2005JMoSt.742..215M} {742, 215}

\bibitem[\protect\citeauthoryear{{Myers}}{{Myers}}{2009}]{myers2009}
{Myers} P.~C.,  2009, \mn@doi [\apj] {10.1088/0004-637X/700/2/1609}, \href {https://ui.adsabs.harvard.edu/abs/2009ApJ...700.1609M} {700, 1609}

\bibitem[\protect\citeauthoryear{{Myers}, {McKee}, {Cunningham}, {Klein}  \& {Krumholz}}{{Myers} et~al.}{2013}]{myers2013}
{Myers} A.~T.,  {McKee} C.~F.,  {Cunningham} A.~J.,  {Klein} R.~I.,   {Krumholz} M.~R.,  2013, \mn@doi [\apj] {10.1088/0004-637X/766/2/97}, \href {https://ui.adsabs.harvard.edu/abs/2013ApJ...766...97M} {766, 97}

\bibitem[\protect\citeauthoryear{{Nony} et~al.,}{{Nony} et~al.}{2018}]{nony2018}
{Nony} T.,  et~al., 2018, \mn@doi [\aap] {10.1051/0004-6361/201833863}, \href {https://ui.adsabs.harvard.edu/abs/2018A&A...618L...5N} {618, L5}

\bibitem[\protect\citeauthoryear{{{\"O}berg}, {van Broekhuizen}, {Fraser}, {Bisschop}, {van Dishoeck}  \& {Schlemmer}}{{{\"O}berg} et~al.}{2005}]{oberg2005}
{{\"O}berg} K.~I.,  {van Broekhuizen} F.,  {Fraser} H.~J.,  {Bisschop} S.~E.,  {van Dishoeck} E.~F.,   {Schlemmer} S.,  2005, \mn@doi [\apjl] {10.1086/428901}, \href {https://ui.adsabs.harvard.edu/abs/2005ApJ...621L..33O} {621, L33}

\bibitem[\protect\citeauthoryear{{Ossenkopf} \& {Henning}}{{Ossenkopf} \& {Henning}}{1994}]{ossenkopf1994}
{Ossenkopf} V.,  {Henning} T.,  1994, \aap, \href {https://ui.adsabs.harvard.edu/abs/1994A&A...291..943O} {291, 943}

\bibitem[\protect\citeauthoryear{{Padoan}, {Pan}, {Juvela}, {Haugb{\o}lle}  \& {Nordlund}}{{Padoan} et~al.}{2020}]{padoan2020}
{Padoan} P.,  {Pan} L.,  {Juvela} M.,  {Haugb{\o}lle} T.,   {Nordlund} {\r{A}}.,  2020, \mn@doi [\apj] {10.3847/1538-4357/abaa47}, \href {https://ui.adsabs.harvard.edu/abs/2020ApJ...900...82P} {900, 82}

\bibitem[\protect\citeauthoryear{{Pagani}, {Daniel}  \& {Dubernet}}{{Pagani} et~al.}{2009}]{pagani2009}
{Pagani} L.,  {Daniel} F.,   {Dubernet} M.~L.,  2009, \mn@doi [\aap] {10.1051/0004-6361:200810570}, \href {https://ui.adsabs.harvard.edu/abs/2009A&A...494..719P} {494, 719}

\bibitem[\protect\citeauthoryear{{Palmeirim} et~al.,}{{Palmeirim} et~al.}{2013}]{palmeirim2013}
{Palmeirim} P.,  et~al., 2013, \mn@doi [\aap] {10.1051/0004-6361/201220500}, \href {https://ui.adsabs.harvard.edu/abs/2013A&A...550A..38P} {550, A38}

\bibitem[\protect\citeauthoryear{{Peretto} \& {Fuller}}{{Peretto} \& {Fuller}}{2010}]{peretto2010}
{Peretto} N.,  {Fuller} G.~A.,  2010, \mn@doi [\apj] {10.1088/0004-637X/723/1/55510.48550/arXiv.1009.0716}, \href {https://ui.adsabs.harvard.edu/abs/2010ApJ...723..555P} {723, 555}

\bibitem[\protect\citeauthoryear{{Peretto} et~al.,}{{Peretto} et~al.}{2013}]{peretto2013}
{Peretto} N.,  et~al., 2013, \mn@doi [\aap] {10.1051/0004-6361/201321318}, \href {https://ui.adsabs.harvard.edu/abs/2013A&A...555A.112P} {555, A112}

\bibitem[\protect\citeauthoryear{{Peretto} et~al.,}{{Peretto} et~al.}{2014}]{peretto2014}
{Peretto} N.,  et~al., 2014, \mn@doi [\aap] {10.1051/0004-6361/201322172}, \href {https://ui.adsabs.harvard.edu/abs/2014A&A...561A..83P} {561, A83}

\bibitem[\protect\citeauthoryear{{Pillai}, {Caselli}, {Kauffmann}, {Zhang}, {Thompson}  \& {Lis}}{{Pillai} et~al.}{2012}]{pillai2012}
{Pillai} T.,  {Caselli} P.,  {Kauffmann} J.,  {Zhang} Q.,  {Thompson} M.~A.,   {Lis} D.~C.,  2012, \mn@doi [\apj] {10.1088/0004-637X/751/2/135}, \href {https://ui.adsabs.harvard.edu/abs/2012ApJ...751..135P} {751, 135}

\bibitem[\protect\citeauthoryear{{Qi} et~al.,}{{Qi} et~al.}{2013}]{qioberg2013}
{Qi} C.,  et~al., 2013, \mn@doi [Science] {10.1126/science.1239560}, \href {https://ui.adsabs.harvard.edu/abs/2013Sci...341..630Q} {341, 630}

\bibitem[\protect\citeauthoryear{{Ragan}, {Heitsch}, {Bergin}  \& {Wilner}}{{Ragan} et~al.}{2012}]{ragan2012a}
{Ragan} S.~E.,  {Heitsch} F.,  {Bergin} E.~A.,   {Wilner} D.,  2012, \mn@doi [\apj] {10.1088/0004-637X/746/2/174}, \href {https://ui.adsabs.harvard.edu/abs/2012ApJ...746..174R} {746, 174}

\bibitem[\protect\citeauthoryear{{Rathborne}, {Jackson}  \& {Simon}}{{Rathborne} et~al.}{2006}]{rathborne2006}
{Rathborne} J.~M.,  {Jackson} J.~M.,   {Simon} R.,  2006, \mn@doi [\apj] {10.1086/500423}, \href {https://ui.adsabs.harvard.edu/abs/2006ApJ...641..389R} {641, 389}

\bibitem[\protect\citeauthoryear{{Redaelli}, {Bovino}, {Giannetti}, {Sabatini}, {Caselli}, {Wyrowski}, {Schleicher}  \& {Colombo}}{{Redaelli} et~al.}{2021}]{redaelli2021}
{Redaelli} E.,  {Bovino} S.,  {Giannetti} A.,  {Sabatini} G.,  {Caselli} P.,  {Wyrowski} F.,  {Schleicher} D.~R.~G.,   {Colombo} D.,  2021, \mn@doi [\aap] {10.1051/0004-6361/202140694}, \href {https://ui.adsabs.harvard.edu/abs/2021A&A...650A.202R} {650, A202}

\bibitem[\protect\citeauthoryear{{Redaelli}, {Bovino}, {Sanhueza}, {Morii}, {Sabatini}, {Caselli}, {Giannetti}  \& {Li}}{{Redaelli} et~al.}{2022}]{redaelli2022}
{Redaelli} E.,  {Bovino} S.,  {Sanhueza} P.,  {Morii} K.,  {Sabatini} G.,  {Caselli} P.,  {Giannetti} A.,   {Li} S.,  2022, \mn@doi [\apj] {10.3847/1538-4357/ac85b4}, \href {https://ui.adsabs.harvard.edu/abs/2022ApJ...936..169R} {936, 169}

\bibitem[\protect\citeauthoryear{{Riener}, {Kainulainen}, {Henshaw}, {Orkisz}, {Murray}  \& {Beuther}}{{Riener} et~al.}{2019}]{riener2019}
{Riener} M.,  {Kainulainen} J.,  {Henshaw} J.~D.,  {Orkisz} J.~H.,  {Murray} C.~E.,   {Beuther} H.,  2019, \mn@doi [\aap] {10.1051/0004-6361/201935519}, \href {https://ui.adsabs.harvard.edu/abs/2019A&A...628A..78R} {628, A78}

\bibitem[\protect\citeauthoryear{{Robitaille} \& {Bressert}}{{Robitaille} \& {Bressert}}{2012}]{robitaille2012}
{Robitaille} T.,  {Bressert} E.,  2012, {APLpy: Astronomical Plotting Library in Python}, Astrophysics Source Code Library, record ascl:1208.017 (\mn@eprint {ascl} {1208.017})

\bibitem[\protect\citeauthoryear{{Rosen}}{{Rosen}}{2022}]{rosen2022}
{Rosen} A.~L.,  2022, \mn@doi [\apj] {10.3847/1538-4357/ac9f3d}, \href {https://ui.adsabs.harvard.edu/abs/2022ApJ...941..202R} {941, 202}

\bibitem[\protect\citeauthoryear{{Rosen} \& {Krumholz}}{{Rosen} \& {Krumholz}}{2020}]{rosen2020}
{Rosen} A.~L.,  {Krumholz} M.~R.,  2020, \mn@doi [\aj] {10.3847/1538-3881/ab9abf}, \href {https://ui.adsabs.harvard.edu/abs/2020AJ....160...78R} {160, 78}

\bibitem[\protect\citeauthoryear{{Rosen}, {Krumholz}, {McKee}  \& {Klein}}{{Rosen} et~al.}{2016}]{rosen2016}
{Rosen} A.~L.,  {Krumholz} M.~R.,  {McKee} C.~F.,   {Klein} R.~I.,  2016, \mn@doi [\mnras] {10.1093/mnras/stw2153}, \href {https://ui.adsabs.harvard.edu/abs/2016MNRAS.463.2553R} {463, 2553}

\bibitem[\protect\citeauthoryear{{Rosen}, {Li}, {Zhang}  \& {Burkhart}}{{Rosen} et~al.}{2019}]{rosen2019}
{Rosen} A.~L.,  {Li} P.~S.,  {Zhang} Q.,   {Burkhart} B.,  2019, \mn@doi [\apj] {10.3847/1538-4357/ab54c6}, \href {https://ui.adsabs.harvard.edu/abs/2019ApJ...887..108R} {887, 108}

\bibitem[\protect\citeauthoryear{{Rosolowsky}, {Pineda}, {Kauffmann}  \& {Goodman}}{{Rosolowsky} et~al.}{2008}]{rosolowsky2008}
{Rosolowsky} E.~W.,  {Pineda} J.~E.,  {Kauffmann} J.,   {Goodman} A.~A.,  2008, \mn@doi [\apj] {10.1086/587685}, \href {https://ui.adsabs.harvard.edu/abs/2008ApJ...679.1338R} {679, 1338}

\bibitem[\protect\citeauthoryear{{Sabatini} et~al.,}{{Sabatini} et~al.}{2021}]{sabatini21}
{Sabatini} G.,  et~al., 2021, \mn@doi [\aap] {10.1051/0004-6361/202140469}, \href {https://ui.adsabs.harvard.edu/abs/2021A&A...652A..71S} {652, A71}

\bibitem[\protect\citeauthoryear{{Sato} et~al.,}{{Sato} et~al.}{2014}]{sato2014}
{Sato} M.,  et~al., 2014, \mn@doi [\apj] {10.1088/0004-637X/793/2/72}, \href {https://ui.adsabs.harvard.edu/abs/2014ApJ...793...72S} {793, 72}

\bibitem[\protect\citeauthoryear{{Schneider}, {Csengeri}, {Bontemps}, {Motte}, {Simon}, {Hennebelle}, {Federrath}  \& {Klessen}}{{Schneider} et~al.}{2010}]{schneider2010}
{Schneider} N.,  {Csengeri} T.,  {Bontemps} S.,  {Motte} F.,  {Simon} R.,  {Hennebelle} P.,  {Federrath} C.,   {Klessen} R.,  2010, \mn@doi [\aap] {10.1051/0004-6361/201014481}, \href {https://ui.adsabs.harvard.edu/abs/2010A&A...520A..49S} {520, A49}

\bibitem[\protect\citeauthoryear{{Schneider} et~al.,}{{Schneider} et~al.}{2012}]{schneider2012}
{Schneider} N.,  et~al., 2012, \mn@doi [\aap] {10.1051/0004-6361/201118566}, \href {https://ui.adsabs.harvard.edu/abs/2012A&A...540L..11S} {540, L11}

\bibitem[\protect\citeauthoryear{{Sch{\"o}ier}, {van der Tak}, {van Dishoeck}  \& {Black}}{{Sch{\"o}ier} et~al.}{2005}]{schoier2005}
{Sch{\"o}ier} F.~L.,  {van der Tak} F.~F.~S.,  {van Dishoeck} E.~F.,   {Black} J.~H.,  2005, \mn@doi [\aap] {10.1051/0004-6361:20041729}, \href {https://ui.adsabs.harvard.edu/abs/2005A&A...432..369S} {432, 369}

\bibitem[\protect\citeauthoryear{{Shimajiri}, {Andr{\'e}}, {Palmeirim}, {Arzoumanian}, {Bracco}, {K{\"o}nyves}, {Ntormousi}  \& {Ladjelate}}{{Shimajiri} et~al.}{2019a}]{shimajiri2019a}
{Shimajiri} Y.,  {Andr{\'e}} P.,  {Palmeirim} P.,  {Arzoumanian} D.,  {Bracco} A.,  {K{\"o}nyves} V.,  {Ntormousi} E.,   {Ladjelate} B.,  2019a, \mn@doi [\aap] {10.1051/0004-6361/201834399}, \href {https://ui.adsabs.harvard.edu/abs/2019A&A...623A..16S} {623, A16}

\bibitem[\protect\citeauthoryear{{Shimajiri}, {Andr{\'e}}, {Ntormousi}, {Men'shchikov}, {Arzoumanian}  \& {Palmeirim}}{{Shimajiri} et~al.}{2019b}]{shimajiri2019b}
{Shimajiri} Y.,  {Andr{\'e}} P.,  {Ntormousi} E.,  {Men'shchikov} A.,  {Arzoumanian} D.,   {Palmeirim} P.,  2019b, \mn@doi [\aap] {10.1051/0004-6361/201935689}, \href {https://ui.adsabs.harvard.edu/abs/2019A&A...632A..83S} {632, A83}

\bibitem[\protect\citeauthoryear{{Smilgys} \& {Bonnell}}{{Smilgys} \& {Bonnell}}{2016}]{smilgys2016}
{Smilgys} R.,  {Bonnell} I.~A.,  2016, \mn@doi [\mnras] {10.1093/mnras/stw791}, \href {https://ui.adsabs.harvard.edu/abs/2016MNRAS.459.1985S} {459, 1985}

\bibitem[\protect\citeauthoryear{{Smilgys} \& {Bonnell}}{{Smilgys} \& {Bonnell}}{2017}]{smilgys2017}
{Smilgys} R.,  {Bonnell} I.~A.,  2017, \mn@doi [\mnras] {10.1093/mnras/stx2396}, \href {https://ui.adsabs.harvard.edu/abs/2017MNRAS.472.4982S} {472, 4982}

\bibitem[\protect\citeauthoryear{{Smith}, {Longmore}  \& {Bonnell}}{{Smith} et~al.}{2009}]{smith2009}
{Smith} R.~J.,  {Longmore} S.,   {Bonnell} I.,  2009, \mn@doi [\mnras] {10.1111/j.1365-2966.2009.15621.x}, \href {https://ui.adsabs.harvard.edu/abs/2009MNRAS.400.1775S} {400, 1775}

\bibitem[\protect\citeauthoryear{{Smith}, {Glover}  \& {Klessen}}{{Smith} et~al.}{2014}]{smith2014}
{Smith} R.~J.,  {Glover} S. C.~O.,   {Klessen} R.~S.,  2014, \mn@doi [\mnras] {10.1093/mnras/stu1915}, \href {https://ui.adsabs.harvard.edu/abs/2014MNRAS.445.2900S} {445, 2900}

\bibitem[\protect\citeauthoryear{{Smith}, {Glover}, {Klessen}  \& {Fuller}}{{Smith} et~al.}{2016}]{smith2016}
{Smith} R.~J.,  {Glover} S. C.~O.,  {Klessen} R.~S.,   {Fuller} G.~A.,  2016, \mn@doi [\mnras] {10.1093/mnras/stv2559}, \href {https://ui.adsabs.harvard.edu/abs/2016MNRAS.455.3640S} {455, 3640}

\bibitem[\protect\citeauthoryear{{Stark}, {van der Tak}  \& {van Dishoeck}}{{Stark} et~al.}{1999}]{stark1999}
{Stark} R.,  {van der Tak} F. F.~S.,   {van Dishoeck} E.~F.,  1999, \mn@doi [\apjl] {10.1086/312182}, \href {https://ui.adsabs.harvard.edu/abs/1999ApJ...521L..67S} {521, L67}

\bibitem[\protect\citeauthoryear{{Tackenberg} et~al.,}{{Tackenberg} et~al.}{2014}]{tackenberg2014}
{Tackenberg} J.,  et~al., 2014, \mn@doi [\aap] {10.1051/0004-6361/201321555}, \href {https://ui.adsabs.harvard.edu/abs/2014A&A...565A.101T} {565, A101}

\bibitem[\protect\citeauthoryear{{Tan}, {Kong}, {Butler}, {Caselli}  \& {Fontani}}{{Tan} et~al.}{2013}]{tan2013}
{Tan} J.~C.,  {Kong} S.,  {Butler} M.~J.,  {Caselli} P.,   {Fontani} F.,  2013, \mn@doi [\apj] {10.1088/0004-637X/779/2/96}, \href {https://ui.adsabs.harvard.edu/abs/2013ApJ...779...96T} {779, 96}

\bibitem[\protect\citeauthoryear{{Tan}, {Beltr{\'a}n}, {Caselli}, {Fontani}, {Fuente}, {Krumholz}, {McKee}  \& {Stolte}}{{Tan} et~al.}{2014}]{tan2014}
{Tan} J.~C.,  {Beltr{\'a}n} M.~T.,  {Caselli} P.,  {Fontani} F.,  {Fuente} A.,  {Krumholz} M.~R.,  {McKee} C.~F.,   {Stolte} A.,  2014, in {Beuther} H.,  {Klessen} R.~S.,  {Dullemond} C.~P.,   {Henning} T.,  eds, Protostars and Planets VI. pp 149--172 (\mn@eprint {arXiv} {1402.0919}), \mn@doi{10.2458/azu_uapress_9780816531240-ch007}

\bibitem[\protect\citeauthoryear{{Tan}, {Kong}, {Zhang}, {Fontani}, {Caselli}  \& {Butler}}{{Tan} et~al.}{2016}]{tan2016}
{Tan} J.~C.,  {Kong} S.,  {Zhang} Y.,  {Fontani} F.,  {Caselli} P.,   {Butler} M.~J.,  2016, \mn@doi [\apj] {10.3847/2041-8205/821/1/L3}, \href {https://ui.adsabs.harvard.edu/abs/2016ApJ...821L...3T} {821, L3}

\bibitem[\protect\citeauthoryear{{Thwala}, {Shafi}, {Colafrancesco}, {Govoni}  \& {Murgia}}{{Thwala} et~al.}{2019}]{thwala2019}
{Thwala} S.~A.,  {Shafi} N.,  {Colafrancesco} S.,  {Govoni} F.,   {Murgia} M.,  2019, \mn@doi [\mnras] {10.1093/mnras/stz347}, \href {https://ui.adsabs.harvard.edu/abs/2019MNRAS.485.1938T} {485, 1938}

\bibitem[\protect\citeauthoryear{{Tig{\'e}} et~al.,}{{Tig{\'e}} et~al.}{2017}]{tige2017}
{Tig{\'e}} J.,  et~al., 2017, \mn@doi [\aap] {10.1051/0004-6361/201628989}, \href {https://ui.adsabs.harvard.edu/abs/2017A&A...602A..77T} {602, A77}

\bibitem[\protect\citeauthoryear{{Trevi{\~n}o-Morales} et~al.,}{{Trevi{\~n}o-Morales} et~al.}{2019}]{trevin2019}
{Trevi{\~n}o-Morales} S.~P.,  et~al., 2019, \mn@doi [\aap] {10.1051/0004-6361/201935260}, \href {https://ui.adsabs.harvard.edu/abs/2019A&A...629A..81T} {629, A81}

\bibitem[\protect\citeauthoryear{{Vastel}, {Caselli}, {Ceccarelli}, {Phillips}, {Wiedner}, {Peng}, {Houde}  \& {Dominik}}{{Vastel} et~al.}{2006}]{vastel2006}
{Vastel} C.,  {Caselli} P.,  {Ceccarelli} C.,  {Phillips} T.,  {Wiedner} M.~C.,  {Peng} R.,  {Houde} M.,   {Dominik} C.,  2006, \mn@doi [\apj] {10.1086/504371}, \href {https://ui.adsabs.harvard.edu/abs/2006ApJ...645.1198V} {645, 1198}

\bibitem[\protect\citeauthoryear{{Vasyunin}, {Semenov}, {Wiebe}  \& {Henning}}{{Vasyunin} et~al.}{2009}]{v2009}
{Vasyunin} A.~I.,  {Semenov} D.~A.,  {Wiebe} D.~S.,   {Henning} T.,  2009, \mn@doi [\apj] {10.1088/0004-637X/691/2/1459}, \href {https://ui.adsabs.harvard.edu/abs/2009ApJ...691.1459V} {691, 1459}

\bibitem[\protect\citeauthoryear{{V{\'a}zquez-Semadeni}, {G{\'o}mez}, {Jappsen}, {Ballesteros-Paredes}, {Gonz{\'a}lez}  \& {Klessen}}{{V{\'a}zquez-Semadeni} et~al.}{2007}]{vs2007}
{V{\'a}zquez-Semadeni} E.,  {G{\'o}mez} G.~C.,  {Jappsen} A.~K.,  {Ballesteros-Paredes} J.,  {Gonz{\'a}lez} R.~F.,   {Klessen} R.~S.,  2007, \mn@doi [\apj] {10.1086/510771}, \href {https://ui.adsabs.harvard.edu/abs/2007ApJ...657..870V} {657, 870}

\bibitem[\protect\citeauthoryear{{V{\'a}zquez-Semadeni}, {Gonz{\'a}lez-Samaniego}  \& {Col{\'\i}n}}{{V{\'a}zquez-Semadeni} et~al.}{2017}]{vs2017}
{V{\'a}zquez-Semadeni} E.,  {Gonz{\'a}lez-Samaniego} A.,   {Col{\'\i}n} P.,  2017, \mn@doi [\mnras] {10.1093/mnras/stw3229}, \href {https://ui.adsabs.harvard.edu/abs/2017MNRAS.467.1313V} {467, 1313}

\bibitem[\protect\citeauthoryear{{V{\'a}zquez-Semadeni}, {Palau}, {Ballesteros-Paredes}, {G{\'o}mez}  \& {Zamora-Avil{\'e}s}}{{V{\'a}zquez-Semadeni} et~al.}{2019}]{vs2019}
{V{\'a}zquez-Semadeni} E.,  {Palau} A.,  {Ballesteros-Paredes} J.,  {G{\'o}mez} G.~C.,   {Zamora-Avil{\'e}s} M.,  2019, \mn@doi [\mnras] {10.1093/mnras/stz2736}, \href {https://ui.adsabs.harvard.edu/abs/2019MNRAS.490.3061V} {490, 3061}

\bibitem[\protect\citeauthoryear{{Virtanen} et~al.,}{{Virtanen} et~al.}{2020}]{scipy2020}
{Virtanen} P.,  et~al., 2020, \mn@doi [Nature Methods] {10.1038/s41592-019-0686-2}, \href {https://ui.adsabs.harvard.edu/abs/2020NatMe..17..261V} {17, 261}

\bibitem[\protect\citeauthoryear{{Wang} et~al.,}{{Wang} et~al.}{2014}]{WANG2014}
{Wang} K.,  et~al., 2014, \mn@doi [\mnras] {10.1093/mnras/stu127}, \href {https://ui.adsabs.harvard.edu/abs/2014MNRAS.439.3275W} {439, 3275}

\bibitem[\protect\citeauthoryear{{Wang}, {Koch}, {Galv{\'a}n-Madrid}, {Lai}, {Liu}, {Lin}  \& {Pattle}}{{Wang} et~al.}{2020}]{wang2020}
{Wang} J.-W.,  {Koch} P.~M.,  {Galv{\'a}n-Madrid} R.,  {Lai} S.-P.,  {Liu} H.~B.,  {Lin} S.-J.,   {Pattle} K.,  2020, \mn@doi [\apj] {10.3847/1538-4357/abc74e}, \href {https://ui.adsabs.harvard.edu/abs/2020ApJ...905..158W} {905, 158}

\bibitem[\protect\citeauthoryear{{Williams}, {Peretto}, {Avison}, {Duarte-Cabral}  \& {Fuller}}{{Williams} et~al.}{2018}]{williams2018}
{Williams} G.~M.,  {Peretto} N.,  {Avison} A.,  {Duarte-Cabral} A.,   {Fuller} G.~A.,  2018, \mn@doi [\aap] {10.1051/0004-6361/201731587}, \href {https://ui.adsabs.harvard.edu/abs/2018A&A...613A..11W} {613, A11}

\bibitem[\protect\citeauthoryear{{Xu} et~al.,}{{Xu} et~al.}{2023}]{xu2023}
{Xu} F.-W.,  et~al., 2023, \mn@doi [\mnras] {10.1093/mnras/stad012}, \href {https://ui.adsabs.harvard.edu/abs/2023MNRAS.520.3259X} {520, 3259}

\bibitem[\protect\citeauthoryear{{Yuan} et~al.,}{{Yuan} et~al.}{2018}]{yuan2018}
{Yuan} J.,  et~al., 2018, \mn@doi [\apj] {10.3847/1538-4357/aa9d40}, \href {https://ui.adsabs.harvard.edu/abs/2018ApJ...852...12Y} {852, 12}

\bibitem[\protect\citeauthoryear{{Zamora-Avil{\'e}s}, {Ballesteros-Paredes}  \& {Hartmann}}{{Zamora-Avil{\'e}s} et~al.}{2017}]{za2017}
{Zamora-Avil{\'e}s} M.,  {Ballesteros-Paredes} J.,   {Hartmann} L.~W.,  2017, \mn@doi [\mnras] {10.1093/mnras/stx1995}, \href {https://ui.adsabs.harvard.edu/abs/2017MNRAS.472..647Z} {472, 647}

\bibitem[\protect\citeauthoryear{{Zernickel}, {Schilke}  \& {Smith}}{{Zernickel} et~al.}{2013}]{zernickel2013}
{Zernickel} A.,  {Schilke} P.,   {Smith} R.~J.,  2013, \mn@doi [\aap] {10.1051/0004-6361/201321425}, \href {https://ui.adsabs.harvard.edu/abs/2013A&A...554L...2Z} {554, L2}

\bibitem[\protect\citeauthoryear{{Zhang}, {Wang}, {Lu}  \& {Jim{\'e}nez-Serra}}{{Zhang} et~al.}{2015}]{zhang2015}
{Zhang} Q.,  {Wang} K.,  {Lu} X.,   {Jim{\'e}nez-Serra} I.,  2015, \mn@doi [\apj] {10.1088/0004-637X/804/2/141}, \href {https://ui.adsabs.harvard.edu/abs/2015ApJ...804..141Z} {804, 141}

\bibitem[\protect\citeauthoryear{{Zhou} et~al.,}{{Zhou} et~al.}{2021}]{zhou2021}
{Zhou} J.-W.,  et~al., 2021, \mn@doi [\mnras] {10.1093/mnras/stab2801}, \href {https://ui.adsabs.harvard.edu/abs/2021MNRAS.508.4639Z} {508, 4639}

\bibitem[\protect\citeauthoryear{{Zhou} et~al.,}{{Zhou} et~al.}{2022}]{zhou2022}
{Zhou} J.-W.,  et~al., 2022, \mn@doi [\mnras] {10.1093/mnras/stac1735}, \href {https://ui.adsabs.harvard.edu/abs/2022MNRAS.514.6038Z} {514, 6038}

\bibitem[\protect\citeauthoryear{{Zhou} et~al.,}{{Zhou} et~al.}{2023}]{zhou2023}
{Zhou} J.-W.,  et~al., 2023, \mn@doi [\mnras] {10.1093/mnras/stac3559}, \href {https://ui.adsabs.harvard.edu/abs/2023MNRAS.519.2391Z} {519, 2391}

\bibitem[\protect\citeauthoryear{{van Dishoeck}, {Phillips}, {Keene}  \& {Blake}}{{van Dishoeck} et~al.}{1992}]{van1992}
{van Dishoeck} E.~F.,  {Phillips} T.~G.,  {Keene} J.,   {Blake} G.~A.,  1992, \aap, \href {https://ui.adsabs.harvard.edu/abs/1992A&A...261L..13V} {261, L13}

\bibitem[\protect\citeauthoryear{{van der Tak}, {Caselli}  \& {Ceccarelli}}{{van der Tak} et~al.}{2005}]{van2005}
{van der Tak} F.~F.~S.,  {Caselli} P.,   {Ceccarelli} C.,  2005, \mn@doi [\aap] {10.1051/0004-6361:20052792}, \href {https://ui.adsabs.harvard.edu/abs/2005A&A...439..195V} {439, 195}

\bibitem[\protect\citeauthoryear{{van der Tak}, {Black}, {Sch{\"o}ier}, {Jansen}  \& {van Dishoeck}}{{van der Tak} et~al.}{2007}]{vandertak2007}
{van der Tak} F.~F.~S.,  {Black} J.~H.,  {Sch{\"o}ier} F.~L.,  {Jansen} D.~J.,   {van Dishoeck} E.~F.,  2007, \mn@doi [\aap] {10.1051/0004-6361:20066820}, \href {https://ui.adsabs.harvard.edu/abs/2007A&A...468..627V} {468, 627}

\bibitem[\protect\citeauthoryear{{van 't Hoff}, {Walsh}, {Kama}, {Facchini}  \& {van Dishoeck}}{{van 't Hoff} et~al.}{2017}]{vanthoff2017}
{van 't Hoff} M.~L.~R.,  {Walsh} C.,  {Kama} M.,  {Facchini} S.,   {van Dishoeck} E.~F.,  2017, \mn@doi [\aap] {10.1051/0004-6361/201629452}, \href {https://ui.adsabs.harvard.edu/abs/2017A&A...599A.101V} {599, A101}

\makeatother
\end{thebibliography}




\appendix

\section{Examples of \nhp\/ (4-3) spectra fitted with SCOUSEPY}
\begin{figure*}
	\includegraphics[width=\textwidth]
    {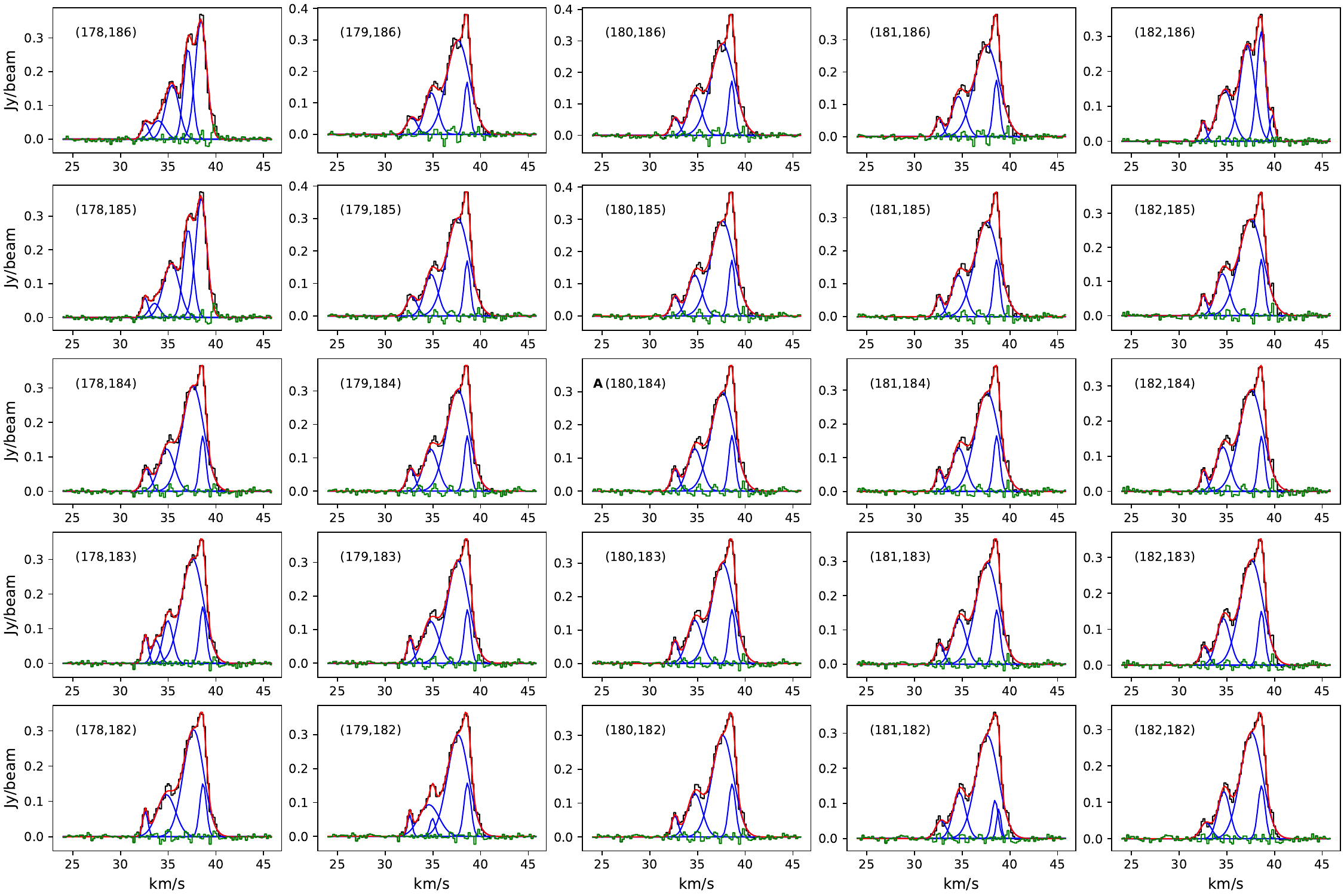}
    \caption{\nhp\/ (4-3) spectra extracted around the local peak in the integrated \nhp\/ emission to the north of MM2 (i.e.\ location A in Fig.~\ref{fig:nhp_spec}). The observed \nhp\/ spectra are shown in black, overlaid with individual best-fitting Gaussian components (blue), total best-fitting models (red), and residuals (green). Each panel is labelled with the pixel coordinates where the spectrum was extracted; the spectrum extracted at location A is additionally labelled with ``A''. 
    }
   \label{fig:fitted_spectra_sample}
\end{figure*}

Figure~\ref{fig:fitted_spectra_sample} shows a sample of the \nhp\/ (4-3) spectra that are extracted around the local peak in the integrated \nhp\/ emission to the north of MM2 (i.e. location A in Fig.~\ref{fig:nhp_spec}).  The observed spectra are overlaid with the SCOUSEPY fits obtained following the procedures described in Section~\ref{sec:gd_scousepy}.

\bsp	
\label{lastpage}
\end{CJK*}
\end{document}